\RequirePackage{fix-cm}
\documentclass[smallextended]{svjour3}       
\smartqed  
\usepackage{graphicx}
\usepackage{amsmath,amssymb,latexsym}

\usepackage{courier}        
\usepackage{graphicx}        
\usepackage{helvet}         

\usepackage{makeidx}         
\usepackage{mathptmx}       
 \usepackage{type1cm}       
 
\graphicspath{{Figures/}}
\usepackage{multicol}        

\usepackage[bottom]{footmisc}

\usepackage[dvipsnames]{xcolor}
\usepackage{graphicx}
\usepackage{natbib}
\usepackage[normalem]{ulem} 

\def\ala#1{$^{#1}$}

\def\alamenos#1{$^{-#1}$}

\newcommand{\alphavir}{\alpha_{\rm vir}}    

\def\diezala#1{10$^{#1}$}

\def\diezalamenos#1{10$^{-#1}$}

\newcommand{\Egrav}{E_{\rm grav}}    %
\newcommand{\Emag}{E_{\rm mag}}    %

\newcommand{\LL}{{\cal L}}

\newcommand{\MM}{{\cal{M}}}
\newcommand{\Msun}{$M_\odot$}    %
\newcommand{\Npdf}{${\rm PDF}_N$}    %
\newcommand{\Npdfmath}{{\rm PDF}_N}    %

\newcommand{\rhopdf}{${\rm PDF}_\rho$}

\newcommand{\sigmav}{\sigma_{\rm v}}

\def\tauintmath{{\tau_{\rm int}}}

\def\aapr{Astron. \&\ Astrophys. Rev.}

\def\apj{ApJ}
\def\apjl{ApJL}
\def\aap{A\& A}
\def\nat{Nature}
\def\araa{ARA\&A}
\def\mnras{MNRAS}

\def\aj{{AJ}}
\def\apjs{ApJS}
\def\bain{{Bull. Astron. Inst. Netherlands}}
\def\mnras{{MNRAS}}
\def\pasj{{PASJ}}
\def\pasp{{PASP}}

\def\pre{{Phys. Rev. E.}}
\def\rmxaa{{Rev. Mex. Astron. Astrofis.}}
\def\ssr{{Space Science Reviews}}

\begin{document}

\title{From diffuse gas to dense molecular cloud cores
}


\author{Javier Ballesteros-Paredes         \and
        Philippe Andr\'e,  \and
        Patrick Hennebelle,   \and
        Ralf S. Klessen,   \and
        Shu-ichiro Inutsuka,  \and
        J.~M.~Diederik~Kruijssen,  \and
        M\'{e}lanie Chevance,   \and 
        Fumitaka Nakamura,  \and
        Angela Adamo \and
        Enrique Vazquez-Semadeni
}


\institute{Javier Ballesteros-Paredes \at 
    Instituto de Radioastronom\'\i a y Astrof\'\i sica, UNAM, Campus Morelia, Antigua Carretera a Patzcuaro 8701. 58090 Morelia, Michoacan, Mexico. \email{j.ballesteros@irya.unam.mx}
\and Philippe Andr\'e \at 
    Laboratoire d'Astrophysique (AIM), CEA/DRF, CNRS, Universit\'e Paris-Saclay, Universit\'e Paris Diderot, Sorbonne Paris Cit\'e,91191 Gif-sur-Yvette, France
\and Patrick Hennebelle \at 
    AIM, CEA, CNRS, Universit\'e Paris-Saclay, Universit\'e Paris Diderot, Sorbonne Paris Cit\'e, 91191, Gif-sur-Yvette, France
\and Ralf S.~Klessen \at 
    Universit\"{a}t Heidelberg, Zentrum f\"{u}r Astronomie, Institut f\"{u}r Theoretische Astrophysik, Albert-Ueberle-Str. 2, 69120 Heidelberg, Germany, 
    Department of Physics, Nagoya University, Furo-cho, Chikusa-ku, Nagoya, Aichi 464-8602, Japan
\and J.~M.~Diederik Kruijssen \at 
    Astronomisches Rechen-Institut, Zentrum f\"ur Astronomie der Universit\"at Heidelberg, M\"onchhofstra\ss{}e 12-14, 69120 Heidelberg, Germany, 
\and M\'elanie Chevance \at 
    Astronomisches Rechen-Institut, Zentrum f\"ur Astronomie der Universit\"at Heidelberg, M\"onchhofstra\ss{}e 12-14, 69120 Heidelberg, Germany 
\and Fumitaka Nakamura \at 
    National Astronomical Observatory of Japan, 2-21-1 Osawa, Mitaka, Tokyo 181-8588, Japan\\
    Department of Astronomy, The University of Tokyo, Hongo, Tokyo 113-0033, Japan\\
    The Graduate University for Advanced Studies (SOKENDAI), 2-21-1 Osawa, Mitaka, Tokyo 181-0015, Japan
\and Angela Adamo \at 
    Department of Astronomy, Oskar Klein Centre, Stockholm University, AlbaNova University Centre, SE-106 91 Stockholm, Sweden  
\and Enrique V\'azquez-Semadeni \at 
    Instituto de Radioastronom\'\i a y Astrof\'\i sica, UNAM, Campus Morelia, Antigua Carretera a Patzcuaro 8701. 58090 Morelia, Michoacan, Mexico. 
}

\date{Received: 2020-01-31 / Accepted: 2020-05-17}

\maketitle

\begin{abstract}
Molecular clouds are a fundamental ingredient of galaxies: they are the channels that transform the diffuse gas into stars. The detailed process of how they do it is not completely understood. We review the current knowledge of molecular clouds and their substructure from 
scales {$\sim~$1~kpc} down to the filament and core scale. We first review the mechanisms of cloud formation from the {
warm} diffuse interstellar medium down to the cold and dense molecular clouds, the process of molecule formation and the role of the thermal and gravitational instabilities. We also discuss the main physical mechanisms through which clouds gather their mass, and note that all of {them} may have a role at various stages of the process.
In order to understand the dynamics of clouds we then give a critical review of the widely used virial theorem, and its relation {to} the measurable properties of molecular clouds. 
Since these properties are the tools we have for understanding the dynamical state of clouds, we critically analyse them{.}
We finally discuss the ubiquitous filamentary structure of molecular clouds and its connection to prestellar cores and star formation.
\end{abstract}

%
%
%






\hyphenation{kruijs-sen}

\def\ltsima{$\; \buildrel < \over \sim \;$}
\def\simlt{\lower.5ex\hbox{\ltsima}}
\def\gtsima{$\; \buildrel > \over \sim \;$}
\def\simgt{\lower.5ex\hbox{\gtsima}}
\def\arcmin{$^\prime$}
\def\arcsec{$^{\prime\prime}$}

\graphicspath{{Figures/}}





\section{Introduction}\label{sec:intro}

{
Since molecular clouds (MCs) are the sites where stars are born, the study of star formation necessarily passes through an understanding on the formation, dynamics,  structure and evolution of molecular clouds. They are called molecular because they match the physical conditions for molecule formation: they are the densest, darkest and coldest regions of the interstellar medium.
}

{
MCs {have temperatures $\sim 10$--20 K, and} span a range of sizes between $\sim$~1 and $\sim 200$~pc \citep[e.g., ][]{Miville-Deschenes+17}. Inside them, smaller and denser structures are found at every level of the hierarchy, down to the resolution limit of the telescopes, a fact that has been interpreted as a fractal nature of MCs  \citep{Scalo90, Falgarone+91, Falgarone+09}. They are typically catalogued by size and mass {\citep[see e.g., ][]{StahlerPalla05} } as: (i) {giant molecular clouds} (GMCs), the biggest clouds, with masses above $10^5$~\Msun, and sizes $\gtrsim~30$~pc, and up to 100$-$200~pc, (ii) MCs, with masses of several $10^2$ to some $10^4$~\Msun\ and sizes of about $10-20$~pc, (iii) clumps, with masses between several $10-10^2$ \Msun\ and sizes between few and some pc, and (IV) cores, with masses below few 10~\Msun\ and sizes of 0.1~pc or less { \citep{Blitz93, HeyerDame15}.} They are nested in a hierarchical structure: while MCs and GMCs are embedded in a warmer {($T \sim 8000$ K)} atomic medium, clumps are nested inside MCs, and cores within the clumps. 
They exhibit a highly filamentary structure, and {
the hierarchy} is such that smaller, denser structures always occupy a {
very small} fraction of {their parent structures'} volume.
}

{
Although {
our} understanding of MCs is continuously improving, still many questions remain open. Among them {are} the
detailed process { 
by} which they are formed, {what defines 
{their inner structure,} 
} the role of filaments and cores in their evolution, the role of the different physical {
processes} (turbulence, magnetic fields, galactic dynamics, etc.) in their dynamics, the meaning of the scaling relations between their mass or their internal velocity dispersion
and their size
In the present paper we 
summarize the current state of
our knowledge about MCs.
}

\subsection{Connection between galactic scales and MC scales}


{
The arrival of facilities capable of carrying out spectroscopy of the molecular and ionised interstellar medium (e.g.\ ALMA, MUSE) has not only brought about a revolution in terms of resolving large samples of Galactic MCs into protostellar cores, but has made a similar step change in terms of resolving large numbers of external galaxies into MCs and HII regions \citep[e.g.][]{Kreckel+18,Sun+18,Utomo+18}. The resulting spatial dynamic range unlocks the interface between MC scales and those of the host galaxy, which in turn enables studies of:
\begin{enumerate}
    \item 
how MC properties change as a function of their large-scale environment \citep[e.g.][]{Sun+18,Sun+20,Schruba+19},
    \item 
how MCs are affected by galactic dynamics \citep[e.g.][]{Meidt+13,Meidt+18,Meidt+19,JeffresonKruijssen18},
    \item 
how MCs combine to constitute the galaxy-scale star formation relation between the gas mass (surface density) and the star formation rate (surface density) \citep[e.g.][]{kennicutt12,Kruijssen+18},
    \item 
how stellar feedback from the young stellar populations born in MCs affects the chemistry and energetics of the host galaxy \citep[e.g.][]{Kreckel+19,McLeod+19},
    \item 
how the evolutionary lifecycle of MCs that connects all of the above stages depends on the galactic environment \citep[e.g.][]{Chevance+20a}.
\end{enumerate}
The latter of these questions warrants a separate discussion and is treated more extensively in Section~\ref{sec:lifecycle} below, and in \citet{Chevance+20b}. We now first briefly summarise the (environmental dependence of) instantaneous statistical properties of MCs and how they constitute the star-forming properties of galaxies.
}

{
Firstly, observations show and theory predicts that MC properties depend on the galactic environment. This specifically concerns their surface and volume densities \citep[e.g.][]{Sun+18}, turbulent pressure and velocity dispersion \citep[e.g.][]{Heyer+09,Field+11,Shetty+12,kruijssen13}, and virial parameter \citep[e.g.][]{Sun+18,Schruba+19}, as well as their characteristic and maximum mass scales \citep[e.g.][]{Hughes+13,reinacampos17}. {{}Observations show and models predict that these} quantities are also correlated: MC densities, velocity dispersions, masses, star formation rate, and cluster formation efficiency typically increase with the gas pressure in the galactic midplane \citep[e.g.][]{VazquezSemadeni94,krumholz05,elmegreen08,padoan2011,kruijssen12,adamo15}.
}

{
In part, the environmental dependence of MC properties reflects a dependence on galactic dynamics. MCs initially condense out of the lower-density interstellar medium (ISM), from which they inherit turbulent and shear-driven motion \citep[e.g.][]{Meidt+18,Meidt+19,Kruijssen+19b}. Galactic dynamics can both stabilise clouds \citep[e.g.][]{Meidt+13} or compress them and induce star formation \citep[e.g.][]{JeffresonKruijssen18}. The external gravitational potential and the ambient medium can lead to enhanced velocity dispersions and {{}`apparent' virial parameters, i.e.\ ones calculated without accounting for the gravitational force of the stars} \citep[e.g.][]{Schruba+19,Sun+20}.
}

{
Somewhat surprisingly, the resulting star formation efficiency per free-fall time seems relatively constant, at $\epsilon_{\rm ff}\sim0.01$ \citep[e.g.][]{Barnes+17,Leroy+17,Utomo+18,Krumholz+19}, in rough agreement with theoretical predictions \citep[e.g.][although see \citealt{Schruba+19} for important areas where observations and theory differ]{federrath12} and the galactic-scale efficiency per dynamical time \citep[e.g.][]{elmegreen87,elmegreen93,elmegreen97,silk97,Kennicutt98}. Observations of individual MCs in the solar neighbourhood suggest that their instantaneous star formation rate is situated above the expectation from the galactic-scale `star formation relation' \citep[e.g.][]{Heiderman+10,Lada+10,gutermuth11} {{}between the molecular gas mass surface density ($\Sigma$) and the star formation rate surface density ($\Sigma_{\rm SFR}$), which is observed to have a power law form of $\Sigma_{\rm SFR}\propto\Sigma^N$ with $N=1{-}1.5$ \citep[e.g.][]{Kennicutt98,Bigiel+08,kennicutt12,Leroy+13}}. However, this difference likely results from the fact that MC studies:
\begin{enumerate}
    \item 
select single clouds that must contain both molecular gas tracers and star formation tracers, thereby restricting them to a specific evolutionary phase and biasing their position relative to the star formation relation \citep[e.g.][]{Schruba+10,Kruijssen+18};
    \item
focus on the star-forming, inner regions of MCs, that achieve higher local star formation efficiencies than the lower-density outskirts of the clouds \citep[e.g.][]{Dobbs+14,longmore2014}.
\end{enumerate}
Combining these biases with the fact that the MC lifecycle is highly dynamic and the instantaneous star formation efficiency is a strong function of an MC's evolutionary stage \citep[see Section~\ref{sec:lifecycle} and e.g.][]{Kruijssen+19,Grudic+19,Chevance+20a}, it is clear that the galactic-scale star formation relation can only arise after averaging over all evolutionary stages of MCs \citep{Feldmann+11,KruijssenLongmore14}. While this explains the large scatter of the relation on sub-kpc scales \citep[e.g.][]{Bigiel+08,Leroy+13}, it raises the question to what extent the galactic-scale star formation relation is purely statistical in nature and trivially arises from applying the central limit theorem to unresolved MC populations \citep[as suggested by][]{lada13}. {{}In other words, does the dynamic range in $\Sigma$ and $\Sigma_{\rm SFR}$ of the galactic-scale star formation relation simply arise from adding up many individual MCs, without any underlying change in the physics of MC evolution and star formation, or do the properties of MCs change across this spectrum of galactic-scale densities? Can the galactic-scale star formation relation teach us anything at all about the MC-scale physics of star formation?} The fact that MC properties are strongly environmentally dependent and change continuously over the full range of large-scale gas surface densities of the host galaxy (see above) suggests that the galactic-scale star formation relation is at least partially physical in nature, {{}rather than being a trivial result of statistical averaging}. To provide a definitive answer to this question, cloud-scale observations of the molecular ISM are necessary, which is now within reach. Irrespectively of the physics or statistics that set the galactic-scale star formation relation, it is clear that the scatter around this relation is a valuable probe of the MC lifecycle.
}

{
Finally, the stellar feedback from the young stellar populations born in MCs drives mass, energy, momentum, and metal enrichment into the surrounding ISM. Recent studies of ionised emission lines show that early feedback mechanisms dominate the dispersal of MCs \citep[e.g.][]{lopez11,lopez14,chevance16,Chevance+20a,Kim+18,Kruijssen+19,McLeod+19b,McLeod+19}. Supernovae detonating in the resulting, cleared environments may contribute to driving galactic winds on spatial scales larger than the disc scale height \citep[e.g.][]{walch2015}. The chemical enrichment from the young stars drives inhomogeneity on spatial scales similar to the gas disc scale height \citep[e.g.][]{Kreckel+19}. It remains an important open question on which timescales these inhomogeneities dissolve by mixing.
}

{We now turn to a brief discussion of the lifecycle of MCs, synthesising the processes discussed above and placing them on an evolutionary timeline. For a more detailed discussion on this topic, we refer to \citet[][this volume]{Chevance+20b}
}

\subsection{Lifecycle of MCs}
\label{sec:lifecycle}

{
Characterising the cloud lifecycle in galaxies is critical to understand the physical processes of star formation and feedback. However, measuring timescales has historically been notoriously difficult in astrophysics and the question of the molecular cloud lifetime has been highly debated, both from theoretical and observational points of view.
In the Milky Way, the lack of observed post-T Tauri stars (with ages $\sim10$~Myr) associated to molecular clouds \citep{Briceno+97, Hartmann+01} has suggested that molecular clouds are transient structures and disperse quickly after star formation. 
The fact that most molecular clouds in the Solar Neighbourhood 
are associated with young stars of ages less than 3\,Myr \citep{BallesterosParedes+99b} is also in favour of short-lived GMCs. Supporting this idea, \citet{Elmegreen00} suggests that star formation occurs on a crossing time, based on the determination of cluster ages in the Large Magellanic Cloud (LMC).
}

{
Extragalactically, measuring the lifetimes of molecular clouds has been even more challenging. A variety of indirect methods has been developed, relying for example on the presence of inter-arm molecular clouds \citep[e.g.][]{ScovilleHersh79, ScovilleWilson04, Koda+09}, on the classification of clouds based on their star formation activity \citep[e.g.][]{Engargiola+03, Blitz+07, Kawamura+09, Miura+12, Corbelli+17}, or on the evolution of clouds along orbital streamlines \citep[e.g.][]{Kruijssen+15, Meidt+15, Henshaw+16, Barnes+17, Jeffreson+18}. Due to differences in the experiment setups and subjective definitions of cloud categories, it remained unclear what part of the large range of values estimated by these different methods (from more than 100\,Myr down to about 1\,Myr) resulted from actual environmental dependence of the cloud lifetime, and what this could tell us about the physical processes regulating star formation and feedback in galaxies. 
}

{
The new statistical approach developed by \cite{KruijssenLongmore14} and \cite{Kruijssen+18} now enables the characterisation of the evolutionary timeline between cloud formation and evolution, star formation and feedback in a systematic way, applicable to a large range of galaxies. This method has been applied to a sample of galaxies in or near the Local Group (e.g. NGC300, \citealt{Kruijssen+19}, M33, \citealt{Hygate+19}; the Large Magellanic Cloud, \citealt{Ward+19}) as well as outside of the Local Group \citep[for a sample of nine galaxies]{Chevance+20a}. These new measurements, which can be extended to the large galaxies surveys at high spatial resolution observed with ALMA (e.g.\ with the PHANGS collaboration, Leroy et al.\ in prep.), now make it possible to quantitatively determine what parameters (such as ISM pressure, galactic dynamic, disc structure) govern cloud lifetime in galaxies. \cite{Chevance+20a} shows that there exist two regimes in galactic molecular gas surface density, where GMC lifetime is regulated by galactic dynamical processes at high ($\geq 8$ M$_{\odot}$ pc$^{2}$) gas surface density \citep[as in][]{JeffresonKruijssen18}, while at low ($\geq 8$ M$_{\odot}$ pc$^{2}$) gas surface density, GMCs decouple from galactic dynamics and their lifetime is governed by local processes, so that they typically live for a free-fall time or a crossing time.
} 

{
The characterisation of the lifecycle of molecular clouds, which can be seen as the building blocks of galaxies \citep{Kruijssen+19}, is further developed in \citet[][this volume]{Chevance+20b}. Measuring the duration of the successive phases of star formation, from cloud assembly, to cloud collapse and cloud destruction by feedback, as a function of the environment (e.g.\ galactic structure, rotation curve, ISM pressure, stellar density, metallicity) provides strong constraints on the physical mechanisms playing a role in these processes, and how they vary throughout galaxy evolution.
}

\section{The formation of molecular clouds}

One of the most fundamental questions regarding the understanding of
the interstellar medium is to understand how molecular clouds
form. Indeed, most of the volume in the Milky Way is filled
by atomic gas which is several times more diffuse than the molecular gas.
How the interstellar gas becomes denser and molecular?
Here the various steps thought to be involved in the process are described.

As we discuss in \citep[][this volume]{Girichidis+2020}, the total mass of the Galactic ISM is about $10^{10} \: {\rm M_{\odot}}$. Most of the volume is occupied by ionized gas, which can extend high above and below the disk midplane. It accounts for about 25\% of the mass. The rest of the mass is split roughly evenly between the atomic and the molecular phases of the ISM \citep{ferriere01}. The atomic component also has a large volume filling factor and extends to large scaleheights in particular in the outer parts of the Milky Way \citep{kalberla2009a}. 

Our discussion here focuses on the dense molecular component of the ISM, which forms by converting atomic hydrogen into H$_2$. We can compute the properties of this component  by combining data from CO observations, which trace clouds with high concentrations of both H$_{2}$ and CO, with measurements of C$^{+}$, which trace so-called ``CO-dark H$_2$ gas'', i.e.\ clouds with high H$_{2}$ fractions but little CO \citep[see e.g.][]{Pineda+13}. On global scales, the distribution of molecular gas shows a  peak within the central few hundred parsec of the Galaxy, a region known as the Central Molecular Zone \cite[CMZ, see e.g.][]{molinari2011a}. It then falls off sharply between 0.5 and 3~kpc, possibly owing to the influence of the Milky Way's central stellar bar, before peaking again at a Galactocentric radius of around 4--6~kpc in a structure known as the Molecular Ring. Outside of the Molecular Ring, the surface density of molecular gas declines exponentially, but it can still be traced out to distances of at least 12--13$\,$kpc \citep{heyer98}. Its vertical scaleheight is very small ($\sim 50\,$pc) and so essentially all molecular gas is closely confined to a dense layer close to the disk midplane, occupying about 1--2\% of the total ISM volume. We note that only about 5\% of the molecular gas mass in the Milky Way is associated with the known molecular cloud complexes \citep{roman-duval2016a}, by far the largest fraction follows a more diffuse and extended distribution \cite[for a complete decomposition of the CO emission in the Galactic midplane, see][]{Miville-Deschenes+17}.

\subsection{Three phase model of the ISM and H$_2$ formation}
\label{sec:3phaseISM}

A very simple model of the phase structure of the ISM was suggested by \citet{field69}. If one assumes that the atomic gas in the ISM is in thermal equilibrium, then there are two thermally stable solutions: a  cold dense phase that is generally called cold neutral medium (CNM), and a warm, diffuse phase termed warm neutral medium (WNM). Gas at intermediate temperatures is thermally unstable. It will either cool down and get denser until it joins the CNM, or heat up and becomes more tenuous until it joins the WNM (see the discussion below in Section \ref{sec:ThermalInstability}). This two-phase model was extended by \citet{mckee77}, who realised that the momentum and energy input from supernovae would create large, ionized bubbles filled with very hot gas. At temperatures around $10^{6}\,$K this gas would cool very slowly compared to the other timescales relevant to the system, and so this component would consitute a third phase known as  the hot ionized medium (HIM).

The chemically most straight-forward path to form H$_{2}$ in the ISM is via the radiative association of two hydrogen atoms: ${\rm H + H} \rightarrow {\rm H_{2}}$. However, the rate coefficients are extremely small, and similar applied to gas phase reactions involving the H$^-$ and H$^+$ radicals. These reactions are only relevant in the early universe in very low metallicity gas \cite[see, e.g.][]{KlessenGlover16}. In the solar neighborhood and essentially in a present-day galaxies, almost all H$_2$ molecules form on dust  \cite[for a comprehensive overview, see][]{tielens2010,draine11}. The association reactions between adsorbed hydrogen atoms occur readily on grain surfaces, and the rate at which H$_2$ forms is only limited by the rate at which H atoms are adsorbed onto the surface. As we discuss in \citep{Girichidis+2020}, the rate for typical Milky Way conditions is 
\begin{equation}
{\bigg(\frac{R_{\rm H_{2}}}{\rm cm^3\ s^{-1}}\bigg) \sim 3 \times 10^{-17}\ \bigg(\frac{n_{\rm H}}{\rm cm^{-3}}\bigg) 
}
\label{eqn:H2-form-rate}
\end{equation}
where $n$ is the total number density of  gas particles and $n_{\rm H}$ is the number density of atomic hydrogen. For purely atomic hydrogen gas, both quantities are identical if we neglect contributions from helium and possibly metals.

While H$_2$ is easily formed on dust, it is also readily destroyed again when exposed to the interstellar radiation field. When molecular hydrogen is photodissociated, the H$_{2}$ molecule first absorbs a UV photon with energy $E > 11.2$~eV and ends up in an excited electronic state. It then undergoes a radiative transition back to the electronic ground state, ending up  either into a bound ro-vibrational level, in which case the molecule survives, or into the vibrational continuum, in which case it dissociates. Because H$_2$ photodissociation is line-based, rather than continuum-based, the rate at which this process occurs is highly sensitive to self-shielding \citep{draine1996a}. Depending on the strength of the interstellar radiation field (expressed in terms of Habing units $G_0$, see \citealt{habing1968a}) this becomes important when the total column density $N$ exceeds a value of 
\begin{equation}
{
\bigg(\frac{N}{\rm cm^{-2}}\bigg) = 10^{20}\ G_{0}\ \bigg(\frac{n}{\rm cm^{-3}}\bigg)^{-1}.
}
\label{eqn:H2-self-shielding}
\end{equation}
Note that also dust extinction contributes to reducing the H$_{2}$ photodissociation rate, however, it typically requires higher column densities than H$_2$ self-shielding, and so it plays only a minor role under normal ISM conditions, \citep[see][this volume]{Girichidis+2020}.

\subsection{The role of thermal instability}\label{sec:ThermalInstability}


\begin{figure}
    \sidecaption
    \includegraphics[width=12cm,angle=0]{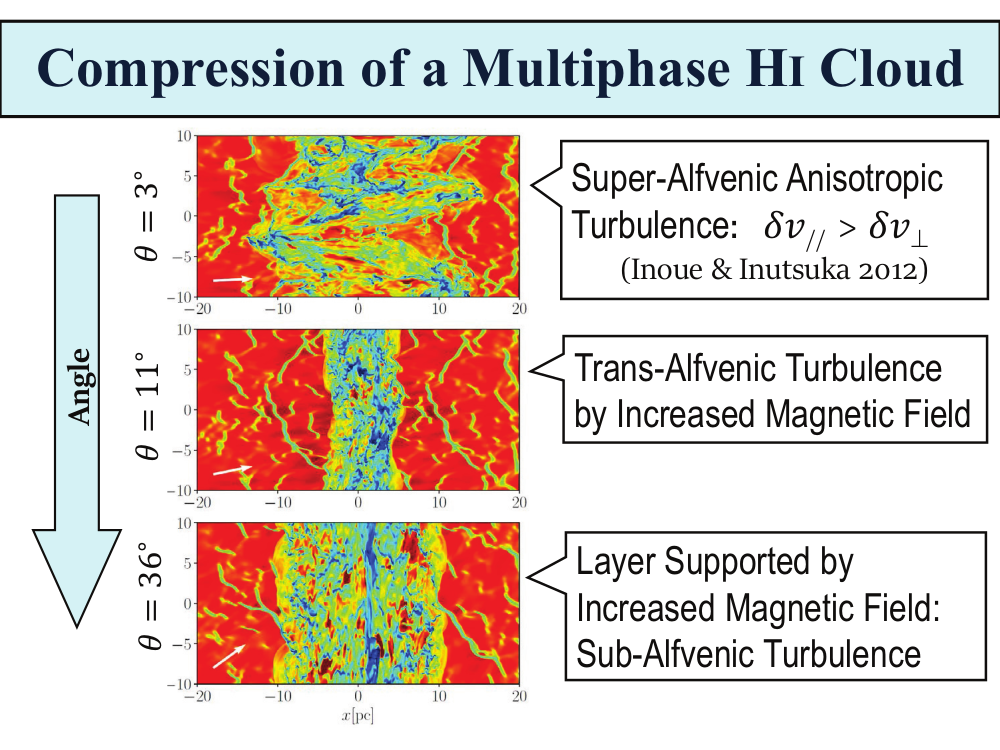}
    \caption{The result of compression of multiphase HI clouds by shock waves  \citep{iwasaki2018}. The column density is shown (red stands for WNM while Blue-green represents CNM). The relative angle ($\theta$) between the shock wave propagation direction and the mean magnetic field is 3 degrees  (upper panel), 11 degrees (middle panel) and 36 degrees (lower panel), respectively.}
    \label{Fig:MagLayer}
\end{figure}


The first step on the way toward getting dense molecular gas is certainly
the transition from the warm neutral medium (with densities of about 1 cm$^{-3}$
and temperature $\simeq 8000$ K) into the cold neutral medium (roughly 100 times
denser and 100 times cooler). The physics of thermal instability \citep{field1965} is
discussed in \citet[][this volume]{Girichidis+2020} Here we simply recalled the
basic aspects. Thermal instability is due to the atomic cooling of the ISM \citep[most important
coolant being H, O and C\,{II}, see][]{wolfire2003} and more precisely to the fact that in a range
of density, typically between a few and a few tens  of particles per cm$^{-3}$, the cooling
function has a relatively low dependence on the temperature. Therefore since
the cooling is proportional to the square of the density (because it arises
through the radiative deexcitation of the coolant which has been exicted through collisions)
while the heating is simply proportional to the density, an instability occurs.
The consequence is that at equilibrium, gas in standard ISM conditions, cannot exist
at intermediate densities, say between 1 and 20-30 cm$^{-3}$. The gas is said to be
thermally unstable. This means that when the WNM enters in the thermally unstable domain
it starts contracting until it reaches the second branch of equilibrium, i.e. the CNM.
This represents a contraction by a factor of about 100.

Numerical simulations have been used to study the non-linear development of
thermal instability triggered either  by shocks propagating in the WNM
\citep{koyama2000,koyama2002,inoue2012}
or by a converging flow of WNM \citep{HennebellePerault99,audit2005,audit2010,heitsch2006,vazquez2006}.
The effect of turbulent driving in the Fourier space, which corresponds to
complex flows entailing all sort of compressive events
\citep{seifried2011}  has also been considered.
 From  all these studies it has been deduced that
 {\it (i)} when sufficiently pushed out of equilibrium WNM   breaks--up into a multi--phase
medium composed of   clumps of CNM surrounded  by the WNM which confines them
; {\it (ii)}  the CNM clumps present statistics  which resemble
the ones  inferred for the CO clumps ; {\it (iii)}
 thermally unstable gas does exist and represents several percents of the gas.
Its existence is due to  the
turbulence  \citep[e.g.][]{Gazol2001}; {\it (iv)} the
various phases are interwoven; {\it (v)} the cold phase has
 supersonic motions  and presents  a
velocity dispersion equal to a fraction of the WNM sound speed.
Indeed the medium presents characteristics of a two phase flows
as well as of a turbulent one.

Simulations including the magnetic field have also been performed.
Although magnetic fields definitely modify the fluid dynamics, the
above conclusions remain qualitatively similar.
There is however an aspect where it possibly makes a significant difference.
An important question is whether
 the clouds of CNM which constitutes the progenitor of molecular clouds
 are created by a single compression event from WNM
or whether  they are more  gradually created from cold dense HI clouds.
\citet{inoue2008}, \citet{inoue2009}, \citet{Heitsch+09b}
and \citet{kortgen2015} have concluded that forming magnetized
CNM tends to be  difficult because magnetic field has a  stabilizing
influence.
\citet{inoue2012} have investigated the scenario in which a dense cloud form
after a series of compression and a
more detailed analyses has been carried out  by \citet{iwasaki2018}.
Figure \ref{Fig:MagLayer} displays  results for compression induced by shock waves
propagating within  magnetized multiphase HI clouds.
The relative angle ($\theta$) between the shock  velocity  and the mean magnetic field is 3 degrees (upper panel),
 11 degrees (middle panel), 36 degrees (lower panel), respectively.
With a small angle a  substantial amount of dense gas formed.
But for larger angles and above a certain critical value, the propagation of shock wave becomes very
inefficient in producing dense gas.
Obviously the value of this critical angle depends on the flow velocity
  and of  the magnetic intensity.
In reality, there is  a distribution of angles between the magnetic and the velocity fields.
 Detailed studies from larger scale simulations revealed
 that  magnetic and velocity fields tend to be
aligned \citep[e.g.][]{iffrig2017}, which would imply that aligned configurations are more frequent
than it was randomly determined.

\subsection{The role of gravity}
In the second step, gravity has been considered and   the
gas condensation has been  further described
\citep{VazquezSemadeni+07,Heitsch+08,hetal2008,baner2009}.

\begin{figure}
     \sidecaption
    {\includegraphics[width=6cm,angle=0]{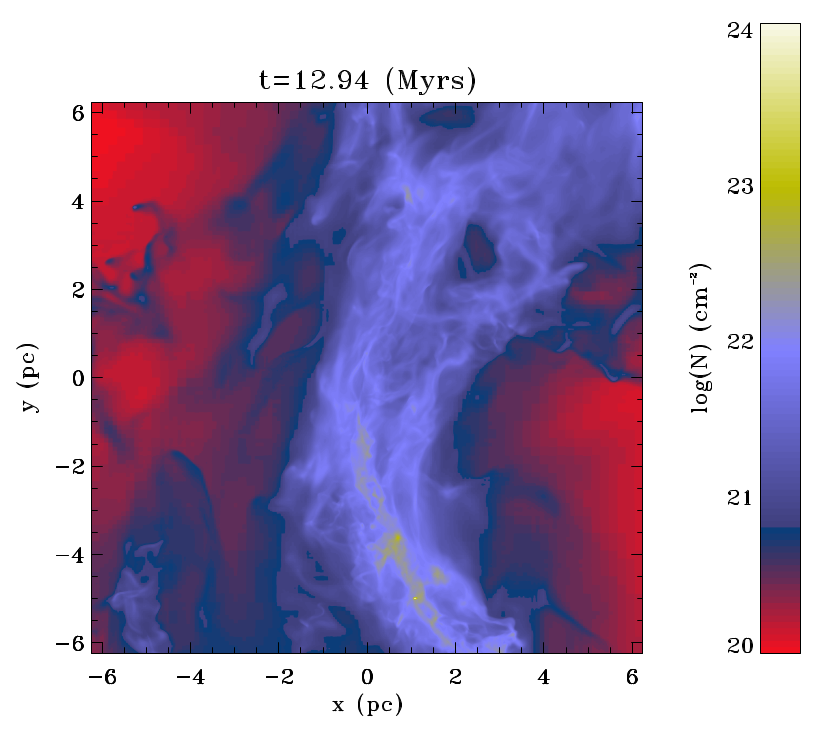}}
    {\includegraphics[width=6cm,angle=0]{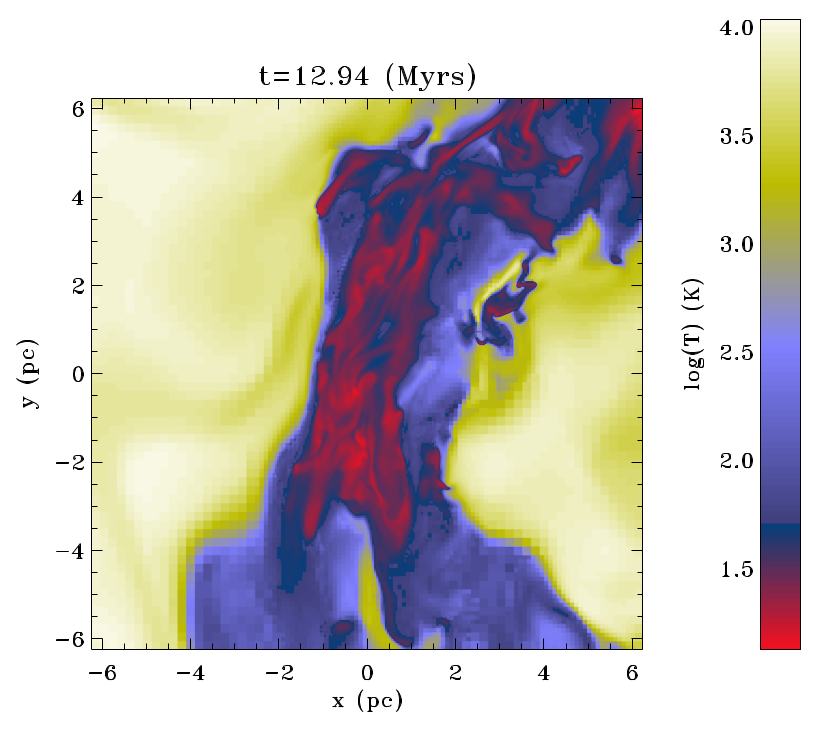}}
    \caption{{Cut of the temperature {\it (right)} and column density {\it (left)}, of a simulation snapshot of molecular
cloud formation} \citep{hetal2008}.}
  \label{multi-phase}
 \end{figure}

Figure~\ref{multi-phase} portrays a snapshot for a numerical simulation of
 a colliding flow \citep[from][]{hetal2008} which has led to the formation of
 a molecular cloud.
As described in the previous section, the dense gas
 has formed from the diffuse gas as well as from the denser structure of the multi-phase medium.
 Due to gravity, few regions undergo gravitational collapse.
Interestingly, in few locations the temperature abruptly jumps
 from about $10^4$ to 10 K. Thus the cloud
depicted in Fig.~\ref{multi-phase} can be truely qualified
as being a multi-phase molecular cloud. In a sense, it is a denser counterpart
of the classical WNM/CNM where CNM has been replaced by denser gas.

In simulations including gravity, the clumps present many
characteristics with the observed CO clumps as for instance
 their density, velocity dispersion
and mass spectrum \citep[e.g.][]{baner2009, Heitsch+09a}.

It has been found that initially, that is to say soon after the dense gas have been assembled, the overall structure of the
cloud remains largely unchanged compared to the case whithout gravity.
One of the most important  difference is obviously that in the
self-gravitating case  the PDF of the gas density extends towards much larger values.
 However, in MCs with conditions appropriate for the solar neighbourhood, the bulk of the mass remains at a density on the order of $n_{\rm H}=100-10^3$ cm$^{-3}$, even at late times, when collapse has already occurred in few places.

The velocity dispersion in particular is already high from the very beginning 
of the cloud lifetime and is likely the result of a accretion-driven 
process as emphasised by \citet{klessen2010}. This turbulence which 
is inherited from the dynamical building of the cloud is likely 
maintained by several processes including continuous accretion 
onto the cloud \citep{klessen2010}, the development of various instabilities such 
as the Kelvin-Helmholz instability for instance \citep{Heitsch+08}, 
gravitational collapse of the densest cloud parts and later 
when stars have started to form by the feedback processes such as 
HII regions \citep{grit2009} and jets \citep{fed2015,offner2018}. 
{In addition, as collapse proceeds, the kinetic energy of clouds tends to follow the gravitational energy \citep{VazquezSemadeni+07}. Thus, it is likely that a non-neggligible contribution to the velocity dispersion comes from gravity itself, since the {ensemble} of observed MCs tend to organize along the virial/free fall collapse lines, with slightly overvirial values \citep[][see also \S\ref{sec:L-Sigma}]{BallesterosParedes+11a}. 
}
%

{The H~I streams collision mechanism depicted above is a general mechanism that should work regardless {of the} detailed mechanism that gave origin to the streams. The main proposed mechanisms for producing such H~I streams are (see \S\ref{sec:MCform}): (i) stellar feedback, as e.g., the expansion of H~II regions or SNe explosions), (ii) the passage of a spiral arm, (iii) a large-scale gravitational instability, and (iv) cloud-cloud collisions. Depending on the origin and length of the streams, the resulting cloud can have more or less mass. For instance, it can be expected that, while bubbles due to stellar activity can produce $10^4$~\Msun, spiral arms or gravitational instability can produce clouds with masses up to $10^5-10^6$~\Msun.
}

An additional factor limiting the formation of the clouds is the relative orientation between the magnetic and the velocity fields. Numerical simulations have shown that the diffuse colliding streams have to move nearly parallel to the magnetic field in order to allow the formation of the molecular clouds in reasonably short timescales (10--20 Myr) \citep[e.g., ][]{Heitsch+09b, Inutsuka+15}. It should be noticed, however, that either the magnetic and the velocity fields in disk galaxies are, at first approximation, circular. Thus, one can conclude that large-scale galactic dynamics do plays a role in the formation of molecular clouds.

\subsection{The formation and role of molecules}

The next step has been the modelling of the
 UV-driven chemistry which has been introduced either
directly during the simulation \citep[e.g.][]{glover2007,glover2012,inoue2012}
 or, at a
more sophisticated level, as a post-treatment of the WNM colliding
flow simulations \citep[e.g.][]{levrier2012}.
  This allows a proper
treatment of the combined influence of density and UV-shielding upon
chemistry  and of the cooling function.
The results
provide a confirmation that the gas temperature is reasonably well
computed in the magneto-hydrodynamical (MHD) simulations.

One important question which has been addressed by these simulations is
the formation timescale of molecular hydrogen. For a long time, this
has remained a mystery because the $H_2$ formation is long: 
{the H$_2$ formation rates on grains are of the order of $t_{\rm form} \sim nR$, where $n$ is the density of the atomic hydrogen, and $R$, the rate coefficient, has typical values of the order of $3\times$~10\ala{17}~cm$^3$~s\alamenos1 \citep{Jura75}. Then, the typical H$_2$ formation timescales are given by}
%
{
\begin{equation}
    \bigg(\frac{t_{\rm form}}{\rm yr}\bigg) = 10^6\ \bigg(\frac{n}{\rm\ 10^3\ cm^{-3}}\bigg)
\end{equation}
}
%
From their simulations, \citet{glover2007}
concluded that H$_2$ forms at relatively large densities, therefore
in a relatively short timescale, in dense clumps induced by turbulence.
As these clumps are transient they eventually mix back with the more diffuse gas
and therefore enriches it in molecular gas.
This result has been confirmed by \citet{valdivia2016}. However, since
multi-phase ISM is considered in their simulations, when H$_2$
spreads from dense clumps to the surrounding medium, as this latter is composed
of warm gas, a fraction of warm (500-1000 K) molecular gas develops and
this may have consequences to form some chemical species (e.g. CH+ see
\citet{valdivia2017}).

 \citet{glover2012a} have investigated the formation of the CO
molecules in turbulent simulations using different methods.
They find \citep[see also][]{shetty2011,gong2018} that all methods tend to produce similar
amount of CO molecules in the dense gas and in good agreement with
observations.
It is worth stressing that so far all the
adopted models  failed to
 reproduce (by almost a factor of 10) the observed CO abundances in
regions poorly shielded from the UV-field  \citep{shetty2011,levrier2012}.

Generally speaking, it has been found \citep[e.g.][]{glover2012} that
not unexpectedly, the gas dynamics is
not sensitive to the details of the chemistry models.

\subsection{Dynamical mechanisms for gathering mass}\label{sec:MCform}

{In addition to the thermal instability, which enables {the gas} to efficiently {
transit} from {a} diffuse, 
warm {phase} into {
dense, cold} clouds, different mechanisms have been proposed for collecting the mass. {
Ultimately}, it is likely that all those mechanism{s} play a role in the formation of MCs and MC complexes. {
Some of} these are, agglomeration {(or coagulation)}, converging flows, {{the passage of}} spiral arms, {cloud-cloud collisions,
shock-wave passage,} and large-scale instabilities. It should be recognized, however, that all cases are, in practice, converging flows, and the differences between them are, on one hand, the physical origin of the inflow{s}, their length, their geometry, and their initial density. 
}

{Each one of the mechanisms producing MCs may also be related to the total mass that they can gather, which will be given by 
}
\begin{equation}
    M = \Delta t\ \oint_S \rho\ u_i\ \hat n_i\ dS,
\end{equation}{}
where $\rho$ is the density of the diffuse interstellar medium, $u_i$ is the component of the velocity of the fluid in the direction $\hat n_i$ perpendicular to the surface $S$,  and $\Delta t$ the time interval that the process lasts.

\subsection{Agglomeration of smaller clouds}\label{sec:agglomeration}

One of the {
earliest} models of cloud formation { 
was the} so-called ``agglomeration'' {or ``coagulation" model}. The idea started with \citet{Oort54}, who proposed that HII regions produced by OB stars in MCs can produce a rocket effect on their parent clouds, ejecting them ballistically through the interstellar medium. Such motions could counteract the loss of kinetic energy produced by further shocks between clouds. This idea was taken later by \citet{FieldSaslaw65}, who made a model for the evolution of the mass spectrum of MCs, assuming ballistic clouds with a typical velocity dispersion. In their model, small clouds could have inelastic shocks, allowing the construction of larger clouds. With simple assumptions about the cross sections of MCs, and their velocity dispersion, the authors could explain the observed shallow slopes of the MC mass spectrum, with  $\gamma\sim 0.5$ in eq. ([\ref{eq:massspec}], see \S\ref{sec:MassSpec}). Similar results were found by \citet{]TaffSavedoff73} and \citet{Hausman82}. This model was revisited later by \citet{Kwan79}, who estimated the ages of clouds constructed in this way. He found a typical {
timescale} of 2$\times 10^8$ years to construct GMCs.

{\citet{BlitzShu80} dismissed the coagulation models, arguing that there were substantial observational evidence to say that GMCs cannot live for more than some times $10^7$ years. Among other reasons, {they argued that: (i)} {{
} Since the abundance of molecular gas had been overestimated, so {should be} the ages of GMCs, and
(ii) 
if GMCs were long-lived, they should also be observed in the inter-arm region, forming OB stars there. However, OB stars appear to be highly correlated to the arms.}
}

{Some years later, \citet{KwanValdes83}, \citet{Tomisaka84} and \citet{KwanValdes87} made actual numerical simulations of the evolution and coalescence of clouds in a galactic disk with a spiral potential and gravity between clouds. Although GMC formation in this model is accelerated by the spiral arms and by the mutual atraction between clouds, still {timescales} larger than $10^8$ years were required to construct GMCs. However, with a similar model, \citet{Tomisaka84} found {timescales} of the order of $4\times 10^7$ yr, opening up the possibility that GMCs were constructed by coagulation, and {addressing} the criticisms raised by \citet{BlitzShu80}.
}

{Those models were  {
due in part to the advent of} more complex numerical simulations of the ISM, {
in which clouds were self-consistently treated as} part of the fluid \citep[e.g., ][]{VazquezSemadeni+95, Passot+95}, rather than {
as discrete, ballistic objects}. The possibility of cloud coagulation, however, has been recently renewed with state of the art numerical simulations of multiphase media in galactic disks \citep{Dobbs08, TaskerTan09, Dobbs+15}
{
which} small clouds coalesce to form larger cloud complexes {as consequence of the converging flows in spiral arms}. 
}

{Although it is likely that actual agglomeration occurs {to some degree}, there are two observational {
constraints on} this possibility: on one hand, all observed molecular clouds have a highly filamentary structure down to the resolution limit \citep[e.g., ][]{Falgarone+91, Andre+14}. In contrast, the small clouds coalescing in the quoted simulations lack  inner structure, exhibiting roundish shapes on scales of 10--
100~pc, 
indicating that those clouds are numerically under resolved. Additionally, the large majority of clouds in the Solar Neighborhood exhibits signs of star formation \citep{BallesterosParedesHartmann07, Kainulainen+09}, with ages typically of 1-3~Myr, and with no stars much older than 5~Myr. This suggests that the star formation events in small clouds are able to rapidly disperse their parent clouds {\citep[e.g.][]{Agertz+13, Dale+14, Hopkins+18, Kruijssen+19}}, limiting the possibility of coalescence. Thus, although the coalescence of smaller clouds into larger ones may occur {to some extent}, this process is probably {
overemphasized} in the quoted simulations due to resolution effects, as well as to the lack of {
stellar} feedback. 
} 

\subsection{Converging flows}\label{sec:convergingflows}

\begin{figure}
\sidecaption
\includegraphics[scale=0.3]{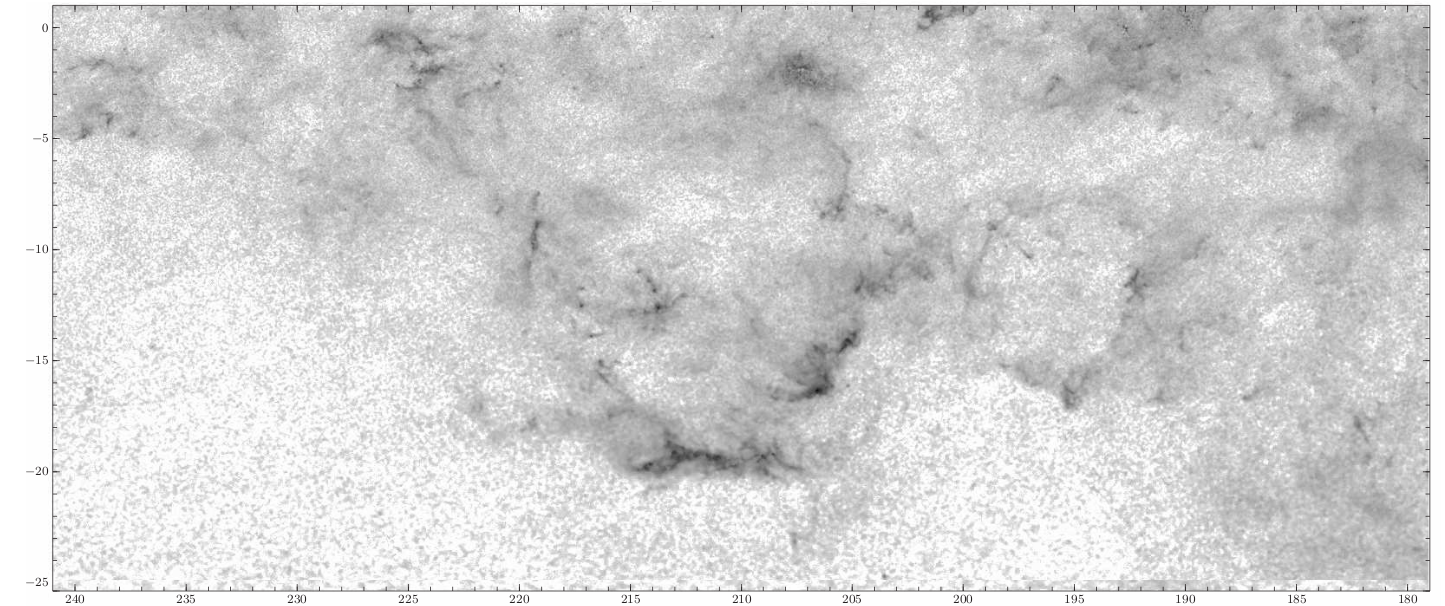}
%
%
\caption{Extinction map towards the Orion molecular complex. Regardless the fact that different features are located at different distances, it is clear that the ISM is full of bubbles at different scales, with MCs located preferencially tracing segments of circles \citep[figure adapted from][]{RowlesFroebrich09}.}
\label{fig:bubbles}       
\end{figure}

{Local compressions like the {
passage of HII region or SN shells} can produce {
supersonic compressions in the diffuse gas that nonlinearly} trigger {the formation of cold, dense atomic gas by} thermal instability.
If enough column density is achieved {
in the cold atomic gas, molecules can begin to form and a molecular cloud begins to appear in the deepest regions of the cold atomic cloud \citep{FrancoCox86, Hartmann+01, Bergin+04, clark2012}}. There is plenty of observational evidence that local MCs are at what would be the edges of bubbles and shells in the ISM {\citep[see Fig.~\ref{fig:bubbles}, and, e.g., ][and references therein]{Heiles79, Heiles84, TenorioTagleBodenheimer88}.}
}

{Numerical simulations of a piece of the interstellar medium including a variety of { 
physical processes such} as shear, rotation, magnetic fields, diffuse and stellar heating, cooling, and self-gravity \citep{VazquezSemadeni+95, Passot+95}, showed that clouds could be { 
understood} as turbulent density fluctuations, i.e., as {
density enhancements resulting from local turbulent compressions in} the diffuse ISM  \citep{HennebellePerault99, BallesterosParedes+99a}. Furthermore, filamentary 
{
{molecular} clouds,} with $M\lesssim$ several $\times 10^4$~\Msun, 
a few~pc wide 
by $\sim 20-40$~pc long, {and densities} $n > 100$~cm$^{-3}$, could be produced in few 
megayears \citep{BallesterosParedes+99b}, once enough material has been accumulated from the diffuse medium along distances of $\sim$~100~pc \citep{Hartmann+01}. {Note, however, that the clouds evolve for significantly longer timescales, $\gtrsim 10$ Myr, in the cold atomic phase, with the transition to a mainly molecular composition occurring shortly before or nearly simultaneously with the onset of star formation 
\citep[][]{FrancoCox86, Hartmann+01, Bergin+04, clark2012, vazquez2018}}. 
}

{One of the achievements of {
the numerical} simulations is that they were able to explain the synchronization of star formation over long distances, as a consequence of the large-scale streams collecting mass along the cross section of the collision, producing simultaneous star formation {
in apparently disconnected regions}. Furthermore, once the stars {
form}, the stellar feedback {
can} disperse the cloud in {
a few more megayears}. As a result, the timescales of MCs, from the time in which the CO appears, to their dispersal time, is around $\sim$~10~Myrs, consistent with observations of stellar clusters \citep{Leisawitz+89} { and of the spatial association of the stars with the gas \citep{Kruijssen+19}}. This could explain the post-T Tauri problem, a {
20-year} observational puzzle, {
namely that} no stars older than 5~Myrs {
are} found associated to molecular gas \citep{Herbig78}
Thus, by rapidly {
assembling the molecular component of} the cloud, rapidly and {
coherently} forming the stars, and rapidly dispersing the clouds, the post-T Tauri puzzle was {
solved}.
}

%

%
{
The converging-flow} mechanism has been successful {
in explaining} the formation of several features observed in nearby ($\lesssim$~1~kpc from the Sun) clouds, {
such as} the ages and age histories of stars and clusters in nearby clouds {\citep{Zamora2012, Hartmann+12, vazquez2017}}, the core-to-core velocity dispersion in clouds \citep{Heitsch+09a}, the near-virial and over-virial distribution of clouds { \citep{camacho2016, BallesterosParedes+18}}, the column density probability distribution functions of clouds \citep{BallesterosParedes+11b}, the position-velocity distribution of gas and stars \citep{Kuznetsova+15, Kuznetsova+18}, etc.

\subsection{Large-scale gravitational instability of the Galactic disk}

{It is not clear how much of the mass in the Milky Way is in large {molecular} complexes, or scattered into small clouds. While it is true that all the cloud mass spectra $dN/d\log{M}$ of CO clouds reported in the literature have power-laws shallower than $-1$, implying thus that most of the mass is in large complexes (see \S\ref{sec:MassSpec}), it has to be recognised that such studies are seriously skewed by distance: there {
is no data for} clouds smaller than $\sim$~10~pc farther than $\sim$~3~kpc from us \citep[e.g., ][see the discussion in \S\ref{sec:MassSpec}]{Miville-Deschenes+17}. Thus, it is clear that we are missing all the small clumps that are far away due to either resolution and/or sensitivity \citep{Miville-Deschenes+17}, as well as by superposition along the line of sight \citep{BallesterosParedes+19}. In fact, \citet{Koda+16} estimated that approximately half of the molecular gas can be in smaller clouds in the interarm regions, rather than organized in large complexes in the spiral arms. {
However, it is unclear to what extent this result is 
} 
affected by the fact that non-circular motions play a key role in the difficulty of reconstructing the spiral structure of the Milky Way \citep{BlitzShu80, Gomez06}. 
}

{In any event, large complexes, with masses $\gtrsim$~\diezala{6}~\Msun\ {
clearly do} exist, and are observed in external galaxies too \citep[see, e.g.,][and references therein]{Dobbs+14} along the spiral arms. For these, it is necessary to have {
large-scale} convergence of flows, {as required by the continuity equation,} and thus it is difficult to construct such complexes only {\it via} {{local}} turbulent streams or random agglomeration. For those complexes, the main mechanism {driving the converging flows 
is likely to be a} large-scale gravitational instability of rotating disks. This occurs when a thin disk of column density $\Sigma$ and sound speed $c_{\rm eff}$ has a Toomre parameter
}
\begin{equation}
    Q\equiv \frac{\kappa c_{\rm eff}}{\pi G \Sigma} \lesssim 1-2
\end{equation}{}
{\citep[][]{GoldreichLyndenbell65}, where {
$\kappa$} is the epicyclic frequency { 
in} the disk. The { 
precise} value of $Q$ for which a disk can become gravitationally unstable depends on whether the disk is isothermal, magnetized, its thickness, etc., but typically these values are around $Q \sim 1-2$ \citep{KimOstriker01, KimOstriker02, Kim+03, Li+05, KimOstriker07}. Numerical simulations have shown that GMC complexes with masses above $10^5-10^6$~\Msun\ can be buil{ 
t} up by gravitational instability in rotating disks \citep{ShettyOstriker08, Dobbs+11b, TaskerTan09}. In addition, it is likely that this mechanism is { 
} responsible for the agglomeration of smaller clouds in the vicinity of such complexes.
{However, we note that if the timescale for assembly of clouds exceeds the timescale on which stellar feedback from new-born stars disperses the gas, { so} the cloud agglomeration process is halted and the mass spectrum of MC complexes is truncated { 
} at the high-mass end \citep[e.g., ][]{reinacampos17}.}
}

\section{The virial theorem and energy budget of MCs.}\label{sec:VT}

\subsection{The virial theorem}

{A { 
fundamental tool} in the study of MCs has been the virial theorem (VT), basically, because it allows us to use the observable quantities, column density, line profiles and sizes, in order to estimate the dynamical state of clouds in terms of their 
gravitational, thermal, kinetic, and magnetic {energies}. 
}

In its scalar version, the VT is obtained from the momentum equation, by dotting it by the position vector, and integrating over a volume of interest, typically, a cloud. { Thus, the VT is a measure of the work done by the external forces on the medium and the resulting kinematic effects on the cloud.} { In the Lagrangian form, it is expressed as }
%
%
 \begin{eqnarray}
%
\frac{1}{2} \frac{d^2 I}{d t^2}  - 2 E_{\rm kin} = & 2 E_{\rm int} - & 2 \tauintmath + {E_{\rm mag}} + \tau_{\rm mag} + W 
\label{eq:LVT}
\end{eqnarray}
{where $I=\int_V \rho r^2 dV$ is the moment of inertia of the cloud 
$E_{\rm kin} = 1/2\int_V \rho u^2 dV$ 
is the kinetic energy of the cloud 
$E_{\rm int} = 3/2\int_V P dV $ is the internal energy, $\tauintmath = - 1/2\oint_S\ x_i\ P\ \hat{n}_i\ dS$ is the pressure surface term, {${E_{\rm mag}}=1/8\pi\int_V B^2 dV$} is the magnetic energy, $\tau_{\rm mag} = 1/4\pi \oint\ x_i\ B_i\ B_j\ \hat{n}_j\ dS$ is the magnetic stress at the surface of the cloud, $W=\int_V \ x_i\ \rho\ \partial \phi / \partial x_i \ dV$ is the gravitational term, which frequently is approximated as the gravitational energy of a homogeneous sphere, $E_g = - 3 GM/5R$, and where  $\phi$ is the gravitational potential.
In the previous equation, $\rho$, $u_i$, $B_i$, $P$, and $\hat{n}$ are { respectively} the density, the $i^{th}$ component of the velocity $u$, the $i^{th}$ component of the magnetic field $B$, the pressure, and a unitary vector perpendicular to the surface $S$ that surrounds the volume $V$, over which the integrals are performed.
In the notation above, 
the Einstein convention { is assumed}, where repeated indexes are summed. 
}

{The measurements that are available through observations are the line profiles and/or the integrated emission, in a given solid angle. From there, the physical quantities that are available through observations are the column density, the area of the cloud, from which one can compute the total mass, and the line widths, from { 
which} one can { in turn} estimate the RMS value of the velocity field, and depending on the line, the magnetic field intensity, and the size of the cloud, which ideally should be computed as an independent measurement of the area, in order to properly understand the scaling properties of clouds, as well as their fractal properties.  From these quantities, { 
and making} assumptions on the geometry of the cloud and its mass distribution, 
estimates of the energies involved in eq.~(\ref{eq:LVT}) { can be obtained}. { 
Finally, this} allows
{ an estimation of} the physical state, { 
and possibly of the evolutionary state}, of MCs. 
}

{There are several assumptions that 
are { frequently} made when computing the energies of the VT, and that are not necessarily true in MCs, and thus, they should be discussed \citep[see][for details]{BallesterosParedes06}. These are:
}

\begin{enumerate}

{\item{} ``The role of { 
nonthermal} motions within a cloud is to provide support against collapse." { 
This is only true if the motions are truly random turbulent motions, microscopic, and driven by an energy source different from gravity}. On the other hand, is should be stressed that, a collapsing cloud will have a 
kinetic energy { of the order of the gravitational energy, very close to the virial value, although 
} by no means { this} implies that the cloud is been supported. In reality, by just computing the kinetic energy, we cannot have tools to distinguish between a collapsing and an expanding cloud.
}

{\item{} {It is also { frequently} assumed that the surface terms in the VT are negligible compared to the volumetric ones. While some times it is recognised that the pressure at the boundary of the clouds could be important, it is less widely recognised that the kinetic pressure, or the magnetic tension at the surface of the cloud could play a role. Both terms could contribute to distorting the cloud, especially in a highly dynamic environment.}
}

\item{} ``The gravitational term of a cloud can be approximated by the gravitational energy of { 
a} homogeneous sphere." { 
That is, that,}
\begin{equation}
    W = \int_V \rho x_i \frac{\partial \Phi}{\partial x_i} dV \simeq -\frac{3}{5}\frac{G\ M^2}{R} = E_{\rm grav}.
    \label{eq:W:Eg}
\end{equation}{}
{ However, it should be noted that the gravitational potential is that generated by the {entire} mass distribution, and not just that interior to the cloud. Therefore,
} it should be stressed, on the one hand, that tidal terms can be important either for the energetic balance of GMCs in the disk \citep{BallesterosParedes+09a, BallesterosParedes+09b, Meidt+18}, as well as  in the Central Molecular Zone, where the effective gravitational potential due to rotation and shear can play an important role.
On the other hand, on small scales, it has been recognised that { protostars still in the process of accretion can gravitationally compete
for} the available material \citep[][see also Lee et al. 2020, this volume]{BonnellBate06, Maschberger+14, BallesterosParedes+15, LeeHennebelle18, Hennebelle+19}, { and that the actual gravitational energy computed from the standard recipe (right part of eq. [\ref{eq:W:Eg}]) can be modified even by a factor up to 10 when considering the actual structure of a stellar cluster \citep[left part of eq (\ref{eq:W:Eg}), see e.g., ][]{Cottaar+12, BallesterosParedes+18}
}from the standard recipe. Conversely, neglecting the actual gravitational potential can { 
result in} substantially wrong estimates of the total gravitational content of the clouds \citep{BallesterosParedes+18}.

{\item{} ``The sign of the second-time derivative of the moment of inertia determines whether the cloud is contracting ($\ddot{I}<0$), expanding ($\ddot{I}>0$), or static ($\ddot{I} =0$)." { 
However}, one can { easily} find counterexamples in each case \citep[see][]{BallesterosParedes06}. 
}

{\item{} ``{ 
Molecular} clouds are { 
near} virial equilibrium, and the \citet{Larson81} relations are the { 
evidence of} that." Although still it is commonly found in the literature that clouds follow the \citet{Larson81} relations, and thus, that they are in virial equilibrium, it has becoming clear that, in reality, they exhibit a wide range of virial parameters \citep[][]{BertoldiMcKee92, Kauffmann+13, Leroy+15}. We will discuss this point in \S\ref{sec:statistics}. 
}

\end{enumerate}

{Even though a variety of numerical simulations of 
(a) galactic disks \citep[e.g., ][]{Dobbs08, TaskerTan09}, (b) parts of the disk \citep[e.g., ][]{deAvillezBreitschwerdt05, ShettyOstriker08, IbanezMejia+16, Seifried+18}, or (c) closed boxes \citep[e.g., ][just to quote a few references]{Federrath+10, Padoan+16, Kim+18}, show a highly dynamical picture of MCs, almost no work in the literature evaluates the second time derivative of the moment of inertia, which presumably, should be relevant. The difficulty resides, in part, in the fact that this term is highly noisy, due to its second-derivative character. { 
Early} 1~kpc$^2$ 2D simulations of the galactic disk with cooling, stellar and diffuse heating, magnetic field, Galactic shear and rotation by \citet{BallesterosParedesVazquezSemadeni97} based on the { 
simulations} of \citet{Passot+95} have computed this term. These authors found that, indeed, the time derivative terms in the Eulerial virial theorem\footnote{It should be noticed that the Eulerian virial theorem involes two additional terms, related to the distribution of mass inside the fixed volume, and the flux of momentum between the volume and its environment \citep[see][]{Parker79, McKeeZweibel92}.}  are dominant, by one or two orders of magnitude, over the remaining energies. There is, however, substantial scatter in these results, likely due to the difficulty of comparing time derivatives to integrated quantities. However, even with such a large scatter, the picture in the ISM is clear: it seems unlikely that MCs obey $\ddot{I}=0$.
}

{
In addition, it should be noted that 
{  
in eq.\ (\ref{eq:LVT}) we have written the kinetic energy term, $E_{\rm kin}$ on the left hand side of the equation, because this term derives directly from the time derivative term in the momentum equation, and so it is part of the kinematic response of the cloud to the work exerted by the forces, rather than an additional energy source of work. That is, the work exerted by the forces causes two types of kinematic responses in the cloud: a change in its internal mass distribution, as measured by its moment of inertia, and kinetic energy that can consist of either an internal velocity dispersion or a bulk motion of the cloud.
Strictly speaking, the virial version of equilibrium or force balance should therefore be written as $\ddot{I}/2 - 2 E_{\rm kin} = 0$, while the equality $E_g + 2\ E_{\rm kim} = 0$, rather than equilibrium, implies that the kinetic energy is been driven by gravity.}}
%

\subsection{Reconstructing the energy budget of MCs}\label{sec:reconstructing}

All the caveats mentioned above should not stop us from computing the { 
various work terms due to the different forces}, and derived quantities, in order to { 
approximately estimate} the dynamical state of MCs. For instance, the fact that assuming $\ddot{I}=0$ is incorrect in an evolving cloud should not stop us { 
from obtaining at least} an order-of-magnitude { 
estimate} of the different energies of clouds. In fact, there are several key concepts that can be derived from the VT, and that allow us to understand better the state of the clouds. In particular, approximations to the Jeans length
\begin{equation}
    \lambda_J \sim \biggl(   \frac{{\pi} c_s^2}{G \rho} \biggr)^{1/2}
    \label{eq:jeanslength}
\end{equation}
(where $c_s$ is the isothermal sound speed, and $G$ is the constant of gravity), and the Jeans mass, $M_J \sim \rho \lambda_J^3$, can be obtained from equating the gravitational and thermal energies, and making assumptions on the geometry.  Also, the ``virial parameter'', defined by \citet{BertoldiMcKee92}, 

{
\begin{equation}
    \alphavir = \frac{5 \sigma_v^2\ R}{G\ M} \simeq \frac{2\ E_{\rm kin}}{E_g}, 
    \label{eq:alphavir}
\end{equation}{}
}
{where $\sigma_{v,\rm 1d}$ is the non-thermal, 1D velocity dispersion},
can give insights on whether a cloud could be collapsing or not, although only in order of magnitude due to different uncertainties \citep[][see \S\ref{sec:alphavir}]{Kauffmann+13, Pan+16, BallesterosParedes+18}. Similarly, it can be noticed easily that the gravitational energy
\begin{equation}
    E_{\rm grav} \sim -\frac{3}{5} \frac{G\ M^2}{R}
    \label{eq:Egrav}
\end{equation}{}
and the magnetic energy,

\begin{equation}
    {E_{\rm mag}}  \sim \frac{1}{8\pi} B^2 R^3 \propto \frac{\Phi_{\rm mag}^2}{R}
    \label{eq:Emag}
\end{equation}{}
{(where $\Phi_{\rm mag} = \pi R^2 B$ is the magnetic flux) have the same dependency with size, since in ideal MHD the magnetic field is anchored to the gas such that the magnetic flux $\Phi_{\rm mag}$ remains constant. Thus, the magnetic and the gravitational energies scale with size in the same way, implying that if a cloud has more gravitational energy (in absolute value), the magnetic field cannot prevent collapse, and if it is smaller, collapse will never occur. 
}%

{In summary, evaluating the energies of MCs can provide insights on what the dynamics of MCs is. There are still some observational biases that have to be taken into account. We will {further} discuss these issues in \S\ref{sec:statistics}.
}

\section{Cloud statistics and scaling relations}\label{sec:statistics}

%
    



%
{Statistical analyses of MC properties have been widely used in the literature to understand the dynamical, kinematic, structural and evolutive properties of MCs. 
{These} { 
can be considered under} two basic approaches: { either} analysing a single cloud, with different tracers and/or at different column densities, such that one can determine what is the internal structure of a single object, { 
or} analysing an ensemble of clouds \citep[or cores within a cloud, with either one or more tracers;
see e.g., ][]{Goodman+98}. The { 
most} relevant properties of MCs, which we will discuss in what follows, are
the mass spectrum of GMCs and/or cores, the column density probability distribution, the scaling relations between the mass, the velocity dispersion and the column density, and their size, the virial parameter, the magnetic field, { and} the fractal nature.
}

\subsection{The mass spectrum of MCs and their cores.}\label{sec:MassSpec}

{ 
The {mass spectrum of the clouds---i.e., the distribution of the masses of MCs and their substructures (clumps and cores)---is one of the fundamental properties of the clouds that requires a theoretical foundation.}}
As a consequence, estimates of the mass spectra of GMCs { 
have been} reported, first from surveys of CO clouds, and later { 
from} dust continuum emission (and a few others in dust extinction). The resulting mass distributions show
that there { 
is} no characteristic mass for MCs, but that they { rather} follow, typically, a power-law with a { 
negative} exponent,
\begin{equation}
    \frac{dN}{d\log{M}} \propto M^{-\gamma}
    \label{eq:massspec}
\end{equation}{}
{The estimates of $\gamma$ for large clouds (MCs and GMCs) based on CO observations give values between $0.2$ and $0.9$ \citep{Sanders+85, Solomon+87, Williams+94, HeyerTerebey98, Heyer+01}. The corresponding estimates for dense, compact, prestellar cores based on continuum emission give values of $\gamma$ between $1$ and $2$, \citep[][see also \S 8.3.3]{Motte+98, TestiSargent98, Johnstone+00, Johnstone+01, Motte+01, BeutherSchilke04, Mookerjea+04, ReidWilson05, ReidWilson06a, ReidWilson06b, Stanke+06, Alves+07, Konyves+15, Ohashi+16, Cheng+18}.
This evidence strongly suggests that compact, self-gravitating cores have a mass distribution closer to the \citet{Salpeter55} stellar initial mass function (IMF), for which an exponent of $-1.35$ has been taken as a standard value.  
}

There are, however, {a} few works that do not quite match the  picture of small cores with { an} IMF-like distribution, and large clouds with flat mass spectra. In the first group, using CO data, values of $\gamma \sim 0.7-0.8$ have been found for {small-scale structures}
rather than for MCs and GMCs  \citep{StutzkiGuesten90, Heithausen+98, Kramer+98}.  In the second group, \citet{Rosolowsky+10} show Salpeter-like values of $\gamma$ for large clouds seen in dust emission. 
%
{ 
These studies}
suggest { instead} that the actual value of the power-law of the mass spectra of clumps and cores could be biased due to the { 
} 
technique { used}. 
Among the possible causes playing a role are opacity effects, %
{CO depletion in dense cores}, and { the} clump { 
identification procedure}. Also, it should be noticed that the study by \citet{Rosolowsky+10} is based on interferometric observations, possibly filtering out large scales that could affect the slope of the mass distribution. 

%
{One more caveat should be mentioned: all the mass spectra derived from continuum emission quoted above, with the exception of \citet{Rosolowsky+10}, exhibit a small ($\lesssim$ one order of magnitude) dynamical range in mass over which a Salpeter-like slope can be fitted. In fact, looking carefully at the mass distributions of these studies \citep[see, e.g., ][]{Motte+01, ReidWilson05, ReidWilson06a, Stanke+06, Alves+07, RomanZuniga+10}, rather than a single power-law, one can { 
see} that lognormal-type shapes can be fitted too, as pointed out by \citet{BallesterosParedes+06} { and} \citet{Konyves+15}.}

{Finally, it should be mentioned that, using ALMA observations, the star forming region W43-MM1 exhibits a top-heavy core mass function \citep{Motte+18}.} 

{Although {
} this { 
study} does not distinguish between prestellar and protostellar cores,
and 
result{s} could be quite sensitive to the assumed temperature,
possibly modifying the slope of the mass spectra,} 
{the inferred slope { 
of} $-0.96$ by \citet{Motte+18} is consistent with a value of $-1$, predicted by some models of gravity-driven 
formation of dense cores \citep[see][and references therein]{VazquezSemadeni+19}.  
}

{With the { 
various} caveats 
mentioned { above}, it is clear that 
the relationship between the mass distribution of prestellar cores and the IMF { remains an open problem}. Further high-resolution observations as well as detailed high-resolution numerical simulations, { 
with carefully controlled comparison schemes,} are required. 
}

\subsection{Volume- and column-density probability distribution functions of MCs}\label{sec:PDF}

{The probability distribution function (PDF) of a variable is a one-point statistics
that { 
measures} the relative fraction of volume (or mass) of a fluid that is in a given range of values of the variable under consideration. These functions are often computed as the histograms of the volume and column density field{s}. The { 
fundamental} hypothesis behind {
studies} of the probability distribution function of the volume- and column-density fields (\rhopdf\ and \Npdf, respectively) is that these are sensitive to the physical {
processes operating in} the interstellar medium \citep[e.g.,][]{VazquezSemadeni94, PassotVazquezSemadeni98, Federrath+08, Kainulainen+09,   Tassis+10, Kainulainen+11, Kritsuk+11, 
BallesterosParedes+11b}. {
 However,} the problem is degenerated, since different physical conditions can produce similar shapes of {
the} PDF \citep[e.g.,] [] {VazquezSemadeniGarcia01, Tassis+10, Pan+19}.}

{It is well known that the density field of the galactic interstellar medium in general, and of molecular clouds in particular, is filled up by a low-density {
substrate}, in which denser structures { are embedded and} occupy only a { 
very} small fraction of the volume, i.e., density enhancements have small filling factors, as can be inferred from the rapidly decreasing \Npdf s of MCs \citep[e.g.,] [] {Kainulainen+09}.
}

{Although different tracers have { 
a} characteristic density at which the emission is produced, it is not possible to estimate, in an unambiguous way, the \rhopdf\ of a MC. This is because the length of a MC {
along the} line of sight is always an unknown, and thus, so it is the exact fraction of the volume occupied by the gas that is producing the emission However, the \Npdf\ is easily calculated by computing the total mass in each line of sight. In general terms, unless a particular arrangement of mass occurs,  the \Npdf\ of the interstellar medium in general, and of the MCs in particular, is also a rapidly decreasing 
function.
}

{{
First reported} by \citet{VazquezSemadeni94}, it is generally accepted that the \rhopdf\ of an isothermal, supersonic, turbulent gas  is lognormal \citep[although departures of the lognormality may occur, depending on the nature of the turbulent field, see][and references therein]{Pan+19}.  The reason is the following: an isothermal shock of Mach number $\MM$ produces a density enhancement $\delta \rho$ over the mean density $\rho$ proportional to $\MM^2$, i.e., $\delta\rho/\rho \propto \MM^2$ Then, in an isothermal turbulent fluid with characteristic Mach number $\MM$, {
the amplitude of the} density fluctuations {
is produced} by {
a random} succession of {
passing} shocks. This produces { a 
random distribution of} multiplicative density enhancements, { 
which 
} 
become additive in the logarithm, and thus, by the central limit theorem, this produces a normal (Gaussian) distribution function in the logarithmic density, i.e., a lognormal function. 
}

{
The corresponding \Npdf\ for an isothermal turbulent field, however, {
does not have} a unique functional form \citep{VazquezSemadeniGarcia01}. It transits from a Gaussian to the 
lognormal as the correlation length of the turbulence increases.
The Gaussian is produced when the size of the cloud in the line of sight is large compared to the correlation length of the turbulence.
In such case, by the central limit theorem again, the stochastic occurrence of density fluctuations in each line of sight produces a Gaussian distribution of 
column densities\footnote{It should be noticed that this result assumes that each line of sight is independent. This may not be the case in the case of strong magnetic fields, large scale gravitating structures, or a large correlation length of the turbulence.}. However, if the correlation length and the line of sight become comparable,
this is no longer valid. In this case, the column density in each line of sight is representative of the mean density of the fluid in the line of sight, and thus, the \Npdf\ inherits the shape of the actual \rhopdf, which, in the case of an isothermal turbulent fluid, is also lognormal, although certainly with a smaller width than the original distribution of the volume density { because of the partial averaging performed in each line of sight}.
}


%
{
We now turn to the observational \Npdf\ reported in the literature. An important warning has to be made { 
first}, however: as pointed out by \citet{Alves+17}, any reported shape of the \Npdf\ from observations has to be made only for those contours that are closed in the map. The inclusion of data from non-closed column density contours when computing the \Npdf\ is necessarily incomplete, and thus, the \Npdf\ at those column densities is underestimated, { causing a spurious drop in the PDF}. In a similar way, { 
some} relationships between the physical properties and the shape of the PDFs of clouds in numerical simulations cannot be extrapolated to real life: in numerical simulations the PDF of volume and column density necessarily have a maximum at { 
some intermediate} value, and decrease towards lower densities due to the { 
finite} size and 
mass of the computational box. { 
That is, the boundary of the simulation is in general equivalent to a non-closed density contour for the density statistics in the numerical box.}
}

{
In summary, PDFs of interstellar gas { 
do} not { necessarily have to} decrease at low densities, and inferring physical properties of MCs by fitting the low-density regime of a PDF may be seriously wrong. In consequence, in what follows we will refer
only to the upper-end column density of the \Npdf s. 
}

{
Observationally, \Npdf s of MCs have been reported to be lognormal functions for clouds that do not exhibit substantial star formation, and {
power laws} for actively{-}star forming clouds \citep{Kainulainen+09}. On the other hand,  { 
} numerical simulations
show that the \Npdf\ from turbulent clouds is, typically, lognormal, and that they transit to a power-law as gravity takes over \citep{Kritsuk+11, BallesterosParedes+11b}, 
{which is indeed observed in 
low-star formation efficiency MC{s} in the Galactic Centre environment \citep{rathborne14}.}
}

{
An interesting result arises from the observations quoted previously in this section: the ratio of active-to-non active clouds in the Solar Neighborhood is about $\sim$~10:1 \citep{Kainulainen+09}. Assuming that the shape of the observed \Npdf s reflects the internal physics that produce their structure, this result suggests, at face value, that only 10\%\ of the clouds are dominated by homogeneous turbulence, which furtheremore, is stirred from the large scales 
while the majority of MCs { 
are } dominated by gravity.
}

\subsection{The mass- and the density-size relations}\label{sec:M-R}

{
{ 
The mass-size (or its equivalent, the mean density-size) relation}
has been used 
to analyse the internal structure (a single cloud at different column density thresholds) as well as a statistical relation between a set of clouds, of clumps, or of cores. Both cases provide different information about the properties of MCs, and thus, have to be { 
discussed} independently. 
}

{
Before going into the details, two points need to be made. First
the size of a { 
projected cloud in the plane of the sky} is { typically} computed as the radius of an equivalent circle that has the same area of the cloud. Thus, usually, the reported size is proportional to the square root of the area of the cloud. We notice that, for fractals, this is a mistake, since the estimate of the size of an object must be 
{ given by a quantity that its independent of the area or volume \citep[e.g., its perimeter, see ][]{Falgarone+91, Stutzki93}
}
{ 
Therefore, if clouds have a fractal structure,
the size calculation as the square root of the area is likely to introduce a spurious systematic bias}. Second, 
the functional form of the \Npdf\ has direct implications on the mass-size relation:  given a \Npdf, the area and the mass of the cloud above some column density threshold $N_{\rm th}$ can be computed from the \Npdf\ \citep[see][]{Lombardi+10} as:}
\begin{equation}
    S(N_{\rm thr}) = S_{\rm tot} \int_{N_{\rm th}}^\infty \Npdfmath \ dN
    \label{eq:AreaPdf}
\end{equation}
and
\begin{equation}
    M(N_{\rm thr}) = S_{\rm tot}\ \mu m_H \int_{N_{\rm th}}^\infty N\  \Npdfmath\ dN
    \label{eq:MassPdf}
\end{equation}
where $S_{\rm tot}$ is the total area of the cloud, $\mu$ is the mean atomic weight and $m_H$ is the mass of the hydrogen atom{ , and $N$ is assumed to be in units of cm$^{-2}$}. Note that eqs. (\ref{eq:AreaPdf}) and (\ref{eq:MassPdf}) imply a definition of a cloud, as all the material that is above some column density threshold, $N_{\rm th}$ in a given region.
From these equations it is clear that the surface and the mass of a cloud are the zero{th} and first moments of the \Npdf. 

{As a general result, it can be demonstrated that (i) the mass-size relation for the inner structure of a single cloud (i.e., for a molecular cloud seen at different thresholds) should be a power-law with slope smaller than 2, frequently modified at large radii by a curve with decreasing slope \citep{BallesterosParedes+12}, as seen in observations \citep[see, e.g., ][]{Kauffmann+10b, Lombardi+10}. In contrast, (ii) the mass-size relation for an ensemble of objects should be a power-law with a slope of 2 \citep[e.g.,][]{Beaumont+12, BallesterosParedes+12}, as observed by \citet{Lombardi+10}.
}

{Both results can be demonstrated from eqs. (\ref{eq:AreaPdf}) and (\ref{eq:MassPdf}) in terms of the rapidly decreasing \Npdf s in MCs.  { For} case (i) (the internal structure of a single cloud), if the \Npdf\ is
a power law, i.e., 
$\Npdfmath\ dN \propto N^{-\beta} dN$,  it can be shown that the mass-size relationship necessarily has slopes smaller than 2 for $\beta > 2$ \citep[see Fig.. 12 in][where $n$ is our $\beta$, and $a$ is the exponent of the mass-size relation]{BallesterosParedes+12}, a result observed in nearby MCs \citep{Kainulainen+09, Kauffmann+10a, Kauffmann+10b, Lombardi+10}. {{Also, f}ollowing the formalism of \citet{Lombardi+10}, it can be shown that the {
apparent flattening of} the mass-size relation 
at large radii \citep[see, e.g., ][]{Kauffmann+10b, Lombardi+10} is due to the departure of the \Npdf\ from the power-law behavior at low column densities, {which has been shown to be due to incomplete sampling of clouds at low column density contours \citep[][see \S\ref{sec:PDF}]{Alves+17}}. 
}
It can be shown also that, if instead of a power-law, the \Npdf\ is assumed to be a lognormal, the increasing slope (in absolute value) of the \Npdf\ will produce a decreasing slope in the mass-size relation \citep{BallesterosParedes+12}. 
}

{
In case (ii) (the mass-size relation for an ensemble of clouds or cores defined by a single column density threshold), \citet{BallesterosParedes+12} pointed out that, since the \Npdf\ is a rapidly decreasing function { of $N$}, {with slopes typically steeper than $-2$ \citep{Kainulainen+09}}, averaging the column density above a threshold { 
causes} all clouds { to} exhibit the same mean column density, close to the threshold used. In fact, \citet{BallesterosParedesMacLow02} noticed that if one were able to define clouds using volume thresholds, the resulting mass-size relation should be a powerlaw with exponent 3, since the mean volume density of clouds defined in such a way should be roughly constant if the \rhopdf\ of MCs is a rapidly decaying distribution function. This fact has been circumstantially confirmed by \citet{Kainulainen+11}, who found a mass-size relation with slope close to 3 ($\sim 2.7$), for  a set of clumps defined in such a way that their volume density { 
was} roughly constant.
}

{
{With these two basic results in mind, 
a crucial} question is, then, whether exponents different from 2 { of the mass-size relation} for an ensemble of clouds or cores indicate intrinsically different physical properties, or whether such differences are the result of the procedure or definition used in the analysis. }
In fact, it is noteworthy that, while cloud surveys performed using dust extinction \citep[e.g.,] [] {Lombardi+10} give slopes very close to 2, CO observations typically show exponents larger than 2 \citep[e.g., ][]{Bolatto+08, Roman-Duval+10, Miville-Deschenes+17}. In { all of} these cases, the clouds are defined through intensity thresholds, which are { essentially} proportional to the column density. { 
In view of the above discussion on the effect of selection, it is the excess over a slope of 2 that requires explanation}. { To do so, \citet{BallesterosParedes+19} argued that, while surveys based on dust extinction are restricted to nearby clouds, 
CO surveys can reach up to very large distances, and the probability of random superposition of different clouds along the line of sight, due to non-circular motions in the Galaxy \citep{BlitzShu80, Gomez06}, increases with distance in any given velocity channel. 
In fact, superposing in the line of sight a random mass factor between \diezalamenos 4 and \diezalamenos 1 of the total mass of the cloud can account for mass-size relationships with slopes similar to those found in the observational surveys \citep{BallesterosParedes+19}. }

%

%
{
{ Finally,} in the case of cores within a single cloud, 
different definitions and methods of core extraction may produce different slopes, and thus, there is no
consensus on what is the slope of the mass-size relation for cores within a single cloud. 
For instance, using getsources \citep{Menshchikov+12} or wavelets \citep{RomanZuniga+10}, slopes between 2.2 and 2.7 are typically found \citep[e.g., ][]{Lada+08, RomanZuniga+10, Konyves+15}, although sometimes {weak} correlations are found \citep{Bresnahan+18}. In the case of Gaussclumps \citep[][]{StutzkiGuesten90}, the mass-size relationships exhibit slopes that could be either larger than 2 \citep{Mookerjea+04, Veltchev+18}, as well as smaller \citep{ZhangLi17}. { 
}
}



\subsection{The velocity dispersion-size relation}\label{sec:dv-r}


%
{
Since the first detections of CO \citep[e.g., ][]{Wilson+70}, it is known that molecular clouds exhibit supersonic linewidths. Given the large Reynolds numbers  in the interstellar medium, ${\cal R} = v_l\ l/\nu \sim 10^5-10^8$ {\citep{Myers83, Miesch+99, ElmegreenScalo04}}, with $v_l$ the characteristic velocity at the scale $l$, and $\nu$ the kinematic viscosity of the fluid, molecular clouds are necessarily turbulent, and thus, naturally, 
%
the dynamical role of such turbulence on the evolution of MCs {has been a mat{t}er} of intense analysis and debate. }


%
{
It should be recognised that in the astrophysical argot, { the term }turbulence{} has been somehow, synonymous of either supersonic turbulence,  {and/or} support against gravity. Certainly, subsonic turbulence cannot account for support against self-gravity better than what thermal pressure does, but it should be noticed that there are regions in the interstellar medium, and within molecular clouds, where the linewidths become sub- or trans-sonic, and that they could be very well { subsonically} turbulent{, similarly to all natural terrestrial turbulent flows}. 
}

{
The velocity structure of MCs is substantially more difficult to determine than the{ir} mass structure: by 
{
considering spectroscopic data from molecular-}line transitions 
we add a 3rd coordinate, and thus, the total emission has to be divided between different velocity channels, lowering the signal-to-noise ratio of the emission. In addition, different molecules frequently exhibit different spatial structure, either because of opacity
{or}
chemistry effects. As pointed out by, e.g.,  \citet{Traficante+18a}, the volume from where the emission comes in one case or another
may play a role in the determination of fundamental physical properties, { such} as 
the virial parameter. 
{ Nonetheless, assuming that the velocity structure measured with a particular tracer is representative of the volume it traces is approximately valid for most molecules, besides CO and its isotopes (Tafalla et al. 2020, in preparation).}
}


%
{
One of the { most} frequent arguments in favour of supersonic turbulence as a universal dynamical ingredient in molecular clouds is the existence of a correlation between the velocity dispersion ($\sigmav$) and the size ($R$) for ensembles of clouds, over $4-5$ orders of magnitude in size \citep[e.g.,] [] {Falgarone+09}. In { his} original work, \citet{Larson81} reported the correlation  
}
\begin{equation}
    \sigma_v \propto R^p,
\end{equation}{}
{with $p\sim 0.38$, over 3 orders of magnitude in size. \citet{Larson81} interpreted this correlation as  incompressible turbulence in molecular clouds, {
 since the exponent is close to that predicted by \citet{Kolmogorov41} for incompressible (i.e., subsonic) turbulence. However, this} is {mgt
inconsistent} with the fact that the velocity dispersion is typically supersonic in the \citet{Larson81} data. 
}

{
The canonical exponent for the velocity dispersion-size relation has been taken as $1/2$, not only because, observationally, this value has been found in a variety of studies \citep[e.g.,][see also Falgarone et al. (2009) and references therein]{Myers83, Solomon+87, OssenkopfMacLow02, HeyerBrunt04}, but because, theoretically, $p\sim 1/2$ is the expected value for a turbulent velocity field dominated by shocks \citep[see, e.g., ][and references therein]{Passot+88, VazquezSemadeni99, McKeeOstriker07}. 
}

\begin{figure}
\sidecaption
\includegraphics[scale=0.4]{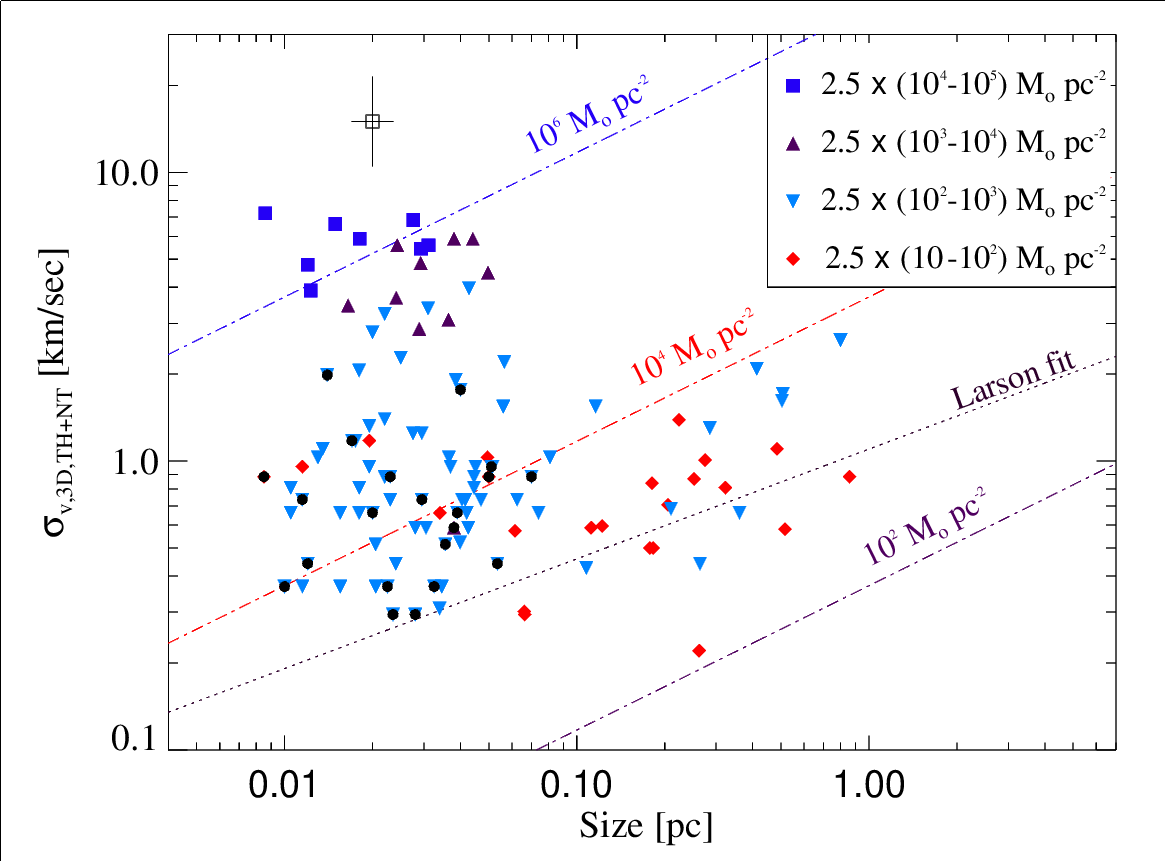}
%
%
\caption{Velocity dispersion-size relation for cores in regions of massive star formation. Clearly, there is not a clear correlation between the size and the velocity dispersion. There is a tendency, with substantial scatter, of cores with large column densities to exhibit large velocity dispersion. Figure from \citep{BallesterosParedes+18}.
\label{fig:dv-r}
}
\end{figure}

Although this result {is apparently well settled \citep[see e.g., ][]{HennebelleFalgarone12, KlessenGlover16},}
the detailed slope of the velocity dispersion-size relation is far from {
being firmly} established. There is still a large list of observational works  where the exponent is typically smaller than 1/2 \citep[e.g.,][to quote a few]{Carr87, CaselliMyers95, Loren89b, Plume+97, Shirley+03, Traficante+18b}. In some cases, $p$ has been found to be even negative \citep[e.g., ][]{Wu+10}, and {
in others,} $p>1/2$, not only for the Central Molecular Zone \citep{Shetty+12, Kauffmann+17}, but even for the ensemble of clouds in the whole galaxy, \citep{Miville-Deschenes+17}. 
Similarly, {\citet{BallesterosParedes+11a}
showed that} {
when massive-clumps are considered together with MCs, the relation is lost, and only a 
a scatter plot remains}. Fig.~\ref{fig:dv-r} shows the velocity dispersion-size diagram for a list of {low-mass} cores in 
{regions of high-mass star formation,}
where there is no clear trend between the velocity dispersion and the size. {{
Nevertheless}, there is a trend to notice:} high column density cores exhibit larger velocity dispersions than low column density cores. This result is expected
 {{ if} either 
} (a) cores are collapsing { (see Sec.\ \ref{sec:L-Sigma})}, or 
(b) localised feedback is present, since it could increase the velocity dispersion within the cores. However, the core selection was based on ammonia emission, which typically is destroyed if UV feedback is present. In addition, the cores 
shown in Fig.~\ref{fig:dv-r} were carefully selected to avoid morphological evidence of feedback \citep{Palau+14}. For these reasons, \citet{BallesterosParedes+18} tend to favour the first option, that the larger velocity dispersion for larger column density cores is due to collapse. We will discuss this point in \S\ref{sec:L-Sigma}.


{
}

%
{
Another approach to the velocity dispersion-size relation in GMCs consists in measuring 
velocity differences {with}in the 
GMC. There are two ways for approaching it: (i) measuring the structure function $S_p(l)$ of the velocity field \citep{HeyerBrunt04, Hily-Blant+08}
\begin{equation}
    S_p(l) = \langle |v(r) - v(r+l)|^p\rangle,
    \label{eq:structurefunction}
\end{equation}
(where $l$ is the spatial displacement between two {
pixels in} the map of the cloud, and $p$ is the order of the structure function), {and (ii) measuring the moments of the PDFs of the centroid velocity differences, as a function of the lag \citep{MieschScalo95, Miesch+99}.}
The results from these studies are {
in general} consistent with the clouds' dynamics {
being} dominated by turbulence. However, it is important to notice that none of these approaches
is equivalent to the velocity dispersion-size relation of clumps and cores within clouds: while the latter measures the motions exclusively in density enhancements (cores or clumps) of size {
$\ell$}, the former two estimate the relative motions between points separated by a distance {
$\ell$} {for the whole field}; i.e., regardless of whether there is a density enhancement or not at each point. For small scales, for instance, the structure function has more weight from the low-density regions than from the high density regions of the cloud, just because the filling factor of the latter is small. {{
This difference in the sampled regions may introduce biases that must be taken into account when comparing results from the two methods}.
} 
}

{}
At the smaller scales, however, there are    velocity dispersion-size relations reported for single {
resolved cores. These are obtained, for example, by performing a succession of pointings at different distances from the core center and plotting the linewidth as a function of distance from the center \citep[e.g.,][]{Pineda+10}}. While some of them exhibit {
power-laws
consistent with  $p\sim{1/2}$ \citep[e.g., ][]{Rosolowsky+08}, there is a {
class} of cores for which velocity dispersion-size relation becomes flat at small scales{, at essentially the speed of sound, implying a subsonic turbulent velocity dispersion}
\citep[e.g., ][]{BarrancoGoodman98, Goodman+98, Caselli+02, Tafalla+04, Pineda+10, Chen+19}, 
These are the 
so-called (velocity) ``coherent cores''. Three different physical interpretations for these cores have been made: some authors {have} interpreted the lack of velocity and density structure 
{
as an indication that these cores are} ``islands of calm'' within a turbulent sea \citep{Goodman+98, Pineda+10}. \citet{Klessen+05}, on the other hand, found that turbulent motions can produce coherent cores by shock compressions, {
so} that a coherent core is the stagnation point of the {
velocity field behind the shock}. Finally, \citet{NaranjoRomero+15} {
have pointed out that the spherical collapse solution for cores in the protostellar stage contains a central region characterized by a flat density profile and an infall speed that decreases linearly with decreasing distance from the center. Therefore, if the nonthermal part of the linewidth in these regions is dominated by the infall speed, it should decrease toward the center, as observed.}}

 {In summary, it is not clear that there is a single value for {
the slope of the velocity dispersion-size relation}; different kind{s} of objects can exhibit different {
slopes}, depending {on} how the {object} selection is made. 

}
\subsection{The Larson ratio-column density diagram}\label{sec:L-Sigma}

{
{ A breakthrough in the interpretation of Larson's scaling relations was advanced by} \citet{Heyer+09}, {who} generalized {
them by noticing that the so-called} Larson ratio\footnote{We call the Larson ratio the square root of the so-called velocity scaling, $C\equiv \sigma_v^2/R$, in the extragalactic literature.} 
$\LL \equiv \sigma_v/R^{0.5}$, {
for Galactic GMCs} depends on the mass column density $\Sigma = \mu m_H N$ as 
}

\begin{equation}
\LL \propto \Sigma^{0.5},
\label{eq:heyer}    
\end{equation}
{ instead of being constant, as would be required by Larson's velocity dispersion-size relation and/or by strongly supersonic turbulence.}
This result has been confirmed by other authors, either for Galactic objects
\citep[e.g., ][]{kruijssen13, Miville-Deschenes+17, Traficante+18c, BallesterosParedes+18}, as well for extragalactic \citep[e.g., ][]{Leroy+15, Sun+18, ImaraFaesi19}, although with substantial scatter. 

{ As pointed out by \citet{Heyer+09}, this scaling is consistent with the clouds being in virial equilibrim, when the cloud sample is not restricted to a constant column density, as it follows directly from the condition $2E_{\rm k} = E_{\rm g}$, where $E_{\rm k} = M \sigma^2/2$, $E_{\rm g} = 3 GM^2/5R$, with $\sigma$ being the one-dimensional velocity dispersion, $M$ the cloud's mass and $R$ its radius, and defining the column density as $\Sigma = M/\pi R^2$. In addition, \citep{BallesterosParedes+11a} further pointed out that the scaling is also consistent with the clouds undergoing free-fall, in which case $E_{\rm k} = |E_{\rm g}|$. It is important to note that the free-fall condition in the $\LL$--$\Sigma$ diagram coincides with what is often referred to as ``marginal binding", even though it corresponds to actually stronger binding than virial equilibrium. For practical purposes, the difference between free-fall and virial equilibrium in the $\LL$ vs.\ $\Sigma$ diagram is within the typical uncertainty of the observations. Nevertheless,} {
at face value}, most of the clouds in the \citet{Heyer+09} sample appear { slightly} overvirial, { and more consistent with free-fall. 
} A similar result is seen in 
extragalactic {GMC} data as well as {
other} Milky Way data { \citep[e.g.,] [] {Sun+18}.
}

{ Furthermore, \citet{BallesterosParedes+11a} showed that the scaling given by eq.\ (\ref{eq:heyer}) holds not only for GMCs, but extends to massive dense cores, for which the Larson velocity dispersion-size relation does not hold. Thus, those authors suggested that the velocity dispersion in both GMCs and dense cores is driven by self-gravity, and specifically consists of infall motions, albeit highly chaotic, due to the presence of turbulence and of multiple collapse centers \citep[see also] [] {Vazquez+19}. However, eq.\ (\ref{eq:heyer} cannot be satisfied simultaneously with the standard Larson relation $\sigma \propto R^{1/2}$ in objects of differet column densities. Instead, clumps and cores of higher column density are expected to exhibit different loci in the $\sigma$-$R$ diagram, as observed \citep[] [see also Fig.\ \ref{fig:dv-r}] {CaselliMyers95, Plume+97, BallesterosParedes+11a}. Note, however, that an alternative interpretation to the different $\sigma$-$R$ scaling in dense cores is that their gravitational contraction is delayed by turbulent pressure \citep{Murray_Chang15, Xu_Laz20}.}

\begin{figure}
\sidecaption
\includegraphics[scale=0.35]{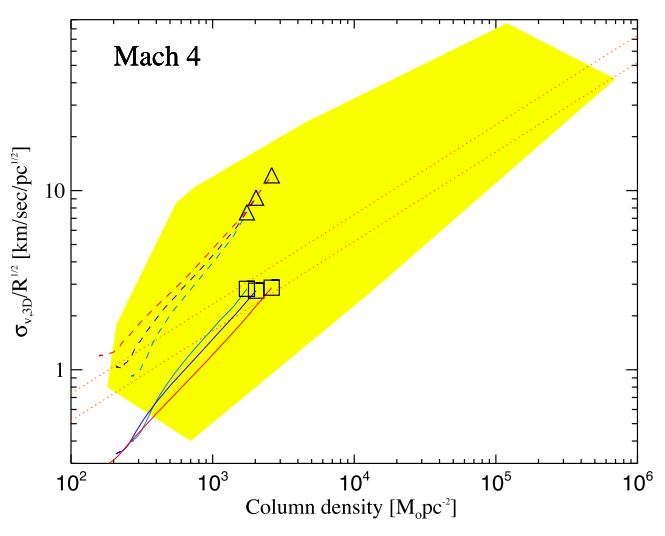}
\includegraphics[scale=0.35]{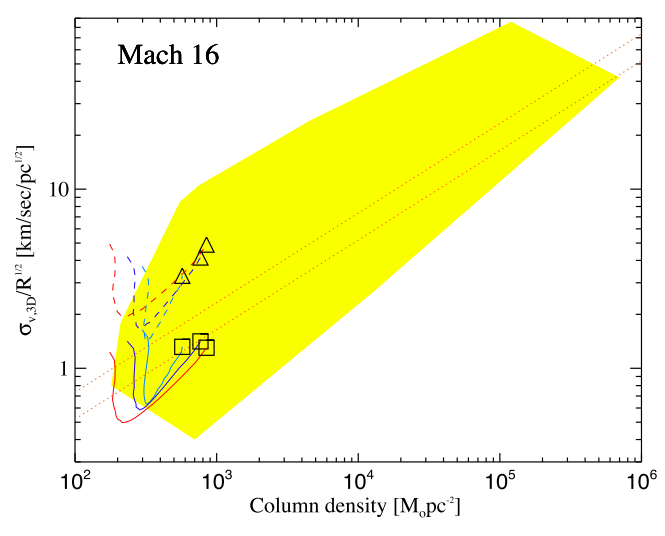}
%
%
\caption{Evolution in the $\LL-\Sigma$ diagram of collapsing cores with initial velocity fluctuations. When the approximation $|Eg| \sim 3GM/5R$ is used, cores frequently appear  overvirial (dashed lines). When the correction to consider the actual gravitational energy is taken into account in the calculations, collapsing cores appear subvirial (solid lines). The yellow region denotes the locus occupied by observed cores in the sample studied by \citet{BallesterosParedes+18}. The dotted lines denote the locus where virial (below) and ``free-fall" (above) cores should fall. \citep[Figure from][]{BallesterosParedes+18}.
\label{fig:L-Sigma}
}
\end{figure}

{ Although the scaling (\ref{eq:heyer}) has been observed to hold for column densities from 10 to $\gtrsim 10^4$ \Msun pc$^{-2}$ \citep[e.g.,] [] {Leroy+15}, significant systematic deviations from it are observed as well. Low-column density objects (the more diffuse clouds, with $\Sigma \sim 10$--100 \Msun pc$^{-2}$) often appear to be strongly over-virial, while high-column density clumps and cores, with $\Sigma \gtrsim 10^3$ \Msun pc$^{-2}$ often appear subvirial or supervirial. The standard interpretation of super-virial low-column density clouds is that they are confined by a large external pressure, since they are gravitationally unbound \citep{Keto_Myers86, Field+11}. However, \citet{camacho2016} found in numerical simulations of cloud formation that such objects are in general out of equilibrium, and are either dispersing or being assembled by externally-driven compressions. On the other hand, for dense cores that appear supervirial, \citet{BallesterosParedes+18} showed that this may be simply an artifact of the fact that}
%
the actual gravitational energy of a structured core can be substantially larger (in absolute value) {
than} the gravitational energy of a\sout{n} homogeneous sphere with the same mass, velocity dispersion, and size {
of the core} \citep{BallesterosParedes+18}. This occurs because the gravitational energy of clouds and cores is typically underestimated by using the approximation $|E_g| \sim 3GM^2/5R$ , and thus, collapsing clouds/cores can appear to be unbound when they are not. In Fig.~\ref{fig:L-Sigma} we show the situation: collapsing cores exhibit overvirial values when the approximation to the gravitational energy is used (dashed lines), but sub-virial values evolving to virial after $\sim$ one free-fall time when the actual gravitational potential is taken into account in the calculation. This result is consistent with observations by \citet{Cesaroni+19}, who show collapsing line profiles in clouds with masses smaller than their virial mass.

%
%

{ Finally, \citet{BallesterosParedes+18} also showed that, if a core is formed by an inertial externally-driven (i.e., not due to self-gravity) compression and it gradually transitions to collapse driven by its own self-gravity, its evolutionary track in the $\LL$-$\Sigma$ diagram starts out overvirial, then becomes subvirial, and finally approaches equipartition (a generic term for $E_{\rm k} \sim E_{\rm g}$). This is because the core first starts with a kinetic energy (the external compression) larger than its gravitational energy. As the core becomes denser and smaller, its gravitational energy increases, and so the ratio $\LL$ decreases, and can in fact become subvirial, because the core may start with an infall speed smaller than the free-fall speed, since it starts infalling from a finite radius, not from infinity. Finally, as the core approaches the free-fall speed and the turbulent speed becomes negligible in comparison, it approaches equipartition. These trajectories are illustrated in Fig.\ \ref{fig:L-Sigma}.
}

\subsection{The virial parameter}\label{sec:alphavir}

\begin{figure}
\sidecaption
\includegraphics[scale=6.5]{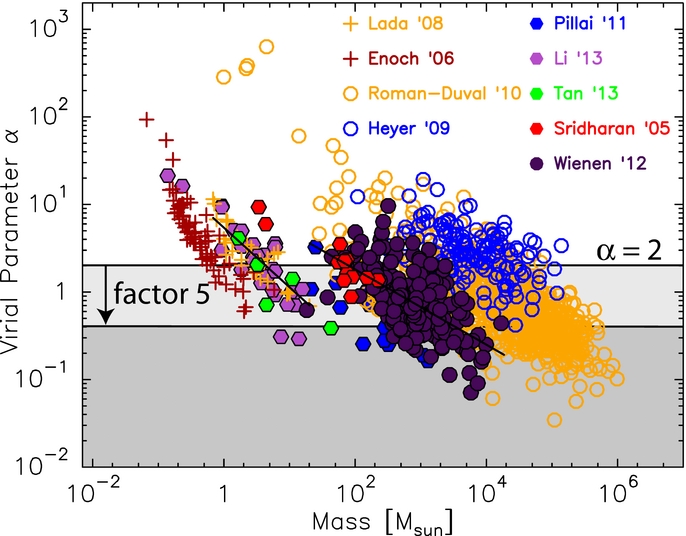}
%
%
\caption{Virial parameter as a function of mass for different samples of cores and clouds in the literature. Figure from \citet{Kauffmann+13}.
}
\label{fig:Kauffmann+13}       
\end{figure}

%
{
The relative importance between self gravity and kinetic energy has been studied since the work of \citet{Larson81}, who {
showed that the} so-called virial parameter {$\alphavir$}  {(see eq. [\ref{eq:alphavir}])}
}
%
%
%
{ {
of MCs has} values around unity.  Nevertheless, subsequent work showed a variety of virial ratios, as seen in Fig.~\ref{fig:Kauffmann+13} \citep[from][]{Kauffmann+13}. This figure shows {data from various studies, and} that, for each {one of them}, $\alphavir \propto M^{-\delta}$,  with $0 \lesssim \delta \lesssim 1$ \citep[e.g., ][]{Carr87, Loren89b, BertoldiMcKee92, Miville-Deschenes+17, Traficante+18b}.  Assuming the standard values for the Larson relations, $\sigma_v\propto R^{1/2}$, and $M\propto R^2$, the virial parameter should be constant. In general, if $M\propto R^q$, and $\sigmav\propto R^p$, $\delta = (2p+1)/q-1$. However, uncertainties may play a crucial role in the determination of the slope of the relation $\alphavir-M$, as discussed by \citet{Kauffmann+13}. For instance, \citet{Miville-Deschenes+17} found a strong $M-R$ correlation, with $q=2.2$, and a weaker correlation $\sigmav-R$ with $p=0.65$. At face value, $\delta\simeq 0$, but in practice, these authors found $\delta \sim -0.5$. A better estimation of $\delta$ is to assume that the  $\sigma-R$ correlation is weak enough to actually not contribute to the $\alphavir-M$ correlation. Indeed, using the reported $q=2.2$, and assuming $p\sim0$ because of the poor Pearson coefficient, one obtains $\delta\sim -0.54$, comparable to the value of $-0.53$ reported by \citet{Miville-Deschenes+17}.

}

{
Although {
values} of $\alphavir < 1$ are {
reported} in a {
number} of recent observations \citep[e.g., ][]{Giannetti+14, Traficante+18a, Zhang+18, Russeil+19,  Nguyen-Luong+20, Li+20}, there is also a {
significant number of studies} where $\alphavir$ is reported to be systematically larger than 1--2 \citep[e.g., ][]{Berne+14, Tsitali+15, HernandezTan15, Barnes+16, Storm+16, Kirk+17, Kerr+19,  Colombo+19, Zhang+20b}. This fact has been interpreted as evidence {
that MCs are often unboud} {\citep[e.g., ][]{Dobbs+11a}}, {
and thus} require pressure confinement {to remain coherent} \citep[e.g., ][]{Field+11, Leroy+15, Sun+18}. }
{
There are, however, {a} few caveats that need to be taken into account on this respect. As commented above (see \S\ref{sec:VT}), not all the kinetic energy is available for support against collapse. Indeed, a pure{ly}-collapsing cloud will exhibit values of $\alphavir>1$ after $\sim$ one free-fall time just because the free-fall velocity is $\sqrt{2}$ times larger than the virial velocity. In fact, it is easy to see that a Bonnor-Ebert core supported by microturbulence, rather than thermal pressure, has an $\alphavir > 1$.  
{On the other hand, {the} velocity dispersion may be overestimated if the molecular lines are optically thick \citep{Phillips+79, GoldsmithLanger99, Hacar+16}.} For instance, in {the} L1517 region, part of the Taurus molecular cloud complex, \citet{Hacar+16} estimate { that the opacity broadening of the line is increasing the actual velocity dispersion by a factor of 4. Taking {this result} at face 
value, the {4-fold over estimation in the} velocity dispersion will give an overestimat{ion by a factor of $\sim 16$} in the virial parameter.} 
}

{
There is another potential problem with the interpretation of the virial parameter:\footnote{{We thank the anonymous referee for pointing out this issue.}} the {
studies by} \citet{Heyer+09} and \citet{Roman-Duval+10} shown in Fig.~\ref{fig:Kauffmann+13} {
are performed using} the same observational dataset, but {using} different {operational} definition of clouds, {with the result that each study produces 
} substantially different {
values} of $\alphavir$. In {
relation to this}, \citet{Traficante+18b} argue
that the volumes from which the dust emission and the line-emission come, if different, may also affect the estimation of $\alphavir$. Thus, cloud definition plays a relevant role in the estimation of the molecular cloud properties, in particular, $\alphavir$ \citep[see also ][]{Khoperskov+16}.} 

{
}

\subsection{Magnetic fields}

{
Diffuse, atomic and dense molecular clouds are partially ionized, and thus, both are subject to magnetic forces. Thus, they are relevant to understand the formation and evolution of MCs (see \S\ref{sec:ThermalInstability}), as well as the formation of stars. However, evaluating the relative dynamical importance of magnetic fields compared to other physical ingredients is quite difficult, either because of the highly non-linear behavoir of the magnetized ISM, as well as because the full information (intensity and direction) {of} magnetic fields is not available from observations.
}

{
In terms of the non-linear behavior of the fields, {as a} first approximation, one can expect that {stronger the magnetic fields}, the smaller the relative importance of gravity, such that collapse could be delayed \citep[see e,g, ][]{Shu+87, NakamuraLi05,HennebelleInutsuka19}. Although this is true in a static medium, the response of the magnetic field in the presence of other agents {such as} turbulence, rotation, or non-linear instabilities is not straightforward. For instance, a moderate magnetic field in a turbulent environment can have a stronger opposition to turbulent compressions than strong magnetic fields, mainly because of two reasons: on one hand, turbulence can {entangle} more efficiently the field, producing a larger and more isotropic magnetic pressure \citep{BallesterosParedesMacLow02}. By the same token, stronger fields can inhibit more efficiently the turbulence, resulting in higher rates of collapse \citep{Zamora-Aviles+18}. As a result, 
smaller density enhancements can be found in turbulent simulations with weaker fields, compared to simulations with stronger fields. 
}

{
In a similar way, in rotating galactic disks, the Coriolis force can inhibit the formation of density enhancements when the magnetic fields are weak. However, {an increase in the magnetic field} intensity can counteract the Coriolis force, allowing for larger density enhancements,  \citep[and larger star formation rates, see, e.g., ][]{Passot+95, KimOstriker01, Kim+02}. Thus, one has to be careful when relating the star formation properties with the magnetization of the field. In fact, \citet{Soler19} found no evidence of relationship between the magnetic fields and the star formation rate for a set of clouds in the Solar Neighborhood. 
}

{
In terms of the observations, on the other hand, {Zeeman} effect can provide us information about the intensity of the magnetic field along the line of sight, but they are quite {time intensive}. On the other hand, polarization measurements are less {time intensive}, and can provide qualitative information about the morphology of the fields (and its relationship to the morphology of the clouds) as projected in the plane of the sky, but obtaining information about the intensity is only {accurate to an} order of magnitude. Nevertheless, even with those difficulties, we can infer valuable information about the relative importance of the magnetic fields in clouds and cores. 
}

\subsection{Morphology of the magnetic field at the cloud scale}\label{sec:BfieldMorphology}

{
Measuring polarization of visible light dates $\sim$~70 years back in time \citep{Hiltner49, DavisGreenstein51}. The detailed mechanisms that produce polarization are summarized elsewhere \citep{Heiles+93, Crutcher12}. One of the main successes of polarization measurements is that it provides us information about the relative orientation of the magnetic fields with respect to the cloud. Thus, having a qualitative idea on how the morphology of clouds and fields could be under different physical situations, the relative importance of magnetic fields in the region could be estimated.  With this idea, \cite{Crutcher12} discussed how the relative morphology of the magnetic field and the cloud should be in the cases of strong and weak magnetic fields, as well as in the cases where gravity is or not relevant. 
}

{
In the case of strong fields, \citet{Crutcher12} argues that (i) magnetic field lines should be smooth. (ii) If gravitationally unbound, density structures should be aligned along the field lines.  (iii) If gravitationally bound, they should be perpendicular to the field lines. (iv) After collapse, the field should have hourglass morphology.  In contrast, in the case of weak fields, (i) turbulence should make the field to appear substantially more tangled. (ii) If gravity is not important, column density should be aligned to magnetic fields, as in the previous case. (iii) If clouds are gravitationally bound, the field should be progressively more ordered, as collapse proceeds, (time-dependency). (iv) As in the previous case, after collapse the hourglass morphology should appear, but the pinch angles should be smaller in the strong-field case, due to the larger resistance of the field to be dragged. 
}
{
With his in mind, \citet{Crutcher12} argues that the polarization maps quoted in the literature, which are quite smooth and show similar directions of the magnetic fields from large (MC) scales to small (core) scales, suggest that magnetic fields are strong enough to not be twisted by turbulence significantly. 
}

\begin{figure}
\sidecaption
\includegraphics[scale=0.45]{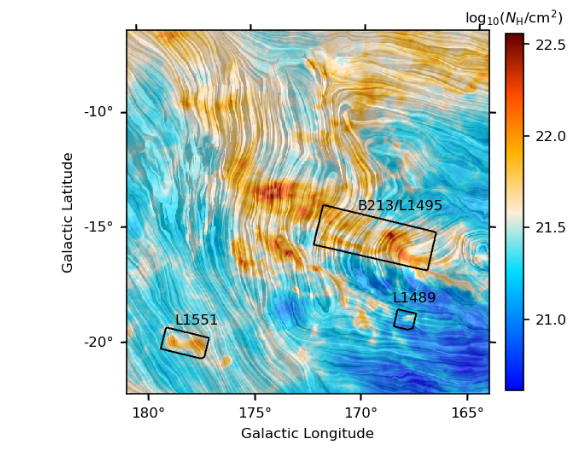}
\includegraphics[scale=0.45]{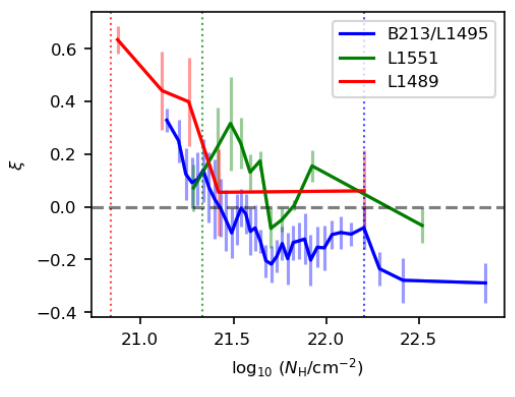}
%
%
\caption{Left panel: plane of the sky magnetic field and column density measured by Planck toward the Taurus molecular cloud. Right panel: relative orientation parameter $\xi$ between the magnetic field in the plane of the sky and the orientation of the filament, for 3 filaments in Taurus \citep[from][]{Soler19}. 
}
\label{fig:Soler19}       
\end{figure}

{
The results discussed by \citet{Crutcher12} were settled in a more quantitative  way using the histogram of relative orientations method \citep[][]{Soler+13,soler2017}. Indeed, \citet{Soler19} analyzed Herschel and Planck observations of local clouds. He found that the relative orientation between the column density of MCs and the magnetic field in the plane of the sky transitions from nearly 0$^\circ$ at low column densities, to 90$^\circ$, at large column densities (see left panel in Fig.~\ref{fig:Soler19}).
Furthermore, \citet{Soler19} found no correlation between the star formation rates in those clouds, and the relative orientation between the column density and the field. Such results suggest,then, that the magnetic fields could be relevant enough to shape the large-scale structure of MCs, but not strong enough to regulate star formation.
}

{
Such morphology has an interesting implication regarding the scaling of the magnetic field: compressions along the magnetic field lines increase the density, but not the magnetic field intensity. Thus, one can expect that in the magnetic field dominated regime, the field does not {correlate} with the density. In contrast, compressions perpendicular to the field lines will increase {both} the density and the magnetic field, and thus, one can expect {the magnetic field to be related to the volume density}. In general, the situation can be expected to be between these limits, and thus, it can be expected that the magnetic field and the density behave as
}

\begin{equation}
B =\begin{cases}
    B_0\ n^\kappa, & \text{if\ $n\geq n_0$}.\\
    B_0,           & \text{if\ $n< n_0$}.
  \end{cases}
  \label{eq:Bmodel}
\end{equation}{}
{for a given volume density $n_0$}, and
{with $0\leq \kappa \leq 1$. {Using Bayesian analysis for a set of clouds with magnetic field measurements, \citet{Crutcher+10} found that $n_0\sim$~300~cm\alamenos 3, and $\kappa\sim 2/3$.}
We will get back to this point in the next section, where this assumption is made as a proxy, in order to interpret the line-of sight intensity of the fields. 
}

\subsection{Magnetic field intensity and collapse}\label{sec:BfieldIntensity}

{
There are two ways to estimate the intensity of magnetic fields. One is with the Chandrasekhar-Fermi method \citep{ChandraFermi53}, the other through the Zeeman effect. Details on the methods can be found in \citet{Crutcher12} and references therein \citep[e.g.][]{crutcher2019}. Here we just make few  comments: in the first case, the method relies on two basic assumptions: (i) that the magnetic field is tangled by isotropic turbulence, and (ii) that there is equipartition between the kinetic and magnetic energy. In such case, the mean magnetic field becomes
}

\begin{equation}
    \langle B\rangle^2 = {4\pi \rho} \ \ 
    \frac{\sigma_v^2}{\sigma_\phi^2} 
\end{equation}{}
{
where $\sigma_v$ is the line-of-sight velocity dispersion, and $\sigma_\phi$ is the dispersion of the orientation angles of the magnetic field.  Although this method can give us order of magnitude estimates of the magnetic field, departures by a factor of two orders of magnitude can be found, specially in the weak fields case \citep{Heitsch+01}. %
{Some modification of the original CF method are also proposed 
for more accurate estimate.}} 

{
On the other hand, the Zeeman effect can give us direct measurements of the projection of the magnetic field along the line of sight. The problem with the Zeeman effect is that it is observationally expensive. 
{Another difficulty is that 
there are only a few species showing observable level of the Zeeman shift, e.g., HI, OH, CN, and CCS. These species have large magnetic dipole moments which cause large frequency splitting.} In addition, since we obtain only the intensity of the field along the line of sight, we obtain only a lower-limit estimate of the intensity of the field. 
}

{
Once we have estimates of the magnetic field, as in the case of the kinetic energy, one could evaluate the relative importance of magnetic fields with respect to gravity through the magnetic-to-gravitational energies ratio. In this case, such ratio is proportional to the inverse of the so-called mass ($M$)-to-magnetic flux {($\Phi_{\rm mag} \sim B\ R^2$)} ratio,
}

{
\begin{equation}
    \frac{\Emag}{\Egrav} \propto \bigg(\frac{\Phi_{\rm mag}}{M}\biggr)^2 = \biggl( \frac{B}{N}  \biggr)^2
\end{equation}{}
}
{
(see eqs.  [\ref{eq:Egrav}] and [\ref{eq:Emag}], where $N\propto M/R^2$ is the column density). In ideal MHD, the magnetic flux is constant, since the magnetic field is frozen to the gas. Thus, there is a critical value for which the magnetic field cannot prevent collapse, and thus, it is common to define
}
{
\begin{equation}
    \mu = \frac{(M/\Phi_{\rm mag})}{(M/\Phi_{\rm mag})_{\rm cr}} \sim \frac{(N/B)}{(N/B)_{\rm cr}}
\end{equation}{}
}
{
such that clouds with $\mu>1$, { (usually called suppercritical)}, will collapse, while clouds with $\mu<1$ {(subcritical)} will be supported by the magnetic field. 
}

{Recalling \S\ref{sec:reconstructing}, in ideal MHD, the magnetic and gravitational energy of a cloud with fixed mass have the same dependency with size, implying that $\mu$ is fixed. \citet{MestelSpitzer56} proposed ambipolar diffusion as a mechanism to increase the gravitational energy without increasing the magnetic energy at the same rate. In this process, while the ions are anchored to the field lines, the neutral particles can drift through them, falling into the gravitational potential well. This results in an increase of the 
mass-to-flux ratio, allowing magnetically subcritical clouds to eventually become supercritical and collapse.}

\begin{figure}
\sidecaption
\includegraphics[scale=0.3]{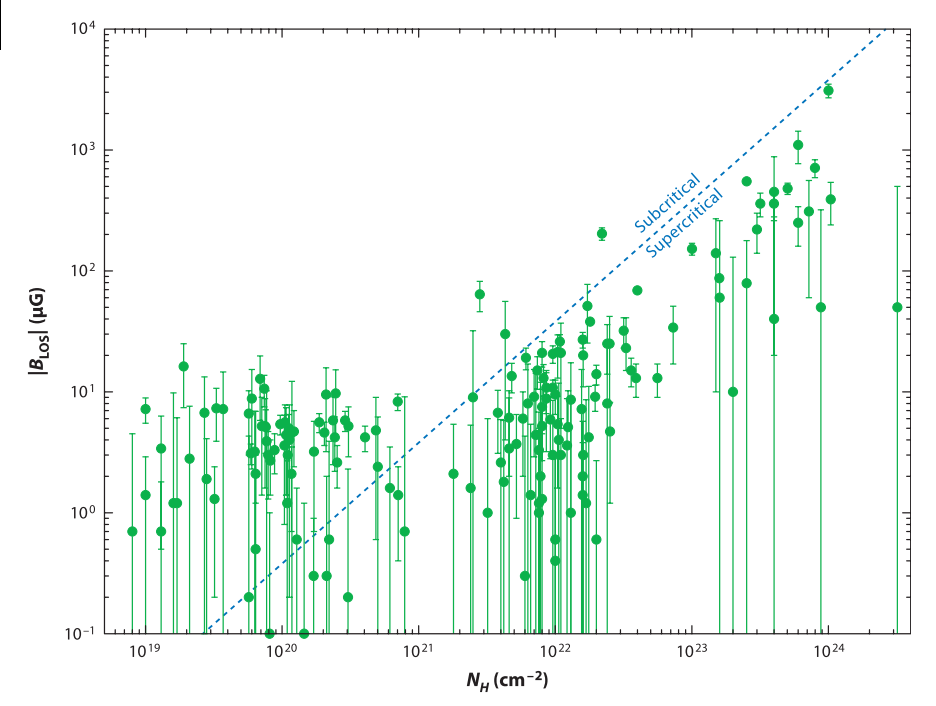}
%
%
\caption{Estimated magnetic field intensities in the line of sight {{\it vs.} column density of H~I and molecular clouds } \citep[from][]{Crutcher12}.
}\label{fig:Crutcher12}       
\end{figure}

{
Estimates of the magnetic field intensity in the line of sight and column densities {from four different surveys \citep[][]{Crutcher+10, HeilesTroland04, TrolandCrutcher08, Falgarone+08}, and compiled by  \citet{Crutcher12},} are shown in Fig.~\ref{fig:Crutcher12}. The dashed line shows the value of the critical ratio $(N/B)_{\rm cr}$. {Data points lying above it imply that these regions are subcritical, and supercritical if they lie below it. As can be seen from this figure, dense molecular cloud cores appear to be supercritical, and low-density H~I clouds appear to be subcritical.
}
}

{\citet{Crutcher+10} pointed out that such distribution of data points could be potentially interpreted as ambipolar diffusion regulated collapse: low column density clouds ($N > 20^{21}$cm$^{-2}$) appear to be supported against gravity by magnetic fields, while large column density clouds have increased their mass-to-flux ratio due to ambipolar diffusion, appearing then supercritical. However, these authors caution that ambipolar diffusion cannot operate because low column density clouds are not self-gravitating, and thus, there is no potential well to which the neutrals will fall in. The results by \citet{Crutcher+10}, along with the morphological results by \citet{Soler19}, suggest that magnetic fields are relevant in diffuse clouds, but that dense molecular clouds in fact should be formed by compressions along the {field lines}, and that, once formed, they are mostly supercritical. 
}

\section{Cloud substructure: from filaments to cores}\label{sec:substructure}
\subsection{Universality of filamentary structures in MCs}\label{sec:fil}

While interstellar clouds have been known to be filamentary for a long time 
\citep[e.g.][and references therein]{Schneider+79, Bally+87, Hartmann02, Myers09}, 
{\it Herschel} imaging surveys have established the ubiquity of 
filaments on almost all length scales 
($\sim 0.5\,$pc to $\sim 100\,$pc) in Galactic molecular clouds and shown that this filamentary structure likely plays a key role in the star formation process 
\citep[e.g.][]{Andre+10, Henning+10, Molinari+10, Hill+11, Wang+15}. 
In particular, filamentary structures appear to dominate the mass budget of MCs at high densities 
($\simgt 10^4\, {\rm cm}^{-3} $)  \citep[][]{Schisano+14,Konyves+15}.

The interstellar filamentary structures detected with {\it Herschel} span broad ranges in length, 
central column density, and mass per unit length \citep[e.g.][]{Schisano+14,Arzoumanian+19}.
In contrast, detailed analysis of the radial column density profiles 
indicates that, at least in nearby molecular clouds, 
 {\it Herschel} filaments are characterized 
by a narrow distribution of inner widths with a typical 
value of $\sim 0.1$~pc and a dispersion of less than a factor of 2  
when the data are averaged over the filament crests  
\citep[][see Fig.~\ref{fig:fil-width}]{Arzoumanian+11,Arzoumanian+19}.
Independent submillimeter continuum studies of filament widths in nearby clouds 
have generally confirmed this result 
\citep[e.g.][]{AlvesdeOliveira+2014,KochRosolowsky15,Salji+15, Rivera-Ingraham+16}, 
even if factor of $\simgt \,$2--4 variations around the mean inner width of $\sim 0.1\,$pc 
have been found along the main axis of a given filament 
\citep[e.g.][]{Juvela+12, Ysard+13}. 
The distribution of local widths found by \citet{Arzoumanian+19} 
for 599 nearby molecular filaments
prior to averaging along the filament crests is well described by a lognormal 
function centered at $0.1\pm 0.01\,$pc 
with a standard deviation of $0.33\pm0.03$ in log$_{10}$(width),  corresponding 
to a factor of $\sim$2 on either side of the median width (see Fig.~\ref{fig:fil-width}b).

Measurements of filament widths obtained in molecular line tracers 
 \citep[e.g.,][]{Pineda+11,Fernandez-Lopez+14, Panopoulou+14, Hacar+18}
have been less consistent with the {\it Herschel} dust continuum results, however.  
For instance, using $^{13}$CO emission, \citet{Panopoulou+14} found a broad distribution of widths in Taurus, with a peak of 0.4~pc. 
\citet{Hacar+18} found a median width of 0.035~pc for Orion ``fibers'' (see \S~6.2, below) in the integral-shaped filament of Orion, 
combining N$_2$H$^+$ ALMA and IRAM 30~m observations. 
These differences can be attributed to the lower dynamic range of  densities 
sampled by observations in any given molecular line tracer compared to dust observations. 
More specifically, $^{12}$CO or $^{13}$CO data only trace low-density gas 
and cannot reliably measure the whole (column) density 
profile (hence the width) of a dense molecular filament. 
Likewise, N$_2$H$^+$ or NH$_3$ data only trace relatively high density gas 
(typically above the critical densities of the observed transitions) 
and cannot reliably measure the whole profile of a filament either. 
In contrast, submillimeter dust continuum images from space 
(with {\it Herschel}) achieve a significantly higher dynamic range and are sensitive to 
both the low density outer parts and the dense inner parts of filaments.

\setlength{\unitlength}{1cm}
\begin{figure*}
\begin{picture} (0,8.3) 
\put(0.0,0){\includegraphics[width=11.5cm,angle=0]{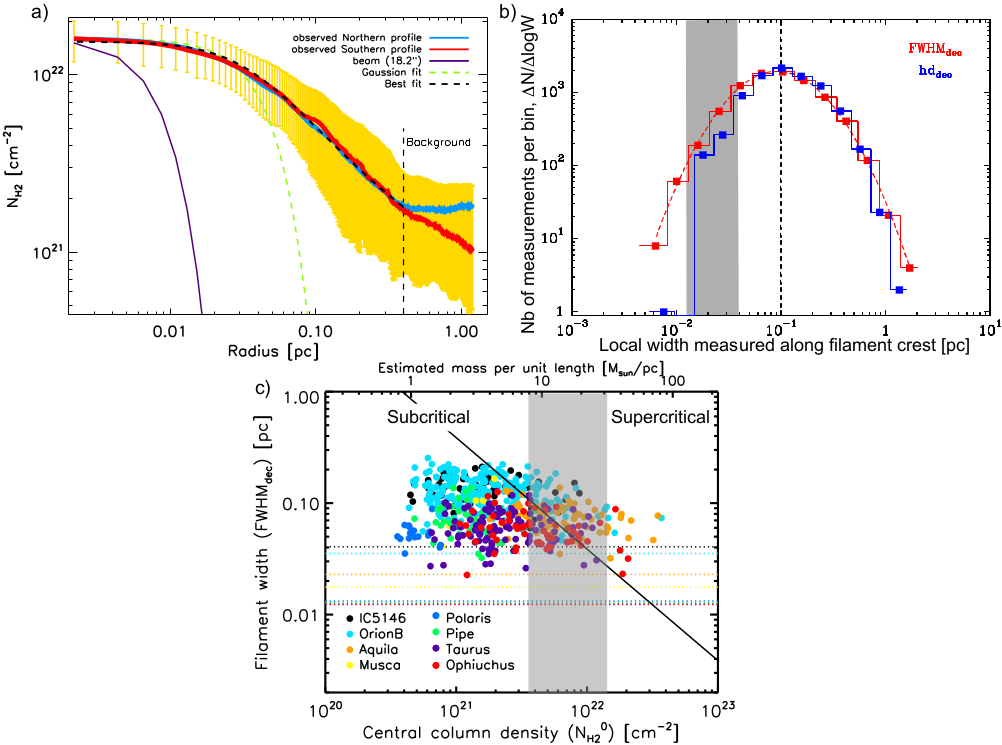} }
\end{picture}
 \caption{
 (a) Mean radial column density profile measured with {\it Herschel} 
perpendicular to the B213/B211 filament in Taurus 
for both the Northern (blue curve) and the Southern part (red curve) of the filament \citep{Palmeirim+13}. 
The yellow area shows the ($\pm 1\sigma$) dispersion of the distribution of radial profiles along the filament.
The inner solid purple curve shows the effective 18\arcsec ~resolution
(0.012~pc at 140~pc) of  the data. 
The dashed black curve is the best-fit Plummer-like model,  
$N_p(r) = N_{\rm H2}^0/[1+ (r/R_{\rm flat})^2]^{\frac{p-1}{2}} $,
with a power-law index $p$=2.0$\pm$0.4 and a diameter $2\times R_{\rm \rm flat} = 0.07 \pm 0.01$~pc.
(b) Distribution of local FWHM widths derived from nearly 9000 independent radial profile measurements along the crests 
of 599 {\it Herschel} filaments  
in 8 nearby molecular clouds \citep{Arzoumanian+19}. 
The red and blue histograms correspond to two methods of estimating the FWHM width (see \citealp{Arzoumanian+19}
for details). The dashed red curve is a lognormal fit to the red histogram. 
The vertical dashed line marks the peak  value of the distribution at $0.1\pm 0.01\,$pc.  
The grey band, between 0.012~pc and 0.041~pc, shows the range of resolutions achieved 
by  {\it Herschel} in the 8 regions. 
(c) Deconvolved FWHM width (averaged along each filament) vs. central column density for the same 599 filaments 
as in panel b). The solid line 
shows the thermal Jeans length as a function of central column density. (Adpated from  \citealp{Arzoumanian+19}.)}
 \label{fig:fil-width}
\end{figure*}

There has also been some controversy about the reliability of the {\it Herschel} result \citep[][]{Smith+14,Panopoulou+17}.
In particular, \citet{Panopoulou+17} pointed out an apparent contradiction between the existence of a characteristic filament width 
and the essentially scale-free nature of the power spectrum of interstellar cloud images 
(well described by a single power law 
from 
{{} 
$\sim 0.01\,$pc to $\sim 50\,$pc -- \citealp{Miville-Deschenes+10,Miville-Deschenes+16}). 
%
However, \citet{Roy+19} showed that there is no contradiction 
given the only modest area filling factors ($\simlt 10\% $) 
and column density contrasts ($\leq 100\% $ in most cases) 
derived by \citet{Arzoumanian+19} for the filaments seen in {\it Herschel} images. 
This is because for realistic filament filling factors and column density contrasts, the filamentary structure contributes 
only a negligible fraction of the image power spectra. 
Another caveat pointed out by \citet{Panopoulou+17} is the presence of potential systematic biases in filament width 
measurements linked to somewhat arbitrary choices of measuring parameters. 
However, based} 
on a number of tests on synthetic data, \citet{Arzoumanian+19} showed 
that their method of measuring filament profiles and widths was reliable and free of significant biases, 
at least when the contrast of the filaments over the local background 
exceeds $\sim 50\%${, which is the case for $\sim 70\%$ of the {\it Herschel} filament population they measured 
and $> 80\%$ of star-forming filaments}.
{{} The median inner diameter of $\sim 0.1\,$pc measured with {\it Herschel} 
may thus reflect the presence of a true common scale in the filamentary structure of 
interstellar clouds. 
Further high-resolution submillimeter continuum studies would nevertheless be required to confirm  
that the same common width applies to low-density, low-contrast ($< 30$--50\%) filaments 
and to investigate whether it also holds beyond the Gould Belt \citep[see][]{Andre+16}.
}


Many theoretical attempts to explain the common inner width of nearby  {\it Herschel}  
filaments 
have been made  \citep[e.g.][]{FischeraMartin12, HennebelleAndre13, Auddy+16, Federrath16, Ntormousi+16}, 
and some of them are discussed in \citep{HennebelleInutsuka19}. 
However, none of the present explanations is fully convincing. 
We must admit that we are still missing something in our detailed physical description 
of filamentary molecular clouds. 

Observationally, three families of molecular filaments may be distinguished according to 
the filament mass per unit length $M_{\rm line} $ compared to the thermal value of the critical mass per unit length 
\citep[e.g.][]{Ostriker64}, $M_{\rm line, crit} = 2\, c_s^2/G \sim 16\, M_\odot \, {\rm pc}^{-1} $ 
for a sound speed $c_{\rm s} \sim 0.2$~km/s, i.e., a typical gas temperature $T \sim 10$~K: 
 {\it thermally supercritical} filaments with $M_{\rm line}  \simgt 2\, M_{\rm line, crit}$,
{\it transcritical} filaments with  $0.5\, M_{\rm line, crit} \simlt M_{\rm line} \simlt 2\, M_{\rm line, crit}$), 
and thermally {\it subcritical} filaments with $M_{\rm line, crit}   \lesssim 0.5\, M_{\rm line, crit}$  
\citep[cf.][]{Arzoumanian+19}.
Again, it is remarkable that all three families of filaments appear to share approximately 
the same inner width $\sim 0.1\,$pc \citep[][see Fig.~\ref{fig:fil-width}b]{Arzoumanian+11,Arzoumanian+19}.  

Using molecular line measurements with the IRAM 30m telescope for a sample of {\it Herschel} filaments
\citep{Arzoumanian+13} showed that thermally subcritical and transcritical filaments have ``transonic''  internal 
velocity dispersions $\sigma_{\rm tot}$ such that $c_{\rm s} \simlt \sigma_{tot} < 2\, c_{\rm s} $. 
Only thermally supercritical filaments have internal velocity dispersions significantly in excess 
of the thermal sound speed  $c_{\rm s} \sim 0.2$~km/s.  
Furthermore, there is a positive correlation between the internal velocity dispersion and the column 
density (or mass per unit length) of thermally supercritical filaments, suggesting 
that these filaments are in approximate virial equipartition (but not necessarily virial equilibrium) 
with $M_{\rm line} \sim M_{\rm line, vir} \equiv 2\, \sigma^2_{\rm tot}/G $ (see Fig.~\ref{fig:vir-fibers}a). 

\setlength{\unitlength}{1cm}
\begin{figure*}
\begin{picture} (0,5)
 \put(0.0,0){\includegraphics[width=11.5cm,angle=0]{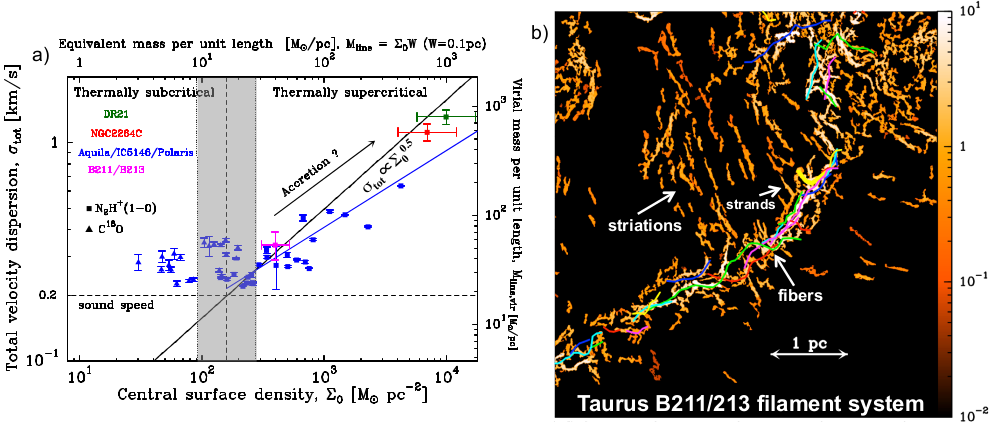} }
\end{picture}
 \caption{
 (a) Total 
 velocity dispersion versus central surface density for a sample of nearby filaments \citep{Arzoumanian+13}.
The horizontal dashed line marks the thermal sound speed $\sim 0.2$ km/s for  $T$~=~10~K. 
The vertical grey band marks the border zone between thermally subcritical and thermally supercritical filaments.
The blue  solid  line shows the best power-law fit $ \sigma_{\rm tot} \propto {\Sigma_0}^{0.36~\pm~0.14}$
to the data points corresponding to supercritical filaments. The latter are in approximate virial equipartition 
with $M_{\rm line} \sim M_{\rm line, vir}$ and $ \sigma_{\rm tot} \propto {\Sigma_0}^{0.5}$. 
(b)  Fine (column) density structure of the B211/B213 filament based on a filtered version of the {\it Herschel} 250 $\mu$m 
    image of \citet{Palmeirim+13}  
    using the algorithm \textsl{getfilaments} \citep{Menshchikov13}. 
    All transverse angular scales larger than 72\arcsec ~(or $\sim 0.05$~pc) 
    were filtered out to enhance the contrast of small-scale structures. 
    The color scale  is in MJy/sr at 250 $\mu$m. 
    The colored curves display the velocity-coherent fibers 
     identified by \citet{Hacar+13}  
     using N$_2$H$^+$/C$^{18}$O observations.  Ê}
 \label{fig:vir-fibers}
\end{figure*}

Most observed molecular filaments must be relatively long-lived structures. 
The fact that most prestellar cores are found to lie within transcritical or supercritical filaments (see \S ~\ref{sec:fil-cores}) 
implies that these filaments must live at least as long as prestellar cores, or at least  $\sim 1$~Myr \citep{Andre+14}. 
A similar lifetime can be inferred for subcritical filaments 
from  a typical sound crossing time $\simgt 5 \times 10^5$~yr, estimated using a nearly sonic internal velocity dispersion $\sim 0.2\, $km/s 
and the typical filament width of $\sim 0.1 $~pc measured with {\it Herschel}. 

\subsection{Striations and fibers}\label{sec:fibers}

A striking feature of wide-field, high-dynamic-range images of molecular clouds in both CO lines from the ground \citep[e.g.][]{Goldsmith+08} 
and submillimeter dust continuum emission from space ({\it Herschel}) is the observation of common patterns in the 
organization of filamentary structures that tend to persist from cloud to cloud. 
A very common filamentary pattern observed with {\it Herschel} is that of a main filament surrounded by a population of 
fainter striations or ``sub-filaments'' approaching the main filament from the side and apparently connected to it (see Fig.~\ref{fig:vir-fibers}b). 
Examples include the Musca filament \citep{Cox+16}, the B211/B213 filament in Taurus \citep{Goldsmith+08,Palmeirim+13},  
the Serpens-South filament in Aquila \citep[e.g.][]{Kirk+13}, and the DR21 ridge in Cygnus~X \citep[][]{Schneider+10,Hennemann+12}.
Interestingly, low-density striations and sub-filaments tend to be parallel 
to the local magnetic field, at least in projection on the plane of the sky \citep[][]{Chapman+11,Palmeirim+13,PlanckXXXV+16}.
The morphology of these striations  and sub-filaments is suggestive of accretion flows, possibly channeled by the local magnetic field, 
and  feeding the main filaments with ambient cloud material. 
Low-density striations are remarkably ordered structures in an otherwise apparently chaotic turbulent medium 
and there is little doubt that magnetic fields are required to explain their properties 
\citep[][]{Heyer+16,TritsisTassis16, Chen+17}. 

Another seemingly common pattern is the presence of significant substructure within many dense molecular filaments, 
observed in the form of velocity-coherent features, called {\it fibers}. 
The presence of fibers was first reported by \citet{Hacar+13}  
in the Taurus B211/B213 filament ($d\sim$140 pc), for which a friends-of-friends algorithm in velocity (FIVE)  
was used to identify at least 20 velocity-coherent components in N$_2$H$^+$ and C$^{18}$O.
Subsequently, similar velocity-coherent components were also detected in N$_2$H$^+$ in other regions, 
including the infrared dark cloud (IRDC) G035.39-00.33 \citep{Henshaw+14}, 
the NGC~1333 protocluster \citep{Hacar+17}, 
IRDC G034.43+00.24 \citep{Barnes+18}, 
the Orion A integral-shaped filament \citep{Hacar+18}, 
and the NGC~6334 main filament \citep{Shimajiri+19a}. 
The velocity-coherent substructures identified in NGC~1333 and Orion~A 
are well separated in the plane of the sky, however, and may differ in nature 
from those observed in  Taurus and  NGC~6334 
which are intertwined. 
Moreover, not all molecular filaments consist of multiple fiber-like substructures. 
The Musca filament, for instance, appears to be a 6-pc-long velocity-coherent sonic filament with much less substructure 
than the Taurus B211/B213 filament and no evidence for multiple velocity-coherent fibers \citep{Hacar+16b,Cox+16}. 

Interestingly, most of the line-identified fibers can also be detected in {\it Herschel} dust continuum maps   
when large-scale emission is filtered out, enhancing the contrast of small-scale structures in the data (cf. Fig.~\ref{fig:vir-fibers}b). 
Moreover, the {\it Herschel} data are suggestive of a direct 
connection between striations and fibers. 
{{} 
In Fig.~\ref{fig:vir-fibers}b, for instance, hair-like strands or spur-like features, which appear to be the tips of larger-scale striations,  
are visible in the immediate vicinity of the Taurus B211/B213 filament, attached to its main body. 
This is consistent with 
the observed striations tracing accretion flows onto the Taurus main filament, possibly 
influencing its fiber-like substructure (see ``accrete and fragment'' scenario below).
}

The exact physical origin of genuine velocity-coherent fibers, i.e.  intertwined filament substructures unaffected 
by  line-of-sight confusion effects, is not well understood 
and remains highly debated \citep[e.g.][]{ZamoraAviles+17, Clarke+17, Clarke+18}. 
In the ``fray and fragment''  scenario introduced by \citet{TafallaHacar15} and supported by \citet{Clarke+17}, 
a main filament forms first by collision of two supersonic turbulent gas flows. 
Then, the main filament fragments into an intertwined system of velocity-coherent fibers, 
due to a combination of vorticity in residual turbulent motions and self-gravity. 
Alternatively, in the ``fray and gather'' scenario \citep{Smith+14,Smith+16}, turbulent compression first generates short, velocity-coherent 
filamentary structures, which are then gathered by large-scale motions within the parent cloud, of either turbulent or gravitational origin. 
{{} 
A variant of these two scenarios is the ``accrete and fragment'' picture, which explicitly links fibers to striations. 
In this picture, dense molecular filaments accrete ambient cloud material along field lines 
through a network of  magnetically-dominated striations (see above). 
The accretion process supplies gravitational energy to the dense filament, which is then converted into 
turbulent kinetic energy in the form of MHD waves \citep{HennebelleAndre13}.
Radial accretion of gas from an inhomogeneous turbulent medium generates 
vorticity primarily in the direction of the filament axis, thus producing  
a system of velocity-coherent fibers in the main filament  \citep[cf.][]{Clarke+17}. 
}
Sterile fiber-like structures would correspond to portions of the accretion flow  
onto the central filament \citep[see][]{Clarke+18}, 
while fertile fibers would be the direct imprint of accretion-driven MHD waves and vorticity within the main filament system. 

\subsection{The characteristic line mass of molecular filaments}\label{sec:linemass}
To understand the dynamics of filamentary molecular clouds we have to understand 
the critical mass per unit length above which a cylindrical cloud cannot be in equilibrium with the isothermal equation of state, as explained below. 
This critical line mass for an isothermal cylindrical cloud does not depend on the density or radius of the cloud and depends only on temperature. 
As the gas temperature in nearby molecular clouds is always on the order of 10~K, 
the critical line mass is almost a constant value in the solar neighborhood. 
Using observations from the {\it Herschel} space observatory, \citep{Andre+10} found a remarkable threshold for star formation process, 
which can be summarized as follows: stars are formed in molecular filaments whose line mass is comparable to or larger than the critical line mass. 
Therefore, the line mass of filaments plays an important role in the evolution of molecular clouds. 
Let us first discuss this basic property of cylindrical geometry. 

Early papers on filament structure such as the classic solution for a self gravitating isothermal filament \citep{Stodolkiewicz63, Ostriker64}, 
assumed that filaments are in cylindrical hydrostatic equilibrium.   
Poisson's equation for self-gravity in a cylinder of infinite length is 
\begin{equation}
     \nabla^2 \Phi 
     = \frac{1}{r}\frac{d}{dr}r\frac{d\Phi}{dr} 
     = 4\pi G \rho .  
                               \label{eq:PoissonCyl}
\end{equation}
If we multiply by $r$ on both sides and integrate from the center 
to the outermost radius $r=R$, we obtain 
\begin{equation}
       R \left. \frac{d\Phi}{dr} \right|_{r=R} 
     = 2 G \int_{0}^{R}2\pi r \rho dr  
     \equiv 2 G M_{\rm line}, 
\end{equation}
where we define the mass per unit length (i.e., the line mass) as 
$M_{\rm line} = \int_{0}^{R}2\pi \rho r dr$. 
This line mass remains constant in a change of the cylinder radius
where $\rho \propto R^{-2}$. 
Thus, the self-gravitational force $F_{\rm g,cyl}$
is 
\begin{equation}
     F_{\rm g,cyl} = \left. \frac{d\Phi}{dr} \right|_{r=R} 
     = 2 \frac{G M_{\rm line}}{R} \propto \frac{1}{R}. 
\end{equation}
On the other hand, if we denote the relation between 
gas pressure $P$ and density $\rho$ as 
$
     P = K \rho^{\gamma_{\rm eff}}, 
$
the pressure gradient force $F_{\rm p}$ scales as 
\begin{equation}
     F_{\rm p} = \frac{1}{\rho}
                 \frac{\partial P}{\partial r}
               \propto  
                 R^{1-2\gamma_{\rm eff}}. 
                                         \label{eq:GradP}
\end{equation}
Therefore, we have 
\begin{equation}
     \frac{ F_{\rm p} }{ F_{\rm g,cyl} } 
     \propto R^{2-2\gamma_{\rm eff}}. 
\end{equation}
This means that, 
if $\gamma_{\rm eff} > 1$, 
the pressure gradient force will dominate self-gravity for a sufficiently small 
radius $R$. 
On the contrary, if if $\gamma_{\rm eff} \leq 1$, 
the radial collapse will continue indefinitely, 
once self-gravity dominates over the pressure force. 
Therefore, we can define the critical ratio of specific heats for the radial stability of 
a self-gravitating cylinder as 
$
        \gamma_{\rm crit,cyl} = 1. 
$
In the case of $\gamma_{\rm eff}~=~1$ 
a cylinder (under sufficiently small ambient pressure)  
will be in hydrostatic equilibrium only when 
its line mass has the special value for which $F_{\rm p}=F_{\rm g,cylinder}$. 
This is the reason why we can define a critical line mass for 
an isothermal cylinder: 
\begin{equation}
     M_{\rm line, crit} \equiv \int_0^{\infty} 2\pi \rho(r)rdr 
     =  \frac{2c_{\rm s}^2}{G} . 
\end{equation}

Using similar arguments, 
we can define the critical $\gamma_{\rm eff}$ for 
a sphere, $\gamma_{\rm crit,sphere} = 4/3$, and 
for a sheet, $\gamma_{\rm crit,sheet} = 0$. 
The thermodynamical property of molecular clouds 
corresponds to $\gamma_{\rm eff} \approx 1$. 
Therefore, the significance of filamentary geometry 
can be understood in terms of 
ISM thermodynamics. 
Importantly, filaments differ from both sheets and spheroids in their global gravitational instability properties. 
For a sheet-like cloud, there is always an equilibrium configuration since the internal pressure gradient can always become strong 
enough to halt the gravitational collapse of the sheet independently of the initial state \citep[e.g.][]{Miyama+87,InutsukaMiyama97}. 
In contrast, the radial collapse of an isothermal cylindrical cloud cannot be halted and no equilibrium is possible
when the line mass exceeds the critical mass per unit length $M_{\rm line, crit}$. Conversely, if the line mass of a filamentary cloud 
is less than $M_{\rm line, crit}$, gravity can never be made to dominate by increasing the external pressure, so that the collapse is 
always halted at some finite cylindrical radius. 
Thus, filaments differ markedly from isothermal spherical clouds which can always be induced to collapse by 
a sufficient increase in external pressure \citep[e.g.][]{Bonnor56,Shu77}.

A major feature of equilibrium filament models is a critical condition for stability; the mass per unit length of the filament.  
For isothermal systems, this is simply $M_{\rm line, crit} = 2\, c_s^2/G$ (where $c_s $ is the isothermal sound speed).  
This quantity plays an analogous role to the virial mass $ M_{vir} = 5 R c_s^2 / G$ for spherical clouds and equilibria.  
Filaments become gravitationally unstable when the critical line mass is exceeded.   

The peculiar behavior of filamentary geometry in isothermal collapse is due to the fact that the isothermal equation of state ($\gamma = 1$) 
is a critical case for the collapse of a filament \citep[e.g.][]{Larson05}: 
For a polytropic equation of state ($P \propto \rho^\gamma$) with $\gamma < 1$, an unstable cylinder can collapse indefinitely toward its axis, 
while if $\gamma > 1$ the pressure gradient increases faster than gravity during contraction and the collapse is always halted at a finite radius. 
For comparison, the critical value is $\gamma = 0$ for sheets and $\gamma = 4/3$ for spheres. 
Indefinite, global gravitational collapse of a structure can occur when $\gamma$ is smaller than the critical value and 
is suppressed when $\gamma$ is larger than the critical value. 
Gravitational fragmentation thus tends to be favored over global collapse when $\gamma$ is close to or larger than the critical value. 

The critical mass per unit length $M_{\rm line, crit}  \approx 16\, M_\odot /{\rm pc} \times (T_{\rm gas}/10\, {\rm K})$ as originally derived, 
depends only on gas temperature $T_{\rm gas}$.
This expression can be readily generalized to include the fact that filaments have non-thermal internal velocity dispersions 
(see \SÊ~\ref{sec:fil} and Fig.~\ref{fig:vir-fibers}). 
In the presence of non-thermal gas motions, the critical mass per unit length becomes $M_{\rm line, vir} = 2\, \sigma_{\rm tot}^2/G $, 
also called the virial mass per unit length, where $\sigma_{\rm tot} = \sqrt{c_s^2 + \sigma_{\rm NT}^2} $ is the total one-dimensional velocity 
dispersion including both thermal and non-thermal components \citep{FiegePudritz00}. 
Clearly, both the equation of state of the gas and filament turbulence play a role in deciding the actual critical line mass. 

\subsection{Nature of MC filaments: equilibrium structures vs. funnel flows }\label{sec:fil-nature}


The mere existence of thermally supercritical filaments with $M_{\rm line} >> M_{\rm line, crit}$ poses a problem, 
since such filaments would be expected to collapse radially to spindles in only about one free-fall 
time (or $< 10^5\, $yr for dense systems such as the NGC~6334 filament -- \citealp{Andre+16}), 
without significant fragmentation along their axis 
according to non-magnetized models for the evolution of isolated self-gravitating cylinders \citep[][see \S~\ref{sec:linemass} above]{InutsukaMiyama92,InutsukaMiyama97}. 
In contrast, {\it Herschel} observations have revealed the presence of numerous $\sim$\,0.1-pc-wide supercritical filaments 
with widespread fragmentation into prestellar cores and an estimated lifetime of $\simgt 1\,$Myr 
\citep[e.g.][]{Arzoumanian+19,Konyves+15,Andre+14,Andre+19}.

Two extreme views for the dynamical state of molecular filaments have been proposed to explain this paradox. 
First, thermally supercritical filaments may be close to virial and/or magneto-hydrostatic equilibrium, 
with support against self-gravity provided by a combination of internal MHD turbulence and magnetic fields. 
Indeed, supercritical filaments are known to be virialized accreting systems with $M_{\rm line} \sim M_{\rm line, vir }$ 
(see above, \citealp{FiegePudritz00}, and \citealp{Arzoumanian+13}). 
While this does not necessarily imply virial {\it equilibrium}, 
accretion-driven MHD waves can possibly maintain an effective virial equilibrium \citep[cf.][]{HennebelleAndre13}.
Assuming rough equipartition between magnetic and kinetic energy,  
thermally supercritical filaments may also be close to magneto-hydrostatic equilibrium since the magnetic critical line mass 
$M^{\rm mag}_{\rm line, crit} \approx 1.66\, c_{\rm s}^2/G + 0.24\, \Phi_{\rm cl}/G^{1/2}$ largely exceeds $M_{\rm line, crit}$ 
when the magnetic flux per unit length $ \Phi_{\rm cl}$ is large \citep{Tomisaka14}. 

A second, alternative picture posits that most dense molecular filaments are far from equilibrium 
and represent dynamical accretion flows onto denser cluster-forming clumps or hubs \citep[e.g.][]{GomezVazquezSemadeni14}.
In this picture, filaments form from gravitational amplification of initial anisotropies as part 
of the ``global hierarchical collapse''  of strongly Jeans-unstable molecular clouds,  
and constitute the collapse flow itself from the larger scales to the small-scale dense hubs
\citep{VazquezSemadeni+19}. Magnetic fields are present but essentially passive. 
Large-scale collapse occurs first along the shortest axis creating sheets and subsequently filaments. 
Gas from sheet-like cloud structures flows onto the 
filaments roughly perpendicular to them, and is then diverted to flow {\it along} the filaments 
onto the hubs. Thus, dense molecular filaments are like rivers and can approach a quasi-stationary state 
without being in equilibrium.

In practice, real molecular filaments may consist of a mix of quasi-equilibrium and non-equilibrium 
structures, and observational constraints on the velocity field and (geometry of) the magnetic field 
may be used to discriminate between the two types of filamentary structures. 
In the quasi-equilibrium view, significant longitudinal velocity gradients are not expected 
and the magnetic field lines should be only slightly distorted, due to local protostellar collapse 
at the positions of dense cores along each filament. 
In contrast, in the funnel flow hypothesis, supersonic longitudinal velocity gradients are 
expected, owing to gravitational acceleration of gas toward the central hub, 
and the magnetic field lines should be dragged by the longitudinal gas  flow, adopting a ``V-shape'' 
within the filament  \citep{Gomez+18}.

Observationally, there is little doubt that gravity is the main player shaping strongly self-gravitating ``hub-filament'' systems, 
where a cluster-forming hub is observed at the center of a converging network of filaments \citep{Myers09}. 
Such systems are particularly prominent in massive star-forming regions 
(e.g. MonR2: \citealp{Didelon+15,Pokhrel+16}; SDC335: \citealp{Peretto+13}), 
but also exist in clouds forming mostly (or only) low- to intermediate-mass stars 
(e.g. B59: \citealp{Peretto+12}; L1688:  \citealp{Ladjelate+20}).
Good candidates for longitudinally collapsing filaments have also been identified, such as the SDC13 system 
of infrared dark filaments \citep{Peretto+14}. 
While the magnetic field {\it inside} dense filaments is poorly constrained 
by current dust polarization observations, it may well be parallel to the filament axis in some cases 
(cf. the G9.62 massive clump -- \citealp{DallOlio+19}), as in the hierarchical collapse picture described above  
or the ``inertial-inflow'' model of \citep{Padoan+19}. 

In most cases, however, transverse velocity gradients {\it across} dense molecular filaments, suggestive of accretion 
{\it onto} rather than along the filaments, appear to dominate over longitudinal velocity gradients 
\citep{Kirk+13, Fernandez-Lopez+14, Dhabal+18, Shimajiri+19b}. 
Some examples of subsonic longitudinal motions, possibly core-forming, 
have also been found \citep{HacarTafalla11}.
Moreover, chains of dense cores with quasi-periodic spacings have been observed toward some 
of these transcritical or supercritical filament systems \citep[][Zhang et al. 2020]{TafallaHacar15, Bracco+17,Shimajiri+19a}.
Such quasi-periodic features are expected in 
quasi-equilibrium filaments because they have a preferred fragmentation lengthscale \citep[cf.][]{InutsukaMiyama92}, 
even in the presence of geometrical bends \citep{Gritschneder+17} or accretion 
from a weakly turbulent  medium \citep{Clarke+16}. 
Filaments formed by large-scale gravitational collapse accrete from inhomogeneous parent structures  
which are themselves collapsing on larger scales and therefore highly ``turbulent'' \citep{VazquezSemadeni+19}. 
They are thus less likely to develop quasi-periodic chains of dense cores \citep[cf.][]{Clarke+17}. 
While current observations suggest that quasi-equilibrium configurations may dominate, 
more work would be needed to draw definitive conclusions on the relative importance 
of the two modes (i.e., equilibrium vs. flow structures).
%

\subsection{The filament-core connection }\label{sec:fil-cores}

The collapse and fragmentation properties of filaments under the assumption of cylindrical symmetry were extensively studied theoretically 
a few decades ago \citep[e.g.][]{Nagasawa87} but have received renewed attention with the {\it Herschel} results. 
The gravitational instability of nearly isothermal filaments is primarily controlled by the value of their mass per unit length $M_{\rm line} \equiv M/L$. 
Above the critical value $M_{\rm line, crit} = 2\, c_s^2/G$ (where $c_s $ is the isothermal sound speed) cylindrical filaments are expected 
to be globally unstable to both radial collapse and fragmentation along their lengths \citep[e.g.][]{InutsukaMiyama92, InutsukaMiyama97}, 
while below $M_{\rm line, crit} $ filaments are gravitationally unbound and thus expected to expand into the surrounding medium 
unless they are confined by some external pressure \citep[e.g.][]{FischeraMartin12}.
 
The fragmentation properties of filaments and sheets differ from those of spheroidal clouds in that there is a preferred scale 
for gravitational fragmentation which directly scales with the scale height of the filamentary or sheet-like medium \citep[e.g.][]{Larson85}. 
In the spherical case, the largest possible scale or mode (i.e., overall collapse of the medium) has the fastest growth rate 
so that global collapse tends to overwhelm the local collapse of finite-sized density perturbations,  
and fragmentation is generally suppressed in the absence of sufficiently large initial density enhancements  \citep[e.g.][]{Tohline82}.
It also well known that spherical collapse quickly becomes strongly centrally concentrated \citep{Larson69,Shu77}, 
which tends to produce a single central density peak as opposed to several condensations \citep[e.g.][]{Whitworth+96}. 
In contrast, sheets have a natural tendency to fragment into filaments \citep[e.g.][]{Miyama+87} and filaments with line masses 
close to $M_{\rm line, crit}$ have a natural tendency to fragment into spheroidal cores \citep[e.g.][]{InutsukaMiyama97}.


The filamentary geometry is thus the most beneficial configuration in molecular clouds for small-scale perturbations 
to collapse locally and grow significantly before global collapse overwhelms them \citep{Pon+11,Pon+12,Toala+12}. 
This is because the isothermal equation of state represents a critical case for filaments, possibly halting or slowing 
down radial collapse \citep[e.g.][see also \SÊ~\ref{sec:linemass} above]{Larson05}, 
and the longitudinal collapse timescale of a filament increases almost linearly with the filament length or aspect ratio \citep[e.g.][]{ClarkeWhitworth15}, 
thus greatly exceeding the local free-fall timescale. 

Indeed,  {\it Herschel} studies of nearby molecular clouds have found that most ($>75\% $) 
prestellar dense cores and young protostars lie in thermally transcritical or supercritical 
filaments \citep[][see also \SÊ~3.2 in Chap. 8]{Andre+10,Konyves+15,Marsh+16}. 
Adopting the typical $\sim 0.1\,$pc inner width measured with {\it Herschel} for molecular filaments, the critical mass per unit length 
is equivalent to a critical threshold or transition at $\sim 160\, M_\odot $/pc$^2$ in gas surface density ($A_V \sim 8)$ 
or $n_{H_2} \sim  2 \times 10^4\, {\rm cm}^{-3} $ in volume density. 
A similar surface density threshold for the formation of prestellar cores (at $A_V \sim \, $5--10) had been suggested earlier based 
on ground-based millimeter and submillimeter studies \citep[e.g.][]{Onishi+98,Johnstone+04,Kirk+06}, but without any clear connection to filaments. 
Interestingly, a comparable threshold in extinction (at $A_V \sim \, $8) has also been observed 
in the spatial distribution of young stellar objects (YSOs) with {\it Spitzer} \citep[e.g.][]{Heiderman+10, Lada+10, Evans+14}.

\noindent
We refer the reader to \S~3 in Chap.~8 for more observational details on the filament--core connection and a summary of dense core properties.

\subsection{Towards the IMF}


As already mentioned in \S ~\ref{sec:MassSpec}, the mass function of GMCs and CO clumps within GMCs 
is known to be rather shallow,  $\Delta N$/$\Delta$log$M \propto M^{-0.6\pm0.2}$ \citep[e.g.][]{Solomon+87,Blitz93}, 
implying that most of the molecular gas mass in the Galaxy resides in the {\it most massive} GMCs, and within the GMCs themselves 
in the most massive CO clumps. 

\setlength{\unitlength}{1cm}
\begin{figure*}
\begin{picture} (0,4.5)
 \put(0.0,0){\includegraphics[width=11.5cm,angle=0]{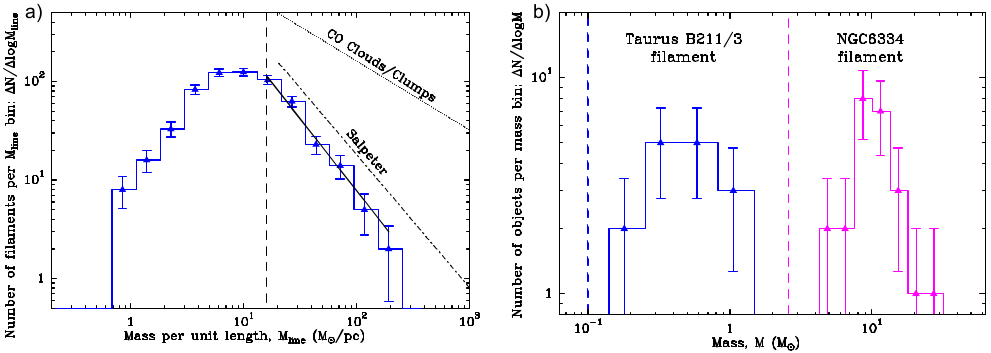} }
\end{picture}
 \caption{
 (a) Filament line mass function (FLMF) derived for 599 {\it Herschel}  filaments in 
 eight nearby clouds \citep[from][]{Andre+19,Arzoumanian+19}. 
Above the critical line mass $M_{\rm line, crit} \sim 16\, M_\odot$/pc (vertical dashed line), 
the filament sample is more than $ 90\%$ complete and 
the FLMF is well fitted by a Salpeter-like power law $\Delta N$/$\Delta$log$M_{\rm line} \propto M_{\rm line}^{-1.6\pm0.1}$ (solid line segment). 
(b) Comparison of the prestellar core mass function (CMF) observed with {\it Herschel} in the low-mass ($M_{\rm line} \sim 50\, M_\odot$/pc)  
B211/B213 filament in Taurus  (blue histogram -- cf. \citealp{Marsh+16}) 
with the CMF derived from ALMA data in the high-density ($M_{\rm line} \sim 500$--1000$\, M_\odot$/pc) NGC~6334 filament (magenta histogram -- \citealp{Shimajiri+19a}). 
The vertical dashed lines mark the $\sim \,$90\% core completeness levels achieved in the two filaments.
(From Palmeirim et al., in prep.)
 }
 \label{fig:flmf-cmf}
\end{figure*}

More recently, a good estimate of the filament mass function (FMF) and filament line mass function (FLMF) 
in nearby molecular clouds has been derived using a comprehensive study of filament properties from 
{\it Herschel} Gould Belt survey observations \citep{Arzoumanian+19,Andre+19}. 
The FLMF is well fit by a power-law distribution in the supercritical mass per unit length regime 
(above $16\, M_\odot$/pc), $\Delta N$/$\Delta$log$M_{\rm line} \propto M_{\rm line}^{-1.6\pm0.1}$ 
(see Fig.~\ref{fig:flmf-cmf}a).
The FMF is very similar in shape to the FLMF and also follows a  
power-law distribution at the high-mass end (for $M_{\rm tot} > 15\, M_\odot $),  
$\Delta N$/$\Delta$log$M_{\rm tot} \propto M_{\rm tot}^{-1.4 \pm0.1}$, 
which is significantly steeper than the GMC mass function. 
Both the FLMF and the FMF are roughly consistent with the Salpeter power-law IMF \citep{Salpeter55}, 
which scales as d$N$/dlog$M_\star$ $\propto$ $M_\star^{-1.35}$ in the same format. 
Thus, molecular filaments may represent the key evolutionary step in the hierarchy of cloud structures at which 
a steep Salpeter-like mass distribution is established. 
Note that the filament mass function differs in a fundamental way from the GMC mass function 
in that most of the filament mass lies in {\it low-mass} filaments. 
In particular, this result implies that most of the mass of star-forming filaments lies 
in thermally transcritical with masses per unit length within a factor 2 of the critical value 
$M_{\rm line, crit} = 2\, c_s^2/G \sim 16\, M_\odot \, {\rm pc}^{-1} $. 

As most prestellar cores are born in transcritical or supercritical filaments (cf. \S ~\ref{sec:fil-cores}) and 
the prestellar core formation efficiency is typically $\sim \,$15\%--25\% in such filaments \citep[e.g.][-- see also Fig.~3a in Chap.~8]{Konyves+15,Konyves+20}, 
the form of the FMF has direct implications for 
the prestellar core mass function (CMF) 
and, by extension, the stellar initial mass function (IMF) itself \citep[cf.][]{Andre+19}.  
Indeed, one expects that there should be a characteristic prestellar core mass corresponding to 
the local Jeans or critical Bonnor-Ebert mass \citep[e.g.][]{Bonnor56} in transcritical filaments. 
For a critical $\sim\, $0.1-pc-wide filament of molecular gas at $\sim \,$10\,K with 
$ M_{\rm line} \approx M_{\rm line, crit} \sim 16\, M_\odot \, {\rm pc}^{-1} $ 
and surface density $\Sigma_{\rm fil}  \approx \Sigma_{\rm gas}^{\rm crit} \sim 160\, M_\odot \, {\rm pc}^{-2} $, 
the critical Bonnor-Ebert mass is:  

\begin{equation}
M_{\rm BE, th}  \sim 1.3\, \frac{c_{\rm s}^4}{G^2 \Sigma_{\rm fil}} \sim 0.5\, M_\odot  \times \left(\frac{T}{10\, \rm K}\right)^2 \times  \left(\frac{ \Sigma_{\rm fil}}{160\, M_\odot \, {\rm pc}^{-2}}\right)^{-1} .
\end{equation}

Thus, one may expect a peak in the prestellar CMF at $\approx 0.5\, M_\odot $. 
This value matches very well with the observed peak of the prestellar CMFs at $\sim \, $0.3--0.7$\,  M_\odot$ 
derived from {\it Herschel} data in nearby molecular clouds 
\citep[][ Ladjelate et al. 2020; Di Francesco et al. 2020; Pezzuto et al. 2020; see also \S~3.3 and Fig.~3b in Chap.~8]{Konyves+15,Konyves+20,Marsh+16}. 
Moreover, the shape of the prestellar CMF at the high-mass end is consistent 
with a Salpeter-like power law (see \S ~\ref{sec:MassSpec} above) and thus resembles the FMF, suggesting 
that it may be directly inherited from the latter. 

The close link between the FMF (or FLMF) and the prestellar CMF may be understood if we recall that 
the thermally supercritical filaments observed with {\it Herschel} in nearby clouds 
have a typical inner width $W_{\rm fil} \sim 0.1\,$pc (see \S ~\ref{sec:fil})
and are virialized with $ M_{\rm line} \sim  \Sigma_{\rm fil} \times W_{\rm fil}  \sim M_{\rm line, vir} \equiv 2\, \sigma^2_{\rm tot}/G $,   
where $\sigma_{\rm tot}$ is equivalent to the effective sound speed (see \S ~\ref{sec:fil-nature}). 
This implies that the effective Bonnor-Ebert mass  $M_{\rm BE, eff}  \sim 1.3\, \sigma_{\rm tot}^4 /(G^2 \Sigma_{\rm fil} $) scales roughly 
as  $\Sigma_{\rm fil}$ or $ M_{\rm line} $ in supercritical filaments. Thus, higher-mass cores may form 
in higher $ M_{\rm line} $ filaments, as indeed suggested by observations  \citep[][see Fig.~\ref{fig:flmf-cmf}b]{Shimajiri+19a}. 
If the CMF produced by a single supercritical filament were a narrow $\delta$ function peaked at $M_{\rm BE, eff} $, 
then there would be a direct correspondence between the FMF and the prestellar CMF \citep[cf.][]{Andre+14}.
In reality, the prestellar CMF generated by a single filament is expected 
to be broader than a $\delta$ function \citep[cf.][]{Inutsuka01} 
and observationally appears to broaden 
as $ M_{\rm line} $ increases \citep[][]{Konyves+20}, 
although for statistical reasons, this is difficult to constrain precisely. 
The global prestellar CMF therefore results from the ``convolution'' of the FMF with the CMFs 
produced by individual filaments \citep{Lee+17}. 
It can be shown, however, that the high-mass end of the global CMF is primarily driven by the power-law shape 
of the FMF and depends only weakly on the widths of the individual CMFs \citep[cf. Appendix B of][]{Andre+19}.  

The global prestellar CMF is itself apparently closely linked to the stellar IMF, or more precisely the stellar system IMF \citep[cf.][]{Chabrier05},  
with to first order only a systematic shift between the two, 
corresponding to a core-to-star formation efficiency $ \epsilon_{\rm core} \sim  0.4^{+0.2}_{-0.1}$ 
(\citealp{Konyves+15}; see also \citealp{Alves+07} and \S ~3.3 of Chap.~8). 
This is suggestive of a one-to-one mapping between prestellar core mass and stellar system mass,
$M_{\star \rm sys} = \epsilon_{\rm core}\, M_{\rm core} $, at least for core masses $\sim \,$0.1--10$\, M_\odot $, 
The existence of a direct {\it physical} connection between the prestellar CMF and the stellar IMF remains uncertain 
and debated, however (see, e.g., Chap.~8). 
Clearly, mechanisms controlling the core-to-star formation efficiency $ \epsilon_{\rm core} $, 
such as feedback from protostellar outflows and rotationally-driven subfragmentation of cores into 
binary/multiple systems, also play important roles in the origin of the IMF.

\section{Summary}\label{sec:conclusions}

{
Molecular clouds are the structures {
through} which galaxies funnel their diffuse, warm gas into stars. In the present work we {have} review{ed} the current knowledge of the mechanisms that control the formation and evolution of molecular clouds in their way to form stars.

We start{ed} reviewing how MCs are formed. Chemically, the H$_2$ molecule is formed in the present{-day} Universe mainly on the surfaces of dust grains. This molecule is highly sensitive to the UV radiation, and thus, extinctions of the order of $A_V\gtrsim 1$ are required to allow the formation of molecular hydrogen. In terms of the physical processes, MCs are formed from the convergence of diffuse, warm atomic  flows{, which may contain some pre-existing molecular gas mixed in}. There is a variety of possibilities on the origin of such flows, namely the passage of a spiral arm, large-scale gravitational instabilities in the disk, the agglomeration of smaller clouds, the expansion of a SN remnant, stellar winds from clusters, etc.  In all of these processes, the convergence of flows itself allows the density to increase. Since the warm H~I is thermal{ly} unstable, the compressed region cools down rapidly, {
further increasing the density}, and allowing gravity to take over. This results in contraction in all directions, producing an even faster growth of the  column density, and thus, a rapid formation of molecules. 

The physical state of MCs is frequently {
described} by means of the virial theorem, as well as {
by its} statistical properties. The virial theorem, which arises from the momentum equation, {
describes the generation of kinetic energy in the cloud and the change in its internal mass distribution due to the work done on the cloud by the various forces acting on it, resulting in} a relationship between the energies involved on the cloud,  which {
can be estimated observationally}. However, such estimations may have important caveats that could lead to {
misinterpretations} about the dynamical state of MCs. Two of the more important caveats are: (a) {
the assumption} that the cloud can be {
considered} to be a homogeneous, isolated sphere, and (b) to interpret the equality $2 E_{\rm kin} = E_{\rm grav}$ as {
implying} equilibrium, when {the} actual equilibrium (force balance) {
is} instead {that} $\ddot I/2 = 2 E_{\rm kin}$. 

In terms of the statistical properties of MCs, a wide set of properties have been studied.  The mass distribution of MCs appears to be flatter than the mass distribution of protostellar cores, the last one having similar slopes to {that of} the stellar initial mass function (IMF), a result that suggest{s} a link between the physics involved in defining the mass of prestellar cores and that of the stars. The column-density probability distribution functions of clouds, on the other hand,  is {
generally agreed} to have {either of} two main shapes {
at large column densities}: lognormal {
or} power-law. The first one has been interpreted in terms of turbulence dominating the dynamics of MCs at the beginning of the formation of MCs. The last one, as a consequence of gravity {
producing} centres of collapse {within the cloud}. The fact that the PDFs of the density of MCs decrease rapidly impl{{y}ies} that most of the{ir} volume and mass  {
are in the low-density regime}.

{Another property of e}nsembles of MCs {is that they 
have a} mass-size relation {that is also a power law}. The exponent is typically around 2, which implies that the ensemble has {a roughly constant} constant column density. This has been interpreted {
as a consequence} of clouds {typically} being defined by column density thresholds{. Moreover,}  since most of the mass is at low column densities, {the clouds in} an ensemble {
} defined {
by means of} a column density threshold will {all} have the {approximately} same mean column density, regardless {of} the{ir} size, mass or star formation activity{
}.

In terms of their velocity structure, larger MCs exhibit larger values of the velocity dispersion, although with substantial scatter. Typically, a power-law {
of the form $\sigma \propto R^\eta$, with $\eta \sim 1/2$ is thought to be valid in general}, although the {
} exponents {
} found {
in observational studies range} between 0 and 1, depending on the performed survey and methodology.  This correlation is frequently {
considered} to be {
evidence} that MCs are dominated by supersonic turbulence.  {
However, a} generalization of the velocity dispersion-size correlation has been 
recently {suggested}, as surveys have allowed to reach a broader dynamical range in column density. In this case, the {so called} Larson
ratio ${\cal{L}} \equiv \sigma_v/R^{1/2}$ is not constant, but {rather} increases with column density {as $\Sigma^{1/2}$, thus} following a virial relation. This correlation also has substantial scatter, and the global trend has been interpreted in terms of gravity {
driving chaotic infall} motions in MCs{, which combine with truly random turbulent motions in different proportions in the clouds as they evolve}.  

The magnetic fields of MCs, at scales of the clouds themselves, {are} found to be preferentially perpendicular to their filamentary structure.  In terms of support against gravity, it has been found that dense cores are {in general} not supported by magnetic fields. 

MCs are also found to be highly filamentary. In the Solar Neighborhood, {{\it Herschel}-identified filaments} appear to exhibit similar width{s}. No theoretical explanation has yet been found for this result. These filaments also exhibit striations and fibers. The  {former} are low column density features { protruding out of the filaments,} mostly aligned with the magnetic field, and thus, perpendicular to the high-column density cloud. The 
{
latter are} velocity-coherent subfilaments that appear in {position-position-velocity cubes from} emission line observations. Whether such sub-filamentary structure is real, or the result of superposition structure in the line of sight, is currently matter of debate.  

Finally, in terms of the dynamics of the filaments, there is a critical {
linear mass} density, above which filaments {in hydrostatic equilibrium} should collapse radially. It is also matter of debate whether filaments are structures in equilibrium, or flows of gas from the low-density regions of MCs through dense, star-forming cores.  In any event, filaments are not homogeneouos, and their {
} dense{r} regions are the so-called molecular cloud cores, the sites where stars are formed.  Most of the prestellar dense cores are formed by fragmentation of filaments that have a mass line density larger than the critical. 

}

\begin{acknowledgements}
The authors acknowledge the hospitality of the International Space Science Institute during the "Workshop on Star Formation", held in Bern, Switzerland, May 2019.
JBP acknowledges UNAM-DGAPA-PAPIIT support through grant number IN-111-219. 
PhA acknowledges support from the French national programs of CNRS/INSU 
on stellar and ISM physics (PNPS and PCMI) and from the European Research Council 
via the ERC Advanced Grant ORISTARS (Grant Agreement no. 291294).
RSK acknowledges financial support from the German Research Foundation (DFG) via the collaborative research center (SFB 881, Project-ID 138713538) “The Milky Way System” (subprojects A1, B1, B2, and B8). He also thanks for funding from the Heidelberg Cluster of Excellence STRUCTURES in the framework of Germany’s Excellence Strategy (grant EXC-2181/1 - 390900948) and for funding from the European Research Council via the ERC Synergy Grant ECOGAL (grant 855130) and the ERC Advanced Grant STARLIGHT (grant 339177).
M.C. and J.M.D.K. gratefully acknowledge funding from the German Research Foundation (DFG) in the form of an Emmy Noether Research Group (grant number KR4801/1-1) and a DFG Sachbeihilfe Grant (grant number KR4801/2-1). 
J.M.D.K. gratefully acknowledges funding from the European Research Council (ERC) under the European Union's Horizon 2020 research and innovation programme via the ERC Starting Grant MUSTANG (grant agreement number 714907), and from Sonderforschungsbereich SFB 881 ``The Milky Way System'' (subproject B2) of the DFG. 
A.A. acknowledges the support of the Swedish Research Council, Vetenskapsr{\aa}det, and the Swedish National Space Agency (SNSA).
\end{acknowledgements}

\bibliographystyle{spbasic}


\begin{thebibliography}{437}
\providecommand{\natexlab}[1]{#1}
\providecommand{\url}[1]{{#1}}
\providecommand{\urlprefix}{URL }
\expandafter\ifx\csname urlstyle\endcsname\relax
  \providecommand{\doi}[1]{DOI~\discretionary{}{}{}#1}\else
  \providecommand{\doi}{DOI~\discretionary{}{}{}\begingroup
  \urlstyle{rm}\Url}\fi
\providecommand{\eprint}[2][]{\url{#2}}

\bibitem[{{Adamo} et~al.(2015){Adamo}, {Kruijssen}, {Bastian}, {Silva-Villa},
  and {Ryon}}]{adamo15}
{Adamo} A, {Kruijssen} JMD, {Bastian} N, et~al. (2015) {Probing the role of the
  galactic environment in the formation of stellar clusters, using M83 as a
  test bench}. \mnras 452:246--260

\bibitem[{{Agertz} et~al.(2013){Agertz}, {Kravtsov}, {Leitner}, and
  {Gnedin}}]{Agertz+13}
{Agertz} O, {Kravtsov} AV, {Leitner} SN, et~al. (2013) {Toward a Complete
  Accounting of Energy and Momentum from Stellar Feedback in Galaxy Formation
  Simulations}. \apj 770(1):25

\bibitem[{{Alves} et~al.(2007){Alves}, {Lombardi}, and {Lada}}]{Alves+07}
{Alves} J, {Lombardi} M, and {Lada} CJ (2007) {The mass function of dense
  molecular cores and the origin of the IMF}. \aap 462(1):L17--L21

\bibitem[{{Alves} et~al.(2017){Alves}, {Lombardi}, and {Lada}}]{Alves+17}
{Alves} J, {Lombardi} M, and {Lada} CJ (2017) {The shapes of column density
  PDFs. The importance of the last closed contour}. \aap 606:L2

\bibitem[{{Alves de Oliveira} et~al.(2014){Alves de Oliveira}, {Schneider},
  {Mer{\'\i}n}, {Prusti}, {Ribas}, {Cox}, {Vavrek}, {K{\"o}nyves},
  {Arzoumanian}, {Puga}, {Pilbratt}, {K{\'o}sp{\'a}l}, {Andr{\'e}}, {Didelon},
  {Men'shchikov}, {Royer}, {Waelkens}, {Bontemps}, {Winston}, and
  {Spezzi}}]{AlvesdeOliveira+2014}
{Alves de Oliveira} C, {Schneider} N, {Mer{\'\i}n} B, et~al. (2014) {Herschel
  view of the large-scale structure in the <ASTROBJ>Chamaeleon</ASTROBJ> dark
  clouds}. \aap 568:A98

\bibitem[{{Andr{\'e}} et~al.(2010){Andr{\'e}}, {Men'shchikov}, {Bontemps},
  {K{\"o}nyves}, {Motte}, {Schneider}, {Didelon}, {Minier}, {Saraceno},
  {Ward-Thompson}, {di Francesco}, {White}, {Molinari}, {Testi}, {Abergel},
  {Griffin}, {Henning}, {Royer}, {Mer{\'{\i}}n}, {Vavrek}, {Attard},
  {Arzoumanian}, {Wilson}, {Ade}, {Aussel}, {Baluteau}, {Benedettini},
  {Bernard}, {Blommaert}, {Cambr{\'e}sy}, {Cox}, {di Giorgio}, {Hargrave},
  {Hennemann}, {Huang}, {Kirk}, {Krause}, {Launhardt}, {Leeks}, {Le Pennec},
  {Li}, {Martin}, {Maury}, {Olofsson}, {Omont}, {Peretto}, {Pezzuto}, {Prusti},
  {Roussel}, {Russeil}, {Sauvage}, {Sibthorpe}, {Sicilia-Aguilar}, {Spinoglio},
  {Waelkens}, {Woodcraft}, and {Zavagno}}]{Andre+10}
{Andr{\'e}} P, {Men'shchikov} A, {Bontemps} S, et~al. (2010) {From filamentary
  clouds to prestellar cores to the stellar IMF: Initial highlights from the
  Herschel Gould Belt Survey}. A{\&}A 518:L102

\bibitem[{{Andr{\'e}} et~al.(2014){Andr{\'e}}, {Di Francesco}, {Ward-Thompson},
  {Inutsuka}, {Pudritz}, and {Pineda}}]{Andre+14}
{Andr{\'e}} P, {Di Francesco} J, {Ward-Thompson} D, et~al. (2014) {From
  Filamentary Networks to Dense Cores in Molecular Clouds: Toward a New
  Paradigm for Star Formation}. in Protostars and Planets VI, ed H Beuther et
  al pp 27--51

\bibitem[{{Andr{\'e}} et~al.(2016){Andr{\'e}}, {Rev{\'e}ret}, {K{\"o}nyves},
  {Arzoumanian}, {Tig{\'e}}, {Gallais}, {Roussel}, {Le Pennec}, {Rodriguez},
  {Doumayrou}, {Dubreuil}, {Lortholary}, {Martignac}, {Talvard}, {Delisle},
  {Visticot}, {Dumaye}, {De Breuck}, {Shimajiri}, {Motte}, {Bontemps},
  {Hennemann}, {Zavagno}, {Russeil}, {Schneider}, {Palmeirim}, {Peretto},
  {Hill}, {Minier}, {Roy}, and {Rygl}}]{Andre+16}
{Andr{\'e}} P, {Rev{\'e}ret} V, {K{\"o}nyves} V, et~al. (2016) {Characterizing
  filaments in regions of high-mass star formation: High-resolution
  submilimeter imaging of the massive star-forming complex NGC 6334 with
  ArT{\'e}MiS}. \aap 592:A54

\bibitem[{{Andr{\'e}} et~al.(2019){Andr{\'e}}, {Arzoumanian}, {K{\"o}nyves},
  {Shimajiri}, and {Palmeirim}}]{Andre+19}
{Andr{\'e}} P, {Arzoumanian} D, {K{\"o}nyves} V, et~al. (2019) {The role of
  molecular filaments in the origin of the prestellar core mass function and
  stellar initial mass function}. \aap 629:L4

\bibitem[{{Arzoumanian} et~al.(2011){Arzoumanian}, {Andr{\'e}}, {Didelon},
  {K{\"o}nyves}, {Schneider}, {Men'shchikov}, {Sousbie}, {Zavagno}, {Bontemps},
  {di Francesco}, {Griffin}, {Hennemann}, {Hill}, {Kirk}, {Martin}, {Minier},
  {Molinari}, {Motte}, {Peretto}, {Pezzuto}, {Spinoglio}, {Ward-Thompson},
  {White}, and {Wilson}}]{Arzoumanian+11}
{Arzoumanian} D, {Andr{\'e}} P, {Didelon} P, et~al. (2011) {Characterizing
  interstellar filaments with Herschel in IC 5146}. A{\&}A 529:L6

\bibitem[{{Arzoumanian} et~al.(2013){Arzoumanian}, {Andr{\'e}}, {Peretto}, and
  {K{\"o}nyves}}]{Arzoumanian+13}
{Arzoumanian} D, {Andr{\'e}} P, {Peretto} N, et~al. (2013) {Formation and
  evolution of interstellar filaments. Hints from velocity dispersion
  measurements}. \aap 553:A119

\bibitem[{{Arzoumanian} et~al.(2019){Arzoumanian}, {Andr{\'e}}, {K{\"o}nyves},
  {Palmeirim}, {Roy}, {Schneider}, {Benedettini}, {Didelon}, {Di Francesco},
  {Kirk}, and {Ladjelate}}]{Arzoumanian+19}
{Arzoumanian} D, {Andr{\'e}} P, {K{\"o}nyves} V, et~al. (2019) {Characterizing
  the properties of nearby molecular filaments observed with Herschel}. \aap
  621:A42

\bibitem[{{Auddy} et~al.(2016){Auddy}, {Basu}, and {Kudoh}}]{Auddy+16}
{Auddy} S, {Basu} S, and {Kudoh} T (2016) {A Magnetic Ribbon Model for
  Star-forming Filaments}. \apj 831:46

\bibitem[{{Audit} and {Hennebelle}(2005)}]{audit2005}
{Audit} E and {Hennebelle} P (2005) {Thermal condensation in a turbulent atomic
  hydrogen flow}. \aap 433:1--13

\bibitem[{{Audit} and {Hennebelle}(2010)}]{audit2010}
{Audit} E and {Hennebelle} P (2010) {On the structure of the turbulent
  interstellar clouds . Influence of the equation of state on the dynamics of
  3D compressible flows}. \aap 511:A76

\bibitem[{{Ballesteros-Paredes}(2006)}]{BallesterosParedes06}
{Ballesteros-Paredes} J (2006) {Six myths on the virial theorem for
  interstellar clouds}. \mnras 372(1):443--449

\bibitem[{{Ballesteros-Paredes} and
  {Hartmann}(2007)}]{BallesterosParedesHartmann07}
{Ballesteros-Paredes} J and {Hartmann} L (2007) {Remarks on Rapid vs. Slow Star
  Formation}. \rmxaa 43:123--136

\bibitem[{{Ballesteros-Paredes} and {Mac
  Low}(2002)}]{BallesterosParedesMacLow02}
{Ballesteros-Paredes} J and {Mac Low} MM (2002) {Physical versus Observational
  Properties of Clouds in Turbulent Molecular Cloud Models}. \apj 570:734--748

\bibitem[{{Ballesteros-Paredes} and
  {V{\'a}zquez-Semadeni}(1997)}]{BallesterosParedesVazquezSemadeni97}
{Ballesteros-Paredes} J and {V{\'a}zquez-Semadeni} E (1997) {Virial balance in
  turbulent MHD two dimensional numerical simulations of the ISM}. In: {Holt}
  SS and {Mundy} LG (eds) American Institute of Physics Conference Series,
  American Institute of Physics Conference Series, vol 393, pp 81--84

\bibitem[{{Ballesteros-Paredes}
  et~al.(1999{\natexlab{a}}){Ballesteros-Paredes}, {Hartmann}, and
  {V{\'a}zquez-Semadeni}}]{BallesterosParedes+99b}
{Ballesteros-Paredes} J, {Hartmann} L, and {V{\'a}zquez-Semadeni} E
  (1999{\natexlab{a}}) {Turbulent Flow-driven Molecular Cloud Formation: A
  Solution to the Post-T Tauri Problem?} \apj 527(1):285--297

\bibitem[{{Ballesteros-Paredes}
  et~al.(1999{\natexlab{b}}){Ballesteros-Paredes}, {V{\'a}zquez-Semadeni}, and
  {Scalo}}]{BallesterosParedes+99a}
{Ballesteros-Paredes} J, {V{\'a}zquez-Semadeni} E, and {Scalo} J
  (1999{\natexlab{b}}) {Clouds as Turbulent Density Fluctuations: Implications
  for Pressure Confinement and Spectral Line Data Interpretation}. \apj
  515(1):286--303

\bibitem[{{Ballesteros-Paredes} et~al.(2006){Ballesteros-Paredes}, {Gazol},
  {Kim}, {Klessen}, {Jappsen}, and {Tejero}}]{BallesterosParedes+06}
{Ballesteros-Paredes} J, {Gazol} A, {Kim} J, et~al. (2006) {The Mass Spectra of
  Cores in Turbulent Molecular Clouds and Implications for the Initial Mass
  Function}. \apj 637(1):384--391

\bibitem[{{Ballesteros-Paredes}
  et~al.(2009{\natexlab{a}}){Ballesteros-Paredes}, {G{\'o}mez}, {Loinard},
  {Torres}, and {Pichardo}}]{BallesterosParedes+09a}
{Ballesteros-Paredes} J, {G{\'o}mez} GC, {Loinard} L, et~al.
  (2009{\natexlab{a}}) {Tidal forces as a regulator of star formation in
  Taurus}. \mnras 395(1):L81--L84

\bibitem[{{Ballesteros-Paredes}
  et~al.(2009{\natexlab{b}}){Ballesteros-Paredes}, {G{\'o}mez}, {Pichardo}, and
  {V{\'a}zquez-Semadeni}}]{BallesterosParedes+09b}
{Ballesteros-Paredes} J, {G{\'o}mez} GC, {Pichardo} B, et~al.
  (2009{\natexlab{b}}) {On the gravitational content of molecular clouds and
  their cores}. \mnras 393(4):1563--1572

\bibitem[{{Ballesteros-Paredes}
  et~al.(2011{\natexlab{a}}){Ballesteros-Paredes}, {Hartmann},
  {V{\'a}zquez-Semadeni}, {Heitsch}, and
  {Zamora-Avil{\'e}s}}]{BallesterosParedes+11a}
{Ballesteros-Paredes} J, {Hartmann} LW, {V{\'a}zquez-Semadeni} E, et~al.
  (2011{\natexlab{a}}) {Gravity or turbulence? Velocity dispersion-size
  relation}. \mnras 411(1):65--70

\bibitem[{{Ballesteros-Paredes}
  et~al.(2011{\natexlab{b}}){Ballesteros-Paredes}, {V{\'a}zquez-Semadeni},
  {Gazol}, {Hartmann}, {Heitsch}, and {Col{\'\i}n}}]{BallesterosParedes+11b}
{Ballesteros-Paredes} J, {V{\'a}zquez-Semadeni} E, {Gazol} A, et~al.
  (2011{\natexlab{b}}) {Gravity or turbulence? - II. Evolving column density
  probability distribution functions in molecular clouds}. \mnras
  416(2):1436--1442

\bibitem[{{Ballesteros-Paredes} et~al.(2012){Ballesteros-Paredes}, {D'Alessio},
  and {Hartmann}}]{BallesterosParedes+12}
{Ballesteros-Paredes} J, {D'Alessio} P, and {Hartmann} L (2012) {On the
  structure of molecular clouds}. \mnras 427:2562--2571

\bibitem[{{Ballesteros-Paredes} et~al.(2015){Ballesteros-Paredes}, {Hartmann},
  {P{\'e}rez-Goytia}, and {Kuznetsova}}]{BallesterosParedes+15}
{Ballesteros-Paredes} J, {Hartmann} LW, {P{\'e}rez-Goytia} N, et~al. (2015)
  {Bondi-Hoyle-Littleton accretion and the upper-mass stellar initial mass
  function}. \mnras 452(1):566--574

\bibitem[{{Ballesteros-Paredes} et~al.(2018){Ballesteros-Paredes},
  {V{\'a}zquez-Semadeni}, {Palau}, and {Klessen}}]{BallesterosParedes+18}
{Ballesteros-Paredes} J, {V{\'a}zquez-Semadeni} E, {Palau} A, et~al. (2018)
  {Gravity or turbulence? - IV. Collapsing cores in out-of-virial disguise}.
  \mnras 479(2):2112--2125

\bibitem[{{Ballesteros-Paredes} et~al.(2019){Ballesteros-Paredes},
  {Rom{\'a}n-Z{\'u}{\~n}iga}, {Salom{\'e}}, {Zamora-Avil{\'e}s}, and
  {Jim{\'e}nez-Donaire}}]{BallesterosParedes+19}
{Ballesteros-Paredes} J, {Rom{\'a}n-Z{\'u}{\~n}iga} C, {Salom{\'e}} Q, et~al.
  (2019) {What is the physics behind the Larson mass-size relation?} \mnras
  490(2):2648--2655

\bibitem[{{Bally} et~al.(1987){Bally}, {Langer}, {Stark}, and
  {Wilson}}]{Bally+87}
{Bally} J, {Langer} WD, {Stark} AA, et~al. (1987) {Filamentary structure in the
  Orion molecular cloud}. \apjl 312:L45--L49

\bibitem[{{Banerjee} et~al.(2009){Banerjee}, {V{\'a}zquez-Semadeni},
  {Hennebelle}, and {Klessen}}]{baner2009}
{Banerjee} R, {V{\'a}zquez-Semadeni} E, {Hennebelle} P, et~al. (2009) {Clump
  morphology and evolution in MHD simulations of molecular cloud formation}.
  \mnras 398:1082--1092

\bibitem[{{Barnes} et~al.(2017){Barnes}, {Longmore}, {Battersby}, {Bally},
  {Kruijssen}, {Henshaw}, and {Walker}}]{Barnes+17}
{Barnes} AT, {Longmore} SN, {Battersby} C, et~al. (2017) {Star formation rates
  and efficiencies in the Galactic Centre}. \mnras 469:2263--2285

\bibitem[{{Barnes} et~al.(2018){Barnes}, {Henshaw}, {Caselli},
  {Jim{\'e}nez-Serra}, {Tan}, {Fontani}, {Pon}, and {Ragan}}]{Barnes+18}
{Barnes} AT, {Henshaw} JD, {Caselli} P, et~al. (2018) {Similar complex
  kinematics within two massive, filamentary infrared dark clouds}. \mnras
  475(4):5268--5289

\bibitem[{{Barnes} et~al.(2016){Barnes}, {Hernandez}, {O'Dougherty}, {Schap},
  and {Muller}}]{Barnes+16}
{Barnes} PJ, {Hernandez} AK, {O'Dougherty} SN, et~al. (2016) {The Galactic
  Census of High- and Medium-mass Protostars. III. $^{12}$CO Maps and Physical
  Properties of Dense Clump Envelopes and Their Embedding GMCs}. \apj 831(1):67

\bibitem[{{Barranco} and {Goodman}(1998)}]{BarrancoGoodman98}
{Barranco} JA and {Goodman} AA (1998) {Coherent Dense Cores. I. NH$_{3}$
  Observations}. \apj 504(1):207--222

\bibitem[{{Beaumont} et~al.(2012){Beaumont}, {Goodman}, {Alves}, {Lombardi},
  {Rom{\'a}n-Z{\'u}{\~n}iga}, {Kauffmann}, and {Lada}}]{Beaumont+12}
{Beaumont} CN, {Goodman} AA, {Alves} JF, et~al. (2012) {A simple perspective on
  the mass-area relationship in molecular clouds}. \mnras 423(3):2579--2586

\bibitem[{{Bergin} et~al.(2004){Bergin}, {Hartmann}, {Raymond}, and
  {Ballesteros-Paredes}}]{Bergin+04}
{Bergin} EA, {Hartmann} LW, {Raymond} JC, et~al. (2004) {Molecular Cloud
  Formation behind Shock Waves}. \apj 612(2):921--939

\bibitem[{{Bern{\'e}} et~al.(2014){Bern{\'e}}, {Marcelino}, and
  {Cernicharo}}]{Berne+14}
{Bern{\'e}} O, {Marcelino} N, and {Cernicharo} J (2014) {IRAM 30 m Large Scale
  Survey of $^{12}$CO(2-1) and $^{13}$CO(2-1) Emission in the Orion Molecular
  Cloud}. \apj 795(1):13

\bibitem[{{Bertoldi} and {McKee}(1992)}]{BertoldiMcKee92}
{Bertoldi} F and {McKee} CF (1992) {Pressure-confined Clumps in Magnetized
  Molecular Clouds}. \apj 395:140

\bibitem[{{Beuther} and {Schilke}(2004)}]{BeutherSchilke04}
{Beuther} H and {Schilke} P (2004) {Fragmentation in MassiveStar Formation}.
  Science 303(5661):1167--1169

\bibitem[{Bigiel et~al.(2008)Bigiel, Leroy, Walter, Brinks, de~Blok, Madore,
  and Thornley}]{Bigiel+08}
Bigiel F, Leroy A, Walter F, et~al. (2008) {the Star Formation Law in Nearby
  Galaxies on Sub-Kpc Scales}. \aj 136(6):2846--2871

\bibitem[{{Blitz}(1993)}]{Blitz93}
{Blitz} L (1993) {Giant molecular clouds}. In: {Levy} E and {Lunine} J (eds)
  Protostars and Planets III, pp 125--161

\bibitem[{{Blitz} and {Shu}(1980)}]{BlitzShu80}
{Blitz} L and {Shu} FH (1980) {The origin and lifetime of giant molecular cloud
  complexes}. \apj 238:148--157

\bibitem[{Blitz et~al.(2007)Blitz, Fukui, Kawamura, Leroy, Mizuno, and
  Rosolowsky}]{Blitz+07}
Blitz L, Fukui Y, Kawamura A, et~al. (2007) {Giant Molecular Clouds in Local
  Group Galaxies}. In: Reipurth B, Jewitt D, and Keil K (eds) Protostars and
  Planets V, University of Arizona Press, Tucson, p~81

\bibitem[{{Bolatto} et~al.(2008){Bolatto}, {Leroy}, {Rosolowsky}, {Walter}, and
  {Blitz}}]{Bolatto+08}
{Bolatto} AD, {Leroy} AK, {Rosolowsky} E, et~al. (2008) {The Resolved
  Properties of Extragalactic Giant Molecular Clouds}. \apj 686(2):948--965

\bibitem[{{Bonnell} and {Bate}(2006)}]{BonnellBate06}
{Bonnell} IA and {Bate} MR (2006) {Star formation through gravitational
  collapse and competitive accretion}. \mnras 370(1):488--494

\bibitem[{{Bonnor}(1956)}]{Bonnor56}
{Bonnor} W (1956) {Boyle's Law and gravitational instability}. \mnras
  116:351--+

\bibitem[{{Bracco} et~al.(2017){Bracco}, {Palmeirim}, {Andr{\'e}}, {Adam},
  {Ade}, {Bacmann}, {Beelen}, {Beno{\^i}t}, {Bideaud}, {Billot}, {Bourrion},
  {Calvo}, {Catalano}, {Coiffard}, {Comis}, {D'Addabbo}, {D{\'e}sert},
  {Didelon}, {Doyle}, {Goupy}, {K{\"o}nyves}, {Kramer}, {Lagache}, {Leclercq},
  {Mac{\'{\i}}as-P{\'e}rez}, {Maury}, {Mauskopf}, {Mayet}, {Monfardini},
  {Motte}, {Pajot}, {Pascale}, {Peretto}, {Perotto}, {Pisano}, {Ponthieu},
  {Rev{\'e}ret}, {Rigby}, {Ritacco}, {Rodriguez}, {Romero}, {Roy}, {Ruppin},
  {Schuster}, {Sievers}, {Triqueneaux}, {Tucker}, and {Zylka}}]{Bracco+17}
{Bracco} A, {Palmeirim} P, {Andr{\'e}} P, et~al. (2017) {Probing changes of
  dust properties along a chain of solar-type prestellar and protostellar cores
  in Taurus with NIKA}. \aap 604:A52

\bibitem[{{Bresnahan} et~al.(2018){Bresnahan}, {Ward-Thompson}, {Kirk},
  {Pattle}, {Eyres}, {White}, {K{\"o}nyves}, {Men'shchikov}, {Andr{\'e}},
  {Schneider}, {Di Francesco}, {Arzoumanian}, {Benedettini}, {Ladjelate},
  {Palmeirim}, {Bracco}, {Molinari}, {Pezzuto}, and {Spinoglio}}]{Bresnahan+18}
{Bresnahan} D, {Ward-Thompson} D, {Kirk} JM, et~al. (2018) {The dense cores and
  filamentary structure of the molecular cloud in Corona Australis: Herschel
  SPIRE and PACS observations from the Herschel Gould Belt Survey}. \aap
  615:A125

\bibitem[{{Briceno} et~al.(1997){Briceno}, {Hartmann}, {Stauffer}, {Gagne},
  {Stern}, and {Caillault}}]{Briceno+97}
{Briceno} C, {Hartmann} LW, {Stauffer} JR, et~al. (1997) {X-Rays Surveys and
  the Post-T Tauri Problem}. \aj 113:740--752

\bibitem[{{Camacho} et~al.(2016){Camacho}, {V{\'a}zquez-Semadeni},
  {Ballesteros-Paredes}, {G{\'o}mez}, {Fall}, and
  {Mata-Ch{\'a}vez}}]{camacho2016}
{Camacho} V, {V{\'a}zquez-Semadeni} E, {Ballesteros-Paredes} J, et~al. (2016)
  {Energy Budget of Forming Clumps in Numerical Simulations of Collapsing
  Clouds}. \apj 833:113

\bibitem[{{Carr}(1987)}]{Carr87}
{Carr} JS (1987) {A Study of Clumping in the Cepheus OB 3 Molecular Cloud}.
  \apj 323:170

\bibitem[{{Caselli} and {Myers}(1995)}]{CaselliMyers95}
{Caselli} P and {Myers} PC (1995) {The Line Width--Size Relation in Massive
  Cloud Cores}. \apj 446:665

\bibitem[{{Caselli} et~al.(2002){Caselli}, {Benson}, {Myers}, and
  {Tafalla}}]{Caselli+02}
{Caselli} P, {Benson} PJ, {Myers} PC, et~al. (2002) {Dense Cores in Dark
  Clouds. XIV. N$_{2}$H$^{+}$ (1-0) Maps of Dense Cloud Cores}. \apj
  572(1):238--263

\bibitem[{{Cesaroni} et~al.(2019){Cesaroni}, {Beuther}, {Ahmadi},
  {Beltr{\'a}n}, {Csengeri}, {Galv{\'a}n-Madrid}, {Gieser}, {Henning},
  {Johnston}, {Klaassen}, {Kuiper}, {Leurini}, {Linz}, {Longmore}, {Lumsden},
  {Maud}, {Moscadelli}, {Mottram}, {Palau}, {Peters}, {Pudritz},
  {S{\'a}nchez-Monge}, {Schilke}, {Semenov}, {Suri}, {Urquhart}, {Winters},
  {Zhang}, and {Zinnecker}}]{Cesaroni+19}
{Cesaroni} R, {Beuther} H, {Ahmadi} A, et~al. (2019) {IRAS 23385+6053: an
  embedded massive cluster in the making}. \aap 627:A68

\bibitem[{{Chabrier}(2005)}]{Chabrier05}
{Chabrier} G (2005) {The Initial Mass Function: from Salpeter 1955 to 2005}.
  In: {E~Corbelli, F~Palla, \& H~Zinnecker} (ed) The Initial Mass Function 50
  Years Later, Astrophysics and Space Science Library, vol 327, pp 41--+

\bibitem[{{Chandrasekhar} and {Fermi}(1953)}]{ChandraFermi53}
{Chandrasekhar} S and {Fermi} E (1953) {Magnetic Fields in Spiral Arms.} \apj
  118:113

\bibitem[{{Chapman} et~al.(2011){Chapman}, {Goldsmith}, {Pineda}, {Clemens},
  {Li}, and {Kr{\v c}o}}]{Chapman+11}
{Chapman} NL, {Goldsmith} PF, {Pineda} JL, et~al. (2011) {The Magnetic Field in
  Taurus Probed by Infrared Polarization}. \apj 741:21

\bibitem[{{Chen} et~al.(2017){Chen}, {Li}, {King}, and {Fissel}}]{Chen+17}
{Chen} CY, {Li} ZY, {King} PK, et~al. (2017) {Fantastic Striations and Where to
  Find Them: The Origin of Magnetically Aligned Striations in Interstellar
  Clouds}. \apj 847:140

\bibitem[{{Chen} et~al.(2019){Chen}, {Pineda}, {Goodman}, {Burkert}, {Offner},
  {Friesen}, {Myers}, {Alves}, {Arce}, {Caselli}, {Chac{\'o}n-Tanarro}, {Chen},
  {Di Francesco}, {Ginsburg}, {Keown}, {Kirk}, {Martin}, {Matzner}, {Punanova},
  {Redaelli}, {Rosolowsky}, {Scibelli}, {Seo}, {Shirley}, {Singh}, and {The GAS
  Collaboration}}]{Chen+19}
{Chen} HHH, {Pineda} JE, {Goodman} AA, et~al. (2019) {Droplets. I.
  Pressure-dominated Coherent Structures in L1688 and B18}. \apj 877(2):93

\bibitem[{{Cheng} et~al.(2018){Cheng}, {Tan}, {Liu}, {Kong}, {Lim}, {Andersen},
  and {Da Rio}}]{Cheng+18}
{Cheng} Y, {Tan} JC, {Liu} M, et~al. (2018) {The Core Mass Function in the
  Massive Protocluster G286.21+0.17 Revealed by ALMA}. \apj 853(2):160

\bibitem[{{Chevance} et~al.(2016){Chevance}, {Madden}, {Lebouteiller},
  {Godard}, {Cormier}, {Galliano}, {Hony}, {Indebetouw}, {Le Bourlot}, {Lee},
  {Le Petit}, {Pellegrini}, {Roueff}, and {Wu}}]{chevance16}
{Chevance} M, {Madden} SC, {Lebouteiller} V, et~al. (2016) {A milestone toward
  understanding PDR properties in the extreme environment of LMC-30 Doradus}.
  \aap 590:A36

\bibitem[{{Chevance} et~al.(2020{\natexlab{a}}){Chevance}, {Kruijssen},
  {Hygate}, {Schruba}, {Longmore}, {Groves}, {Henshaw}, {Herrera}, {Hughes},
  {Jeffreson}, {Lang}, {Leroy}, {Meidt}, {Pety}, {Razza}, {Rosolowsky},
  {Schinnerer}, {Bigiel}, {Blanc}, {Emsellem}, {Faesi}, {Glover}, {Haydon},
  {Ho}, {Kreckel}, {Lee}, {Liu}, {Querejeta}, {Saito}, {Sun}, {Usero}, and
  {Utomo}}]{Chevance+20a}
{Chevance} M, {Kruijssen} JMD, {Hygate} APS, et~al. (2020{\natexlab{a}}) {The
  lifecycle of molecular clouds in nearby star-forming disc galaxies}. \mnras
  493(2):2872--2909

\bibitem[{{Chevance} et~al.(2020{\natexlab{b}}){Chevance}, {Kruijssen},
  {Vazquez-Semadeni}, {Nakamura}, {Klessen}, {Ballesteros-Paredes}, {Inutsuka},
  {Adamo}, and {Hennebelle}}]{Chevance+20b}
{Chevance} M, {Kruijssen} JMD, {Vazquez-Semadeni} E, et~al.
  (2020{\natexlab{b}}) {The Molecular Cloud Lifecycle}. \ssr 216(4):50

\bibitem[{{Clark} et~al.(2012){Clark}, {Glover}, {Klessen}, and
  {Bonnell}}]{clark2012}
{Clark} PC, {Glover} SCO, {Klessen} RS, et~al. (2012) {How long does it take to
  form a molecular cloud?} \mnras 424:2599--2613

\bibitem[{{Clarke} and {Whitworth}(2015)}]{ClarkeWhitworth15}
{Clarke} SD and {Whitworth} AP (2015) {Investigating the global collapse of
  filaments using smoothed particle hydrodynamics}. \mnras 449(2):1819--1825

\bibitem[{{Clarke} et~al.(2016){Clarke}, {Whitworth}, and {Hubber}}]{Clarke+16}
{Clarke} SD, {Whitworth} AP, and {Hubber} DA (2016) {Perturbation growth in
  accreting filaments}. \mnras 458:319--324

\bibitem[{{Clarke} et~al.(2017){Clarke}, {Whitworth}, {Duarte-Cabral}, and
  {Hubber}}]{Clarke+17}
{Clarke} SD, {Whitworth} AP, {Duarte-Cabral} A, et~al. (2017) {Filamentary
  fragmentation in a turbulent medium}. \mnras 468:2489--2505

\bibitem[{{Clarke} et~al.(2018){Clarke}, {Whitworth}, {Spowage},
  {Duarte-Cabral}, {Suri}, {Jaffa}, {Walch}, and {Clark}}]{Clarke+18}
{Clarke} SD, {Whitworth} AP, {Spowage} RL, et~al. (2018) {Synthetic C$^{18}$O
  observations of fibrous filaments: the problems of mapping from PPV to PPP}.
  \mnras 479(2):1722--1746

\bibitem[{{Colombo} et~al.(2019){Colombo}, {Rosolowsky}, {Duarte-Cabral},
  {Ginsburg}, {Glenn}, {Zetterlund}, {Hernand ez}, {Dempsey}, and
  {Currie}}]{Colombo+19}
{Colombo} D, {Rosolowsky} E, {Duarte-Cabral} A, et~al. (2019) {The integrated
  properties of the molecular clouds from the JCMT CO(3-2) High-Resolution
  Survey}. \mnras 483(4):4291--4340

\bibitem[{Corbelli et~al.(2017)Corbelli, Braine, Bandiera, Brouillet, Combes,
  Druard, Gratier, Mata, Schuster, Xilouris, and Palla}]{Corbelli+17}
Corbelli E, Braine J, Bandiera R, et~al. (2017) {From molecules to young
  stellar clusters: the star formation cycle across the disk of M 33}. \aap
  601:A146

\bibitem[{{Cottaar} et~al.(2012){Cottaar}, {Meyer}, {Andersen}, and
  {Espinoza}}]{Cottaar+12}
{Cottaar} M, {Meyer} MR, {Andersen} M, et~al. (2012) {Is the massive young
  cluster Westerlund I bound?} \aap 539:A5

\bibitem[{{Cox} et~al.(2016){Cox}, {Arzoumanian}, {Andr{\'e}}, {Rygl},
  {Prusti}, {Men'shchikov}, {Royer}, {K{\'o}sp{\'a}l}, {Palmeirim}, {Ribas},
  {K{\"o}nyves}, {Bernard}, {Schneider}, {Bontemps}, {Merin}, {Vavrek}, {Alves
  de Oliveira}, {Didelon}, {Pilbratt}, and {Waelkens}}]{Cox+16}
{Cox} NLJ, {Arzoumanian} D, {Andr{\'e}} P, et~al. (2016) {Filamentary structure
  and magnetic field orientation in Musca}. \aap 590:A110

\bibitem[{{Crutcher}(2012)}]{Crutcher12}
{Crutcher} RM (2012) {Magnetic Fields in Molecular Clouds}. \araa 50:29--63

\bibitem[{{Crutcher} and {Kemball}(2019)}]{crutcher2019}
{Crutcher} RM and {Kemball} AJ (2019) {Review of Zeeman Effect Observations of
  Regions of Star Formation K Zeeman Effect, Magnetic Fields, Star formation,
  Masers, Molecular clouds}. Frontiers in Astronomy and Space Sciences 6:66

\bibitem[{{Crutcher} et~al.(2010){Crutcher}, {Wandelt}, {Heiles}, {Falgarone},
  and {Troland}}]{Crutcher+10}
{Crutcher} RM, {Wandelt} B, {Heiles} C, et~al. (2010) {Magnetic Fields in
  Interstellar Clouds from Zeeman Observations: Inference of Total Field
  Strengths by Bayesian Analysis}. \apj 725(1):466--479

\bibitem[{{Dale} et~al.(2014){Dale}, {Ngoumou}, {Ercolano}, and
  {Bonnell}}]{Dale+14}
{Dale} JE, {Ngoumou} J, {Ercolano} B, et~al. (2014) {Before the first
  supernova: combined effects of H II regions and winds on molecular clouds}.
  \mnras 442(1):694--712

\bibitem[{{Dall'Olio} et~al.(2019){Dall'Olio}, {Vlemmings}, {Persson}, {Alves},
  {Beuther}, {Girart}, {Surcis}, {Torrelles}, and {Van
  Langevelde}}]{DallOlio+19}
{Dall'Olio} D, {Vlemmings} WHT, {Persson} MV, et~al. (2019) {ALMA reveals the
  magnetic field evolution in the high-mass star forming complex G9.62+0.19}.
  \aap 626:A36

\bibitem[{{Davis} and {Greenstein}(1951)}]{DavisGreenstein51}
{Davis} J Leverett and {Greenstein} JL (1951) {The Polarization of Starlight by
  Aligned Dust Grains.} \apj 114:206

\bibitem[{{de Avillez} and {Breitschwerdt}(2005)}]{deAvillezBreitschwerdt05}
{de Avillez} MA and {Breitschwerdt} D (2005) {Global dynamical evolution of the
  ISM in star forming galaxies. I. High resolution 3D simulations: Effect of
  the magnetic field}. \aap 436(2):585--600

\bibitem[{{Dhabal} et~al.(2018){Dhabal}, {Mundy}, {Rizzo}, {Storm}, and
  {Teuben}}]{Dhabal+18}
{Dhabal} A, {Mundy} LG, {Rizzo} MJ, et~al. (2018) {Morphology and Kinematics of
  Filaments in the Serpens and Perseus Molecular Clouds}. \apj 853(2):169

\bibitem[{{Didelon} et~al.(2015){Didelon}, {Motte}, {Tremblin}, {Hill}, {Hony},
  {Hennemann}, {Hennebelle}, {Anderson}, {Galliano}, {Schneider}, {Rayner},
  {Rygl}, {Louvet}, {Zavagno}, {K{\"o}nyves}, {Sauvage}, {Andr{\'e}},
  {Bontemps}, {Peretto}, {Griffin}, {Gonz{\'a}lez}, {Lebouteiller},
  {Arzoumanian}, {Bernard}, {Benedettini}, {Di Francesco}, {Men'shchikov},
  {Minier}, {Nguy{\^e}n Luong}, {Palmeirim}, {Pezzuto}, {Rivera-Ingraham},
  {Russeil}, {Ward-Thompson}, and {White}}]{Didelon+15}
{Didelon} P, {Motte} F, {Tremblin} P, et~al. (2015) {From forced collapse to H
  ii region expansion in Mon R2: Envelope density structure and age
  determination with Herschel}. \aap 584:A4

\bibitem[{{Dobbs}(2008)}]{Dobbs08}
{Dobbs} CL (2008) {GMC formation by agglomeration and self gravity}. \mnras
  391(2):844--858

\bibitem[{{Dobbs} et~al.(2011{\natexlab{a}}){Dobbs}, {Burkert}, and
  {Pringle}}]{Dobbs+11b}
{Dobbs} CL, {Burkert} A, and {Pringle} JE (2011{\natexlab{a}}) {The properties
  of the interstellar medium in disc galaxies with stellar feedback}. \mnras
  417(2):1318--1334

\bibitem[{{Dobbs} et~al.(2011{\natexlab{b}}){Dobbs}, {Burkert}, and
  {Pringle}}]{Dobbs+11a}
{Dobbs} CL, {Burkert} A, and {Pringle} JE (2011{\natexlab{b}}) {Why are most
  molecular clouds not gravitationally bound?} \mnras 413(4):2935--2942

\bibitem[{{Dobbs} et~al.(2014){Dobbs}, {Krumholz}, {Ballesteros-Paredes},
  {Bolatto}, {Fukui}, {Heyer}, {Low}, {Ostriker}, and
  {V{\'a}zquez-Semadeni}}]{Dobbs+14}
{Dobbs} CL, {Krumholz} MR, {Ballesteros-Paredes} J, et~al. (2014) {Formation of
  Molecular Clouds and Global Conditions for Star Formation}. In: {Beuther} H,
  {Klessen} RS, {Dullemond} CP, et~al. (eds) Protostars and Planets VI, p~3

\bibitem[{{Dobbs} et~al.(2015){Dobbs}, {Pringle}, and
  {Duarte-Cabral}}]{Dobbs+15}
{Dobbs} CL, {Pringle} JE, and {Duarte-Cabral} A (2015) {The frequency and
  nature of `cloud-cloud collisions' in galaxies}. \mnras 446(4):3608--3620

\bibitem[{{Draine}(2011)}]{draine11}
{Draine} BT (2011) {Physics of the Interstellar and Intergalactic Medium}.
  Princeton University Press

\bibitem[{{Draine} and {Bertoldi}(1996)}]{draine1996a}
{Draine} BT and {Bertoldi} F (1996) {Structure of Stationary Photodissociation
  Fronts}. \apj 468:269

\bibitem[{{Elmegreen}(1987)}]{elmegreen87}
{Elmegreen} BG (1987) {Supercloud formation by nonaxisymmetric gravitational
  instabilities in sheared magnetic galaxy disks}. \apj 312:626--639

\bibitem[{{Elmegreen}(1993)}]{elmegreen93}
{Elmegreen} BG (1993) {Disk instabilities and star formation.} In: {Franco} J,
  {Ferrini} F, and {Tenorio-Tagle} G (eds) Star Formation, Galaxies and the
  Interstellar Medium, p 337

\bibitem[{{Elmegreen}(1997)}]{elmegreen97}
{Elmegreen} BG (1997) {Theory of Starbursts in Nuclear Rings}. In: {Franco} J,
  {Terlevich} R, and {Serrano} A (eds) Revista Mexicana de Astronomia y
  Astrofisica Conference Series, Revista Mexicana de Astronomia y Astrofisica,
  vol. 27, vol~6, p 165

\bibitem[{{Elmegreen}(2000)}]{Elmegreen00}
{Elmegreen} BG (2000) {Star Formation in a Crossing Time}. \apj 530(1):277--281

\bibitem[{{Elmegreen}(2008)}]{elmegreen08}
{Elmegreen} BG (2008) {Variations in Stellar Clustering with Environment:
  Dispersed Star Formation and the Origin of Faint Fuzzies}. \apj
  672:1006--1012

\bibitem[{{Elmegreen} and {Scalo}(2004)}]{ElmegreenScalo04}
{Elmegreen} BG and {Scalo} J (2004) {Interstellar Turbulence I: Observations
  and Processes}. \araa 42(1):211--273

\bibitem[{Engargiola et~al.(2003)Engargiola, Plambeck, Rosolowsky, and
  Blitz}]{Engargiola+03}
Engargiola G, Plambeck R, Rosolowsky E, et~al. (2003) {Giant Molecular Clouds
  in M33 - I. BIMA All Disk Survey}. \apjs 149(2):343--363

\bibitem[{{Evans} et~al.(2014){Evans}, {Heiderman}, and
  {Vutisalchavakul}}]{Evans+14}
{Evans} NJ II, {Heiderman} A, and {Vutisalchavakul} N (2014) {Star Formation
  Relations in Nearby Molecular Clouds}. \apj 782:114

\bibitem[{{Falgarone} et~al.(1991){Falgarone}, {Phillips}, and
  {Walker}}]{Falgarone+91}
{Falgarone} E, {Phillips} TG, and {Walker} CK (1991) {The Edges of Molecular
  Clouds: Fractal Boundaries and Density Structure}. The Astrophysical Journal
  378:186

\bibitem[{{Falgarone} et~al.(2008){Falgarone}, {Troland}, {Crutcher}, and
  {Paubert}}]{Falgarone+08}
{Falgarone} E, {Troland} TH, {Crutcher} RM, et~al. (2008) {CN Zeeman
  measurements in star formation regions}. \aap 487(1):247--252

\bibitem[{{Falgarone} et~al.(2009){Falgarone}, {Pety}, and
  {Hily-Blant}}]{Falgarone+09}
{Falgarone} E, {Pety} J, and {Hily-Blant} P (2009) {Intermittency of
  interstellar turbulence: extreme velocity-shears and CO emission on
  milliparsec scale}. Astronomy and Astrophysics 507(1):355--368

\bibitem[{{Federrath}(2015)}]{fed2015}
{Federrath} C (2015) {Inefficient star formation through turbulence, magnetic
  fields and feedback}. \mnras 450:4035--4042

\bibitem[{{Federrath}(2016)}]{Federrath16}
{Federrath} C (2016) {On the universality of interstellar filaments: theory
  meets simulations and observations}. \mnras 457:375--388

\bibitem[{{Federrath} and {Klessen}(2012)}]{federrath12}
{Federrath} C and {Klessen} RS (2012) {The Star Formation Rate of Turbulent
  Magnetized Clouds: Comparing Theory, Simulations, and Observations}. \apj
  761:156

\bibitem[{{Federrath} et~al.(2008){Federrath}, {Klessen}, and
  {Schmidt}}]{Federrath+08}
{Federrath} C, {Klessen} RS, and {Schmidt} W (2008) {The Density Probability
  Distribution in Compressible Isothermal Turbulence: Solenoidal versus
  Compressive Forcing}. \apjl 688(2):L79

\bibitem[{{Federrath} et~al.(2010){Federrath}, {Roman-Duval}, {Klessen},
  {Schmidt}, and {Mac Low}}]{Federrath+10}
{Federrath} C, {Roman-Duval} J, {Klessen} RS, et~al. (2010) {Comparing the
  statistics of interstellar turbulence in simulations and observations.
  Solenoidal versus compressive turbulence forcing}. \aap 512:A81

\bibitem[{{Feldmann} et~al.(2011){Feldmann}, {Gnedin}, and
  {Kravtsov}}]{Feldmann+11}
{Feldmann} R, {Gnedin} NY, and {Kravtsov} AV (2011) {How Universal is the
  $\Sigma_{SFR} - \Sigma _{H_2}$ Relation?} \apj 732:115

\bibitem[{{Fern{\'a}ndez-L{\'o}pez} et~al.(2014){Fern{\'a}ndez-L{\'o}pez},
  {Arce}, {Looney}, {Mundy}, {Storm}, {Teuben}, {Lee}, {Segura-Cox}, {Isella},
  {Tobin}, {Rosolowsky}, {Plunkett}, {Kwon}, {Kauffmann}, {Ostriker}, {Tassis},
  {Shirley}, and {Pound}}]{Fernandez-Lopez+14}
{Fern{\'a}ndez-L{\'o}pez} M, {Arce} HG, {Looney} L, et~al. (2014) {CARMA Large
  Area Star Formation Survey: Observational Analysis of Filaments in the
  Serpens South Molecular Cloud}. \apjl 790:L19

\bibitem[{{Ferri{\'{e}}re}(2001)}]{ferriere01}
{Ferri{\'{e}}re} KM (2001) {The interstellar environment of our galaxy}.
  Reviews of Modern Physics 73:1031--1066

\bibitem[{{Fiege} and {Pudritz}(2000)}]{FiegePudritz00}
{Fiege} JD and {Pudritz} RE (2000) {Polarized Submillimeter Emission from
  Filamentary Molecular Clouds}. \apj 544:830--837

\bibitem[{{Field}(1965)}]{field1965}
{Field} GB (1965) {Thermal Instability.} \apj 142:531

\bibitem[{{Field} and {Saslaw}(1965)}]{FieldSaslaw65}
{Field} GB and {Saslaw} WC (1965) {A Statistical Model of the Formation of
  Stars and Interstellar Clouds.} \apj 142:568

\bibitem[{{Field} et~al.(1969){Field}, {Goldsmith}, and {Habing}}]{field69}
{Field} GB, {Goldsmith} DW, and {Habing} HJ (1969) {Cosmic-Ray Heating of the
  Interstellar Gas}. \apjl 155:L149

\bibitem[{{Field} et~al.(2011){Field}, {Blackman}, and {Keto}}]{Field+11}
{Field} GB, {Blackman} EG, and {Keto} ER (2011) {Does external pressure explain
  recent results for molecular clouds?} \mnras 416(1):710--714

\bibitem[{{Fischera} and {Martin}(2012)}]{FischeraMartin12}
{Fischera} J and {Martin} PG (2012) {Physical properties of interstellar
  filaments}. \aap 542:A77

\bibitem[{{Franco} and {Cox}(1986)}]{FrancoCox86}
{Franco} J and {Cox} DP (1986) {Molecular clouds in galaxies with different Z:
  fragmentation of diffuse clouds driven by opacity.} \pasp 98:1076--1079

\bibitem[{{Gazol} et~al.(2001){Gazol}, {V{\'a}zquez-Semadeni},
  {S{\'a}nchez-Salcedo}, and {Scalo}}]{Gazol2001}
{Gazol} A, {V{\'a}zquez-Semadeni} E, {S{\'a}nchez-Salcedo} FJ, et~al. (2001)
  {The Temperature Distribution in Turbulent Interstellar Gas}. \apjl
  557(2):L121--L124

\bibitem[{{Giannetti} et~al.(2014){Giannetti}, {Wyrowski}, {Brand}, {Csengeri},
  {Fontani}, {Walmsley}, {Nguyen Luong}, {Beuther}, {Schuller}, {G{\"u}sten},
  and {Menten}}]{Giannetti+14}
{Giannetti} A, {Wyrowski} F, {Brand} J, et~al. (2014) {ATLASGAL-selected
  massive clumps in the inner Galaxy. I. CO depletion and isotopic ratios}.
  \aap 570:A65

\bibitem[{{Girichidis et al.}(2020)}]{Girichidis+2020}
{Girichidis et al} J (2020) {Phisical processes in star formation}. \ssr

\bibitem[{{Glover} and {Clark}(2012{\natexlab{a}})}]{glover2012a}
{Glover} SCO and {Clark} PC (2012{\natexlab{a}}) {Approximations for modelling
  CO chemistry in giant molecular clouds: a comparison of approaches}. \mnras
  421(1):116--131

\bibitem[{{Glover} and {Clark}(2012{\natexlab{b}})}]{glover2012}
{Glover} SCO and {Clark} PC (2012{\natexlab{b}}) {Is molecular gas necessary
  for star formation?} \mnras 421(1):9--19

\bibitem[{{Glover} and {Mac Low}(2007)}]{glover2007}
{Glover} SCO and {Mac Low} MM (2007) {Simulating the Formation of Molecular
  Clouds. II. Rapid Formation from Turbulent Initial Conditions}. \apj
  659(2):1317--1337

\bibitem[{{Goldreich} and {Lynden-Bell}(1965)}]{GoldreichLyndenbell65}
{Goldreich} P and {Lynden-Bell} D (1965) {I. Gravitational stability of
  uniformly rotating disks}. \mnras 130:97

\bibitem[{{Goldsmith} and {Langer}(1999)}]{GoldsmithLanger99}
{Goldsmith} PF and {Langer} WD (1999) {Population Diagram Analysis of Molecular
  Line Emission}. \apj 517(1):209--225

\bibitem[{{Goldsmith} et~al.(2008){Goldsmith}, {Heyer}, {Narayanan}, {Snell},
  {Li}, and {Brunt}}]{Goldsmith+08}
{Goldsmith} PF, {Heyer} M, {Narayanan} G, et~al. (2008) {Large-Scale Structure
  of the Molecular Gas in Taurus Revealed by High Linear Dynamic Range Spectral
  Line Mapping}. \apj 680:428-445

\bibitem[{{G{\'o}mez}(2006)}]{Gomez06}
{G{\'o}mez} GC (2006) {Errors in Kinematic Distances and Our Image of the Milky
  Way Galaxy}. \aj 132(6):2376--2382

\bibitem[{{G{\'o}mez} and
  {V{\'a}zquez-Semadeni}(2014)}]{GomezVazquezSemadeni14}
{G{\'o}mez} GC and {V{\'a}zquez-Semadeni} E (2014) {Filaments in Simulations of
  Molecular Cloud Formation}. \apj 791:124

\bibitem[{{G{\'o}mez} et~al.(2018){G{\'o}mez}, {V{\'a}zquez-Semadeni}, and
  {Zamora-Avil{\'e}s}}]{Gomez+18}
{G{\'o}mez} GC, {V{\'a}zquez-Semadeni} E, and {Zamora-Avil{\'e}s} M (2018) {The
  magnetic field structure in molecular cloud filaments}. \mnras 480:2939--2944

\bibitem[{{Gong} et~al.(2018){Gong}, {Ostriker}, and {Kim}}]{gong2018}
{Gong} M, {Ostriker} EC, and {Kim} CG (2018) {The X $_{CO}$ Conversion Factor
  from Galactic Multiphase ISM Simulations}. \apj 858(1):16

\bibitem[{{Goodman} et~al.(1998){Goodman}, {Barranco}, {Wilner}, and
  {Heyer}}]{Goodman+98}
{Goodman} AA, {Barranco} JA, {Wilner} DJ, et~al. (1998) {Coherence in Dense
  Cores. II. The Transition to Coherence}. \apj 504:223--246

\bibitem[{{Gritschneder} et~al.(2009){Gritschneder}, {Naab}, {Walch},
  {Burkert}, and {Heitsch}}]{grit2009}
{Gritschneder} M, {Naab} T, {Walch} S, et~al. (2009) {Driving Turbulence and
  Triggering Star Formation by Ionizing Radiation}. \apjl 694(1):L26--L30

\bibitem[{{Gritschneder} et~al.(2017){Gritschneder}, {Heigl}, and
  {Burkert}}]{Gritschneder+17}
{Gritschneder} M, {Heigl} S, and {Burkert} A (2017) {Oscillating Filaments. I.
  Oscillation and Geometrical Fragmentation}. \apj 834:202

\bibitem[{{Grudi{\'c}} et~al.(2019){Grudi{\'c}}, {}, {Hopkins}, {Lee},
  {Murray}, {Faucher-Gigu{\`e}re}, and {Johnson}}]{Grudic+19}
{Grudi{\'c}}, {} MY, {Hopkins} PF, et~al. (2019) {On the nature of variations
  in the measured star formation efficiency of molecular clouds}. \mnras
  488(2):1501--1518

\bibitem[{{Gutermuth} et~al.(2011){Gutermuth}, {Pipher}, {Megeath}, {Myers},
  {Allen}, and {Allen}}]{gutermuth11}
{Gutermuth} RA, {Pipher} JL, {Megeath} ST, et~al. (2011) {A Correlation between
  Surface Densities of Young Stellar Objects and Gas in Eight Nearby Molecular
  Clouds}. \apj 739:84

\bibitem[{{Habing}(1968)}]{habing1968a}
{Habing} HJ (1968) {The interstellar radiation density between 912 A and 2400
  A}. \bain 19:421

\bibitem[{{Hacar} and {Tafalla}(2011)}]{HacarTafalla11}
{Hacar} A and {Tafalla} M (2011) {Dense core formation by fragmentation of
  velocity-coherent filaments in L1517}. \aap 533:A34

\bibitem[{{Hacar} et~al.(2013){Hacar}, {Tafalla}, {Kauffmann}, and
  {Kov{\'a}cs}}]{Hacar+13}
{Hacar} A, {Tafalla} M, {Kauffmann} J, et~al. (2013) {Cores, filaments, and
  bundles: hierarchical core formation in the L1495/B213 Taurus region}. \aap
  554:A55

\bibitem[{{Hacar} et~al.(2016{\natexlab{a}}){Hacar}, {Alves}, {Burkert}, and
  {Goldsmith}}]{Hacar+16}
{Hacar} A, {Alves} J, {Burkert} A, et~al. (2016{\natexlab{a}}) {Opacity
  broadening and interpretation of suprathermal CO linewidths: Macroscopic
  turbulence and tangled molecular clouds}. \aap 591:A104

\bibitem[{{Hacar} et~al.(2016{\natexlab{b}}){Hacar}, {Kainulainen}, {Tafalla},
  {Beuther}, and {Alves}}]{Hacar+16b}
{Hacar} A, {Kainulainen} J, {Tafalla} M, et~al. (2016{\natexlab{b}}) {The Musca
  cloud: A 6 pc-long velocity-coherent, sonic filament}. \aap 587:A97

\bibitem[{{Hacar} et~al.(2017){Hacar}, {Tafalla}, and {Alves}}]{Hacar+17}
{Hacar} A, {Tafalla} M, and {Alves} J (2017) {Fibers in the NGC 1333
  proto-cluster}. \aap 606:A123

\bibitem[{{Hacar} et~al.(2018){Hacar}, {Tafalla}, {Forbrich}, {Alves},
  {Meingast}, {Grossschedl}, and {Teixeira}}]{Hacar+18}
{Hacar} A, {Tafalla} M, {Forbrich} J, et~al. (2018) {An ALMA study of the Orion
  Integral Filament. I. Evidence for narrow fibers in a massive cloud}. \aap
  610:A77

\bibitem[{{Hartmann}(2002)}]{Hartmann02}
{Hartmann} L (2002) {Flows, Fragmentation, and Star Formation. I. Low-Mass
  Stars in Taurus}. \apj 578:914--924

\bibitem[{{Hartmann} et~al.(2001){Hartmann}, {Ballesteros-Paredes}, and
  {Bergin}}]{Hartmann+01}
{Hartmann} L, {Ballesteros-Paredes} J, and {Bergin} EA (2001) {Rapid Formation
  of Molecular Clouds and Stars in the Solar Neighborhood}. \apj
  562(2):852--868

\bibitem[{{Hartmann} et~al.(2012){Hartmann}, {Ballesteros-Paredes}, and
  {Heitsch}}]{Hartmann+12}
{Hartmann} L, {Ballesteros-Paredes} J, and {Heitsch} F (2012) {Rapid star
  formation and global gravitational collapse}. \mnras 420(2):1457--1461

\bibitem[{{Hausman}(1982)}]{Hausman82}
{Hausman} MA (1982) {Theoretical models of the mass spectrum of interstellar
  clouds}. \apj 261:532--542

\bibitem[{{Heiderman} et~al.(2010){Heiderman}, {Evans}, {Allen}, {Huard}, and
  {Heyer}}]{Heiderman+10}
{Heiderman} A, {Evans} NJ II, {Allen} LE, et~al. (2010) {The Star Formation
  Rate and Gas Surface Density Relation in the Milky Way: Implications for
  Extragalactic Studies}. \apj 723:1019--1037

\bibitem[{{Heiles}(1979)}]{Heiles79}
{Heiles} C (1979) {H I shells and supershells}. \apj 229:533--537

\bibitem[{{Heiles}(1984)}]{Heiles84}
{Heiles} C (1984) {HI shells, supershells, shell-like objects, and ``worms''.}
  \apjs 55:585--595

\bibitem[{{Heiles} and {Troland}(2005)}]{HeilesTroland04}
{Heiles} C and {Troland} TH (2005) {The Millennium Arecibo 21 Centimeter
  Absorption-Line Survey. IV. Statistics of Magnetic Field, Column Density, and
  Turbulence}. \apj 624(2):773--793

\bibitem[{{Heiles} et~al.(1993){Heiles}, {Goodman}, {McKee}, and
  {Zweibel}}]{Heiles+93}
{Heiles} C, {Goodman} AA, {McKee} CF, et~al. (1993) {Magnetic Fields in
  Star-Forming Regions - Observations}. In: {Levy} EH and {Lunine} JI (eds)
  Protostars and Planets III, p 279

\bibitem[{{Heithausen} et~al.(1998){Heithausen}, {Bensch}, {Stutzki},
  {Falgarone}, and {Panis}}]{Heithausen+98}
{Heithausen} A, {Bensch} F, {Stutzki} J, et~al. (1998) {The IRAM key project:
  small-scale structure of pre-star forming regions. Combined mass spectra and
  scaling laws}. \aap 331:L65--L68

\bibitem[{{Heitsch} et~al.(2001){Heitsch}, {Zweibel}, {Mac Low}, {Li}, and
  {Norman}}]{Heitsch+01}
{Heitsch} F, {Zweibel} EG, {Mac Low} MM, et~al. (2001) {Magnetic Field
  Diagnostics Based on Far-Infrared Polarimetry: Tests Using Numerical
  Simulations}. \apj 561(2):800--814

\bibitem[{{Heitsch} et~al.(2006){Heitsch}, {Slyz}, {Devriendt}, {Hartmann}, and
  {Burkert}}]{heitsch2006}
{Heitsch} F, {Slyz} AD, {Devriendt} JEG, et~al. (2006) {The Birth of Molecular
  Clouds: Formation of Atomic Precursors in Colliding Flows}. \apj
  648:1052--1065

\bibitem[{{Heitsch} et~al.(2008){Heitsch}, {Hartmann}, and
  {Burkert}}]{Heitsch+08}
{Heitsch} F, {Hartmann} LW, and {Burkert} A (2008) {Fragmentation of Shocked
  Flows: Gravity, Turbulence, and Cooling}. \apj 683(2):786--795

\bibitem[{{Heitsch} et~al.(2009{\natexlab{a}}){Heitsch}, {Ballesteros-Paredes},
  and {Hartmann}}]{Heitsch+09a}
{Heitsch} F, {Ballesteros-Paredes} J, and {Hartmann} L (2009{\natexlab{a}})
  {Gravitational Collapse and Filament Formation: Comparison with the Pipe
  Nebula}. \apj 704(2):1735--1742

\bibitem[{{Heitsch} et~al.(2009{\natexlab{b}}){Heitsch}, {Stone}, and
  {Hartmann}}]{Heitsch+09b}
{Heitsch} F, {Stone} JM, and {Hartmann} LW (2009{\natexlab{b}}) {Effects of
  Magnetic Field Strength and Orientation on Molecular Cloud Formation}. \apj
  695(1):248--258

\bibitem[{{Hennebelle} and {Andr{\'e}}(2013)}]{HennebelleAndre13}
{Hennebelle} P and {Andr{\'e}} P (2013) {Ion-neutral friction and
  accretion-driven turbulence in self-gravitating filaments}. \aap 560:A68

\bibitem[{{Hennebelle} and {Falgarone}(2012)}]{HennebelleFalgarone12}
{Hennebelle} P and {Falgarone} E (2012) {Turbulent molecular clouds}. \aapr
  20:55

\bibitem[{{Hennebelle} and {Inutsuka}(2019)}]{HennebelleInutsuka19}
{Hennebelle} P and {Inutsuka} Si (2019) {The role of magnetic field in
  molecular cloud formation and evolution}. Frontiers in Astronomy and Space
  Sciences 6:5

\bibitem[{{Hennebelle} and {P{\'e}rault}(1999)}]{HennebellePerault99}
{Hennebelle} P and {P{\'e}rault} M (1999) {Dynamical condensation in a
  thermally bistable flow. Application to interstellar cirrus}. \aap
  351:309--322

\bibitem[{{Hennebelle} et~al.(2008){Hennebelle}, {Banerjee},
  {V{\'a}zquez-Semadeni}, {Klessen}, and {Audit}}]{hetal2008}
{Hennebelle} P, {Banerjee} R, {V{\'a}zquez-Semadeni} E, et~al. (2008) {From the
  warm magnetized atomic medium to molecular clouds}. \aap 486:L43--L46

\bibitem[{{Hennebelle} et~al.(2019){Hennebelle}, {Lee}, and
  {Chabrier}}]{Hennebelle+19}
{Hennebelle} P, {Lee} YN, and {Chabrier} G (2019) {How First Hydrostatic Cores,
  Tidal Forces, and Gravoturbulent Fluctuations Set the Characteristic Mass of
  Stars}. \apj 883(2):140

\bibitem[{{Hennemann} et~al.(2012){Hennemann}, {Motte}, {Schneider}, {Didelon},
  {Hill}, {Arzoumanian}, {Bontemps}, {Csengeri}, {Andr{\'e}}, {Konyves},
  {Louvet}, {Marston}, {Men'shchikov}, {Minier}, {Nguyen Luong}, {Palmeirim},
  {Peretto}, {Sauvage}, {Zavagno}, {Anderson}, {Bernard}, {Di Francesco},
  {Elia}, {Li}, {Martin}, {Molinari}, {Pezzuto}, {Russeil}, {Rygl}, {Schisano},
  {Spinoglio}, {Sousbie}, {Ward-Thompson}, and {White}}]{Hennemann+12}
{Hennemann} M, {Motte} F, {Schneider} N, et~al. (2012) {The spine of the swan:
  a Herschel study of the DR21 ridge and filaments in Cygnus X}. \aap 543:L3

\bibitem[{{Henning} et~al.(2010){Henning}, {Linz}, {Krause}, {Ragan},
  {Beuther}, {Launhardt}, {Nielbock}, and {Vasyunina}}]{Henning+10}
{Henning} T, {Linz} H, {Krause} O, et~al. (2010) {The seeds of star formation
  in the filamentary infrared-dark cloud G011.11-0.12}. \aap 518:L95

\bibitem[{{Henshaw} et~al.(2014){Henshaw}, {Caselli}, {Fontani},
  {Jim{\'e}nez-Serra}, and {Tan}}]{Henshaw+14}
{Henshaw} JD, {Caselli} P, {Fontani} F, et~al. (2014) {The dynamical properties
  of dense filaments in the infrared dark cloud G035.39-00.33}. \mnras
  440:2860--2881

\bibitem[{{Henshaw} et~al.(2016){Henshaw}, {Longmore}, {Kruijssen}, {Davies},
  {Bally}, {Barnes}, {Battersby}, {Burton}, {Cunningham}, {Dale}, {Ginsburg},
  {Immer}, {Jones}, {Kendrew}, {Mills}, {Molinari}, {Moore}, {Ott}, {Pillai},
  {Rathborne}, {Schilke}, {Schmiedeke}, {Testi}, {Walker}, {Walsh}, and
  {Zhang}}]{Henshaw+16}
{Henshaw} JD, {Longmore} SN, {Kruijssen} JMD, et~al. (2016) {Molecular gas
  kinematics within the central 250 pc of the Milky Way}. \mnras 457:2675--2702

\bibitem[{{Herbig}(1978)}]{Herbig78}
{Herbig} GH (1978) {Can Post-T Tauri Stars Be Found?} In: {Mirzoyan} LV (ed)
  Problems of Physics and Evolution of the Universe., Publishing House of the
  Armenian Academy of Sciences, p 171

\bibitem[{{Hernandez} and {Tan}(2015)}]{HernandezTan15}
{Hernandez} AK and {Tan} JC (2015) {The Giant Molecular Cloud Environments of
  Infrared Dark Clouds}. \apj 809(2):154

\bibitem[{{Heyer} and {Dame}(2015)}]{HeyerDame15}
{Heyer} M and {Dame} TM (2015) {Molecular Clouds in the Milky Way}. \araa
  53:583--629

\bibitem[{{Heyer} et~al.(2009){Heyer}, {Krawczyk}, {Duval}, and
  {Jackson}}]{Heyer+09}
{Heyer} M, {Krawczyk} C, {Duval} J, et~al. (2009) {Re-Examining Larson's
  Scaling Relationships in Galactic Molecular Clouds}. \apj 699(2):1092--1103

\bibitem[{{Heyer} et~al.(2016){Heyer}, {Goldsmith}, {Y{\i}ld{\i}z}, {Snell},
  {Falgarone}, and {Pineda}}]{Heyer+16}
{Heyer} M, {Goldsmith} PF, {Y{\i}ld{\i}z} UA, et~al. (2016) {Striations in the
  Taurus molecular cloud: Kelvin-Helmholtz instability or MHD waves?} \mnras
  461:3918--3926

\bibitem[{{Heyer} and {Brunt}(2004)}]{HeyerBrunt04}
{Heyer} MH and {Brunt} CM (2004) {The Universality of Turbulence in Galactic
  Molecular Clouds}. \apjl 615(1):L45--L48

\bibitem[{{Heyer} and {Terebey}(1998)}]{HeyerTerebey98}
{Heyer} MH and {Terebey} S (1998) {The Anatomy of the Perseus Spiral Arm:
  $^{12}$CO and IRAS Imaging Observations of the W3-W4-W5 Cloud Complex}. \apj
  502(1):265--277

\bibitem[{{Heyer} et~al.(1998){Heyer}, {Brunt}, {Snell}, {Howe}, {Schloerb},
  and {Carpenter}}]{heyer98}
{Heyer} MH, {Brunt} C, {Snell} RL, et~al. (1998) {The Five College Radio
  Astronomy Observatory CO Survey of the Outer Galaxy}. \apjs 115:241

\bibitem[{{Heyer} et~al.(2001){Heyer}, {Carpenter}, and {Snell}}]{Heyer+01}
{Heyer} MH, {Carpenter} JM, and {Snell} RL (2001) {The Equilibrium State of
  Molecular Regions in the Outer Galaxy}. \apj 551(2):852--866

\bibitem[{{Hill} et~al.(2011){Hill}, {Motte}, {Didelon}, {Bontemps}, {Minier},
  {Hennemann}, {Schneider}, {Andr{\'e}}, {Men'shchikov}, {Anderson},
  {Arzoumanian}, {Bernard}, {di Francesco}, {Elia}, {Giannini}, {Griffin},
  {K{\"o}nyves}, {Kirk}, {Marston}, {Martin}, {Molinari}, {Nguyen Luong},
  {Peretto}, {Pezzuto}, {Roussel}, {Sauvage}, {Sousbie}, {Testi},
  {Ward-Thompson}, {White}, {Wilson}, and {Zavagno}}]{Hill+11}
{Hill} T, {Motte} F, {Didelon} P, et~al. (2011) {Filaments and ridges in Vela C
  revealed by Herschel: from low-mass to high-mass star-forming sites}. \aap
  533:A94

\bibitem[{{Hiltner}(1949)}]{Hiltner49}
{Hiltner} WA (1949) {Polarization of Light from Distant Stars by Interstellar
  Medium}. Science 109(2825):165

\bibitem[{{Hily-Blant} et~al.(2008){Hily-Blant}, {Falgarone}, and
  {Pety}}]{Hily-Blant+08}
{Hily-Blant} P, {Falgarone} E, and {Pety} J (2008) {Dissipative structures of
  diffuse molecular gas. III. Small-scale intermittency of intense
  velocity-shears}. \aap 481(2):367--380

\bibitem[{{Hopkins} et~al.(2018){Hopkins}, {Wetzel}, {Kere{\v{s}}},
  {Faucher-Gigu{\`e}re}, {Quataert}, {Boylan-Kolchin}, {Murray}, {Hayward},
  {Garrison-Kimmel}, {Hummels}, {Feldmann}, {Torrey}, {Ma},
  {Angl{\'e}s-Alc{\'a}zar}, {Su}, {Orr}, {Schmitz}, {Escala}, {Sanderson},
  {Grudi{\'c}}, {Hafen}, {Kim}, {Fitts}, {Bullock}, {Wheeler}, {Chan},
  {Elbert}, and {Narayanan}}]{Hopkins+18}
{Hopkins} PF, {Wetzel} A, {Kere{\v{s}}} D, et~al. (2018) {FIRE-2 simulations:
  physics versus numerics in galaxy formation}. \mnras 480(1):800--863

\bibitem[{{Hughes} et~al.(2013){Hughes}, {Meidt}, {Colombo}, {Schinnerer},
  {Pety}, {Leroy}, {Dobbs}, {Garc{\'\i}a-Burillo}, {Thompson}, and
  {Dumas}}]{Hughes+13}
{Hughes} A, {Meidt} SE, {Colombo} D, et~al. (2013) {A Comparative Study of
  Giant Molecular Clouds in M51, M33, and the Large Magellanic Cloud}. \apj
  779(1):46

\bibitem[{{Hygate} et~al.(2019){Hygate}, {Kruijssen}, {Chevance}, {Walter},
  {Schruba}, {Kim}, {Haydon}, and {Longmore}}]{Hygate+19}
{Hygate} APS, {Kruijssen} JMD, {Chevance} M, et~al. (2019) {The cloud-scale
  physics of star-formation and feedback in M33}. \mnras~submitted

\bibitem[{{Ib{\'a}{\~n}ez-Mej{\'\i}a} et~al.(2016){Ib{\'a}{\~n}ez-Mej{\'\i}a},
  {Mac Low}, {Klessen}, and {Baczynski}}]{IbanezMejia+16}
{Ib{\'a}{\~n}ez-Mej{\'\i}a} JC, {Mac Low} MM, {Klessen} RS, et~al. (2016)
  {Gravitational Contraction versus Supernova Driving and the Origin of the
  Velocity Dispersion-Size Relation in Molecular Clouds}. \apj 824(1):41

\bibitem[{{Iffrig} and {Hennebelle}(2017)}]{iffrig2017}
{Iffrig} O and {Hennebelle} P (2017) {Structure distribution and turbulence in
  self-consistently supernova-driven ISM of multiphase magnetized galactic
  discs}. \aap 604:A70

\bibitem[{{Imara} and {Faesi}(2019)}]{ImaraFaesi19}
{Imara} N and {Faesi} CM (2019) {ALMA Observations of Giant Molecular Clouds in
  the Starburst Dwarf Galaxy Henize 2-10}. \apj 876(2):141

\bibitem[{{Inoue} and {Inutsuka}(2008)}]{inoue2008}
{Inoue} T and {Inutsuka} Si (2008) {Two-Fluid Magnetohydrodynamic Simulations
  of Converging H I Flows in the Interstellar Medium. I. Methodology and Basic
  Results}. \apj 687:303--310

\bibitem[{{Inoue} and {Inutsuka}(2009)}]{inoue2009}
{Inoue} T and {Inutsuka} Si (2009) {Two-Fluid Magnetohydrodynamics Simulations
  of Converging H I Flows in the Interstellar Medium. II. Are Molecular Clouds
  Generated Directly from a Warm Neutral Medium?} \apj 704:161--169

\bibitem[{{Inoue} and {Inutsuka}(2012)}]{inoue2012}
{Inoue} T and {Inutsuka} Si (2012) {Formation of Turbulent and Magnetized
  Molecular Clouds via Accretion Flows of H I Clouds}. \apj 759:35

\bibitem[{{Inutsuka}(2001)}]{Inutsuka01}
{Inutsuka} Si (2001) {The Mass Function of Molecular Cloud Cores}. \apjl
  559:L149--L152

\bibitem[{{Inutsuka} and {Miyama}(1992)}]{InutsukaMiyama92}
{Inutsuka} Si and {Miyama} SM (1992) {Self-similar solutions and the stability
  of collapsing isothermal filaments}. \apj 388:392--399

\bibitem[{{Inutsuka} and {Miyama}(1997)}]{InutsukaMiyama97}
{Inutsuka} Si and {Miyama} SM (1997) {A Production Mechanism for Clusters of
  Dense Cores}. \apj 480:681

\bibitem[{{Inutsuka} et~al.(2015){Inutsuka}, {Inoue}, {Iwasaki}, and
  {Hosokawa}}]{Inutsuka+15}
{Inutsuka} Si, {Inoue} T, {Iwasaki} K, et~al. (2015) {The formation and
  destruction of molecular clouds and galactic star formation. An origin for
  the cloud mass function and star formation efficiency}. \aap 580:A49

\bibitem[{{Iwasaki} et~al.(2018){Iwasaki}, {Tomida}, {Inoue}, and
  {Inutsuka}}]{iwasaki2018}
{Iwasaki} K, {Tomida} K, {Inoue} T, et~al. (2018) {The Early Stage of Molecular
  Cloud Formation by Compression of Two-phase Atomic Gases}. arXiv e-prints

\bibitem[{{Jeffreson} and {Kruijssen}(2018)}]{JeffresonKruijssen18}
{Jeffreson} SMR and {Kruijssen} JMD (2018) {A general theory for the lifetimes
  of giant molecular clouds under the influence of galactic dynamics}. \mnras
  476:3688--3715

\bibitem[{{Jeffreson} et~al.(2018){Jeffreson}, {Kruijssen}, {Krumholz}, and
  {Longmore}}]{Jeffreson+18}
{Jeffreson} SMR, {Kruijssen} JMD, {Krumholz} MR, et~al. (2018) {On the physical
  mechanisms governing the cloud lifecycle in the Central Molecular Zone of the
  Milky Way}. \mnras 478:3380--3385

\bibitem[{{Johnstone} et~al.(2000){Johnstone}, {Wilson}, {Moriarty-Schieven},
  {Giannakopoulou-Creighton}, and {Gregersen}}]{Johnstone+00}
{Johnstone} D, {Wilson} CD, {Moriarty-Schieven} G, et~al. (2000) {Large-Area
  Mapping at 850 Microns. I. Optimum Image Reconstruction from Chop
  Measurements}. \apjs 131(2):505--518

\bibitem[{{Johnstone} et~al.(2001){Johnstone}, {Fich}, {Mitchell}, and
  {Moriarty-Schieven}}]{Johnstone+01}
{Johnstone} D, {Fich} M, {Mitchell} GF, et~al. (2001) {Large Area Mapping at
  850 Microns. III. Analysis of the Clump Distribution in the Orion B Molecular
  Cloud}. \apj 559(1):307--317

\bibitem[{{Johnstone} et~al.(2004){Johnstone}, {Di Francesco}, and
  {Kirk}}]{Johnstone+04}
{Johnstone} D, {Di Francesco} J, and {Kirk} H (2004) {An Extinction Threshold
  for Protostellar Cores in Ophiuchus}. \apjl 611:L45--L48

\bibitem[{{Jura}(1975)}]{Jura75}
{Jura} M (1975) {Interstellar clouds containing optically thin H$_{2}$.} \apj
  197:575--580

\bibitem[{{Juvela} et~al.(2012){Juvela}, {Ristorcelli}, {Pagani}, {Doi},
  {Pelkonen}, {Marshall}, {Bernard}, {Falgarone}, {Malinen}, {Marton},
  {McGehee}, {Montier}, {Motte}, {Paladini}, {T{\'o}th}, {Ysard}, {Zahorecz},
  and {Zavagno}}]{Juvela+12}
{Juvela} M, {Ristorcelli} I, {Pagani} L, et~al. (2012) {Galactic cold cores.
  III. General cloud properties}. \aap 541:A12

\bibitem[{{Kainulainen} et~al.(2009){Kainulainen}, {Beuther}, {Henning}, and
  {Plume}}]{Kainulainen+09}
{Kainulainen} J, {Beuther} H, {Henning} T, et~al. (2009) {Probing the evolution
  of molecular cloud structure. From quiescence to birth}. \aap 508:L35--L38

\bibitem[{{Kainulainen} et~al.(2011){Kainulainen}, {Beuther}, {Banerjee},
  {Federrath}, and {Henning}}]{Kainulainen+11}
{Kainulainen} J, {Beuther} H, {Banerjee} R, et~al. (2011) {Probing the
  evolution of molecular cloud structure. II. From chaos to confinement}. \aap
  530:A64

\bibitem[{{Kalberla} and {Kerp}(2009)}]{kalberla2009a}
{Kalberla} PMW and {Kerp} J (2009) {The Hi Distribution of the Milky Way}.
  \araa 47(1):27--61

\bibitem[{{Kauffmann} et~al.(2010{\natexlab{a}}){Kauffmann}, {Pillai},
  {Shetty}, {Myers}, and {Goodman}}]{Kauffmann+10a}
{Kauffmann} J, {Pillai} T, {Shetty} R, et~al. (2010{\natexlab{a}}) {The
  Mass-Size Relation from Clouds to Cores. I. A New Probe of Structure in
  Molecular Clouds}. \apj 712(2):1137--1146

\bibitem[{{Kauffmann} et~al.(2010{\natexlab{b}}){Kauffmann}, {Pillai},
  {Shetty}, {Myers}, and {Goodman}}]{Kauffmann+10b}
{Kauffmann} J, {Pillai} T, {Shetty} R, et~al. (2010{\natexlab{b}}) {The
  Mass-size Relation from Clouds to Cores. II. Solar Neighborhood Clouds}. \apj
  716(1):433--445

\bibitem[{{Kauffmann} et~al.(2013){Kauffmann}, {Pillai}, and
  {Goldsmith}}]{Kauffmann+13}
{Kauffmann} J, {Pillai} T, and {Goldsmith} PF (2013) {Low Virial Parameters in
  Molecular Clouds: Implications for High-mass Star Formation and Magnetic
  Fields}. \apj 779(2):185

\bibitem[{{Kauffmann} et~al.(2017){Kauffmann}, {Pillai}, {Zhang}, {Menten},
  {Goldsmith}, {Lu}, and {Guzm{\'a}n}}]{Kauffmann+17}
{Kauffmann} J, {Pillai} T, {Zhang} Q, et~al. (2017) {The Galactic Center
  Molecular Cloud Survey. I. A steep linewidth-size relation and suppression of
  star formation}. \aap 603:A89

\bibitem[{Kawamura et~al.(2009)Kawamura, Mizuno, Minamidani, {D.
  Fillipovi{\'{c}}}, Staveley-Smith, Kim, Mizuno, Onishi, Mizuno, and
  Fukui}]{Kawamura+09}
Kawamura A, Mizuno Y, Minamidani T, et~al. (2009) {the Second Survey of the
  Molecular Clouds in the Large Magellanic Cloud By Nanten. Ii. Star
  Formation}. \apjs 184(1):1--17

\bibitem[{{Kennicutt}(1998)}]{Kennicutt98}
{Kennicutt} J Robert~C (1998) {The Global Schmidt Law in Star-forming
  Galaxies}. The Astrophysical Journal 498(2):541--552

\bibitem[{{Kennicutt} and {Evans}(2012)}]{kennicutt12}
{Kennicutt} RC and {Evans} NJ (2012) {Star Formation in the Milky Way and
  Nearby Galaxies}. \araa 50:531--608

\bibitem[{{Kerr} et~al.(2019){Kerr}, {Kirk}, {Di Francesco}, {Keown}, {Chen},
  {Rosolowsky}, {Offner}, {Friesen}, {Pineda}, {Shirley}, {Redaelli},
  {Caselli}, {Punanova}, {Seo}, {Alves}, {Chac{\'o}n-Tanarro}, and
  {Chen}}]{Kerr+19}
{Kerr} R, {Kirk} H, {Di Francesco} J, et~al. (2019) {The Green Bank Ammonia
  Survey: A Virial Analysis of Gould Belt Clouds in Data Release 1}. \apj
  874(2):147

\bibitem[{{Keto} and {Myers}(1986)}]{Keto_Myers86}
{Keto} ER and {Myers} PC (1986) {CO Observations of Southern High-Latitude
  Clouds}. \apj 304:466

\bibitem[{{Khoperskov} et~al.(2016){Khoperskov}, {Vasiliev}, {Ladeyschikov},
  {Sobolev}, and {Khoperskov}}]{Khoperskov+16}
{Khoperskov} SA, {Vasiliev} EO, {Ladeyschikov} DA, et~al. (2016) {Giant
  molecular cloud scaling relations: the role of the cloud definition}. \mnras
  455(2):1782--1795

\bibitem[{{Kim} et~al.(2018){Kim}, {Kim}, and {Ostriker}}]{Kim+18}
{Kim} JG, {Kim} WT, and {Ostriker} EC (2018) {Modeling UV Radiation Feedback
  from Massive Stars. II. Dispersal of Star-forming Giant Molecular Clouds by
  Photoionization and Radiation Pressure}. \apj 859(1):68

\bibitem[{{Kim} and {Ostriker}(2001)}]{KimOstriker01}
{Kim} WT and {Ostriker} EC (2001) {Amplification, Saturation, and Q Thresholds
  for Runaway: Growth of Self-Gravitating Structures in Models of Magnetized
  Galactic Gas Disks}. \apj 559(1):70--95

\bibitem[{{Kim} and {Ostriker}(2002)}]{KimOstriker02}
{Kim} WT and {Ostriker} EC (2002) {Formation and Fragmentation of Gaseous Spurs
  in Spiral Galaxies}. \apj 570(1):132--151

\bibitem[{{Kim} and {Ostriker}(2007)}]{KimOstriker07}
{Kim} WT and {Ostriker} EC (2007) {Gravitational Runaway and Turbulence Driving
  in Star-Gas Galactic Disks}. \apj 660(2):1232--1245

\bibitem[{{Kim} et~al.(2002){Kim}, {Ostriker}, and {Stone}}]{Kim+02}
{Kim} WT, {Ostriker} EC, and {Stone} JM (2002) {Three-dimensional Simulations
  of Parker, Magneto-Jeans, and Swing Instabilities in Shearing Galactic Gas
  Disks}. \apj 581(2):1080--1100

\bibitem[{{Kim} et~al.(2003){Kim}, {Ostriker}, and {Stone}}]{Kim+03}
{Kim} WT, {Ostriker} EC, and {Stone} JM (2003) {Magnetorotationally Driven
  Galactic Turbulence and the Formation of Giant Molecular Clouds}. \apj
  599(2):1157--1172

\bibitem[{{Kirk} et~al.(2006){Kirk}, {Johnstone}, and {Di Francesco}}]{Kirk+06}
{Kirk} H, {Johnstone} D, and {Di Francesco} J (2006) {The Large- and
  Small-Scale Structures of Dust in the Star-forming Perseus Molecular Cloud}.
  \apj 646:1009--1023

\bibitem[{{Kirk} et~al.(2017){Kirk}, {Friesen}, {Pineda}, {Rosolowsky},
  {Offner}, {Matzner}, {Myers}, {Di Francesco}, {Caselli}, {Alves},
  {Chac{\'o}n-Tanarro}, {Chen}, {Chun-Yuan Chen}, {Keown}, {Punanova}, {Seo},
  {Shirley}, {Ginsburg}, {Hall}, {Singh}, {Arce}, {Goodman}, {Martin}, and
  {Redaelli}}]{Kirk+17}
{Kirk} H, {Friesen} RK, {Pineda} JE, et~al. (2017) {The Green Bank Ammonia
  Survey: Dense Cores under Pressure in Orion A}. \apj 846(2):144

\bibitem[{{Kirk, H.} et~al.(2013){Kirk, H.}, {Myers}, {Bourke}, {Gutermuth},
  {Hedden}, and {Wilson}}]{Kirk+13}
{Kirk, H}, {Myers} PC, {Bourke} TL, et~al. (2013) {Filamentary Accretion Flows
  in the Embedded Serpens South Protocluster}. \apj 766:115

\bibitem[{{Klessen} and {Glover}(2016)}]{KlessenGlover16}
{Klessen} RS and {Glover} SCO (2016) {Physical Processes in the Interstellar
  Medium}. Saas-Fee Advanced Course 43:85

\bibitem[{{Klessen} and {Hennebelle}(2010)}]{klessen2010}
{Klessen} RS and {Hennebelle} P (2010) {Accretion-driven turbulence as
  universal process: galaxies, molecular clouds, and protostellar disks}. \aap
  520:A17

\bibitem[{{Klessen} et~al.(2005){Klessen}, {Ballesteros-Paredes},
  {V{\'a}zquez-Semadeni}, and {Dur{\'a}n-Rojas}}]{Klessen+05}
{Klessen} RS, {Ballesteros-Paredes} J, {V{\'a}zquez-Semadeni} E, et~al. (2005)
  {Quiescent and Coherent Cores from Gravoturbulent Fragmentation}. \apj
  620(2):786--794

\bibitem[{{Koch} and {Rosolowsky}(2015)}]{KochRosolowsky15}
{Koch} EW and {Rosolowsky} EW (2015) {Filament identification through
  mathematical morphology}. \mnras 452:3435--3450

\bibitem[{Koda et~al.(2009)Koda, Scoville, Sawada, {La Vigne}, Vogel, Potts,
  Carpenter, Corder, Wright, White, Zauderer, Patience, Sargent, Bock, Hawkins,
  Hodges, Kemball, Lamb, Plambeck, Pound, Scott, Teuben, and Woody}]{Koda+09}
Koda J, Scoville N, Sawada T, et~al. (2009) {DYNAMICALLY DRIVEN EVOLUTION OF
  THE INTERSTELLAR MEDIUM IN M51}. \apj 700(2):L132--L136

\bibitem[{{Koda} et~al.(2016){Koda}, {Scoville}, and {Heyer}}]{Koda+16}
{Koda} J, {Scoville} N, and {Heyer} M (2016) {Evolution of Molecular and Atomic
  Gas Phases in the Milky Way}. \apj 823(2):76

\bibitem[{{Kolmogorov}(1941)}]{Kolmogorov41}
{Kolmogorov} A (1941) {The Local Structure of Turbulence in Incompressible
  Viscous Fluid for Very Large Reynolds' Numbers}. Akademiia Nauk SSSR Doklady
  30:301--305

\bibitem[{{K{\"o}nyves} et~al.(2015){K{\"o}nyves}, {Andr{\'e}}, {Men'shchikov},
  {Palmeirim}, {Arzoumanian}, {Schneider}, {Roy}, {Didelon}, {Maury}, and
  {Shimajiri}}]{Konyves+15}
{K{\"o}nyves} V, {Andr{\'e}} P, {Men'shchikov} A, et~al. (2015) {A census of
  dense cores in the Aquila cloud complex: SPIRE/PACS observations from the
  Herschel Gould Belt survey}. \aap 584:A91

\bibitem[{{K{\"o}nyves} et~al.(2020){K{\"o}nyves}, {Andr{\'e}}, {Arzoumanian},
  {Schneider}, {Men'shchikov}, {Bontemps}, {Ladjelate}, {Didelon}, {Pezzuto},
  {Benedettini}, {Bracco}, {Di Francesco}, {Goodwin}, {Rygl}, {Shimajiri},
  {Spinoglio}, {Ward-Thompson}, and {White}}]{Konyves+20}
{K{\"o}nyves} V, {Andr{\'e}} P, {Arzoumanian} D, et~al. (2020) {Properties of
  the dense core population in Orion B as seen by the Herschel Gould Belt
  survey}. \aap 635:A34

\bibitem[{{K{\"o}rtgen} and {Banerjee}(2015)}]{kortgen2015}
{K{\"o}rtgen} B and {Banerjee} R (2015) {Impact of magnetic fields on molecular
  cloud formation and evolution}. \mnras 451:3340--3353

\bibitem[{{Koyama} and {Inutsuka}(2000)}]{koyama2000}
{Koyama} H and {Inutsuka} Si (2000) {Molecular Cloud Formation in
  Shock-compressed Layers}. \apj 532:980--993

\bibitem[{{Koyama} and {Inutsuka}(2002)}]{koyama2002}
{Koyama} H and {Inutsuka} Si (2002) {An Origin of Supersonic Motions in
  Interstellar Clouds}. \apjl 564:L97--L100

\bibitem[{{Kramer} et~al.(1998){Kramer}, {Stutzki}, {Rohrig}, and
  {Corneliussen}}]{Kramer+98}
{Kramer} C, {Stutzki} J, {Rohrig} R, et~al. (1998) {Clump mass spectra of
  molecular clouds}. \aap 329:249--264

\bibitem[{{Kreckel} et~al.(2018){Kreckel}, {Faesi}, {Kruijssen}, {Schruba},
  {Groves}, {Leroy}, {Bigiel}, {Blanc}, {Chevance}, {Herrera}, {Hughes},
  {McElroy}, {Pety}, {Querejeta}, {Rosolowsky}, {Schinnerer}, {Sun}, {Usero},
  and {Utomo}}]{Kreckel+18}
{Kreckel} K, {Faesi} C, {Kruijssen} JMD, et~al. (2018) {A 50 pc Scale View of
  Star Formation Efficiency across NGC 628}. \apjl 863:L21

\bibitem[{{Kreckel} et~al.(2019){Kreckel}, {Ho}, {Blanc}, {Groves}, {Santoro},
  {Schinnerer}, {Bigiel}, {Chevance}, {Congiu}, {Emsellem}, {Faesi}, {Glover},
  {Grasha}, {Kruijssen}, {Lang}, {Leroy}, {Meidt}, {McElroy}, {Pety},
  {Rosolowsky}, {Saito}, {Sandstrom}, {Sanchez-Blazquez}, and
  {Schruba}}]{Kreckel+19}
{Kreckel} K, {Ho} IT, {Blanc} GA, et~al. (2019) {Mapping Metallicity Variations
  across Nearby Galaxy Disks}. \apj 887(1):80

\bibitem[{{Kritsuk} et~al.(2011){Kritsuk}, {Norman}, and {Wagner}}]{Kritsuk+11}
{Kritsuk} AG, {Norman} ML, and {Wagner} R (2011) {On the Density Distribution
  in Star-forming Interstellar Clouds}. \apj 727(1):L20

\bibitem[{{Kruijssen}(2012)}]{kruijssen12}
{Kruijssen} JMD (2012) {On the fraction of star formation occurring in bound
  stellar clusters}. \mnras 426:3008--3040

\bibitem[{{Kruijssen} and {Longmore}(2013)}]{kruijssen13}
{Kruijssen} JMD and {Longmore} SN (2013) {Comparing molecular gas across cosmic
  time-scales: the Milky Way as both a typical spiral galaxy and a
  high-redshift galaxy analogue}. \mnras 435:2598--2603

\bibitem[{Kruijssen and Longmore(2014)}]{KruijssenLongmore14}
Kruijssen JMD and Longmore SN (2014) {An uncertainty principle for star
  formation - I. why galactic star formation relations break down below a
  certain spatial scale}. \mnras 439(4):3239--3252

\bibitem[{{Kruijssen} et~al.(2015){Kruijssen}, {Dale}, and
  {Longmore}}]{Kruijssen+15}
{Kruijssen} JMD, {Dale} JE, and {Longmore} SN (2015) {The dynamical evolution
  of molecular clouds near the Galactic Centre - I. Orbital structure and
  evolutionary timeline}. \mnras 447:1059--1079

\bibitem[{{Kruijssen} et~al.(2018){Kruijssen}, {Schruba}, {Hygate}, {Hu},
  {Haydon}, and {Longmore}}]{Kruijssen+18}
{Kruijssen} JMD, {Schruba} A, {Hygate} APS, et~al. (2018) {An uncertainty
  principle for star formation - II. A new method for characterizing the
  cloud-scale physics of star formation and feedback across cosmic history}.
  \mnras 479:1866--1952

\bibitem[{{Kruijssen} et~al.(2019{\natexlab{a}}){Kruijssen}, {Dale},
  {Longmore}, {Walker}, {Henshaw}, {Jeffreson}, {Petkova}, {Ginsburg},
  {Barnes}, {Battersby}, {Immer}, {Jackson}, {Keto}, {Krieger}, {Mills},
  {S{\'a}nchez-Monge}, {Schmiedeke}, {Suri}, and {Zhang}}]{Kruijssen+19b}
{Kruijssen} JMD, {Dale} JE, {Longmore} SN, et~al. (2019{\natexlab{a}}) {The
  dynamical evolution of molecular clouds near the Galactic Centre - II.
  Spatial structure and kinematics of simulated clouds}. \mnras
  484(4):5734--5754

\bibitem[{{Kruijssen} et~al.(2019{\natexlab{b}}){Kruijssen}, {Schruba},
  {Chevance}, {Longmore}, {Hygate}, {Haydon}, {McLeod}, {Dalcanton}, {Tacconi},
  and {van Dishoeck}}]{Kruijssen+19}
{Kruijssen} JMD, {Schruba} A, {Chevance} M, et~al. (2019{\natexlab{b}}) {Fast
  and inefficient star formation due to short-lived molecular clouds and rapid
  feedback}. \nat 569(7757):519--522

\bibitem[{{Krumholz} and {McKee}(2005)}]{krumholz05}
{Krumholz} MR and {McKee} CF (2005) {A General Theory of Turbulence-regulated
  Star Formation, from Spirals to Ultraluminous Infrared Galaxies}. \apj
  630:250--268

\bibitem[{{Krumholz} et~al.(2019){Krumholz}, {McKee}, and {Bland
  -Hawthorn}}]{Krumholz+19}
{Krumholz} MR, {McKee} CF, and {Bland -Hawthorn} J (2019) {Star Clusters Across
  Cosmic Time}. \araa 57:227--303

\bibitem[{{Kuznetsova} et~al.(2015){Kuznetsova}, {Hartmann}, and
  {Ballesteros-Paredes}}]{Kuznetsova+15}
{Kuznetsova} A, {Hartmann} L, and {Ballesteros-Paredes} J (2015) {Signatures of
  Star Cluster Formation by Cold Collapse}. \apj 815(1):27

\bibitem[{{Kuznetsova} et~al.(2018){Kuznetsova}, {Hartmann}, and
  {Ballesteros-Paredes}}]{Kuznetsova+18}
{Kuznetsova} A, {Hartmann} L, and {Ballesteros-Paredes} J (2018) {Kinematics
  and structure of star-forming regions: insights from cold collapse models}.
  \mnras 473(2):2372--2377

\bibitem[{{Kwan}(1979)}]{Kwan79}
{Kwan} J (1979) {The mass spectrum of interstellar clouds.} \apj 229:567--577

\bibitem[{{Kwan} and {Valdes}(1983)}]{KwanValdes83}
{Kwan} J and {Valdes} F (1983) {Spiral gravitational potentials and the mass
  growth of molecular clouds}. \apj 271:604--610

\bibitem[{{Kwan} and {Valdes}(1987)}]{KwanValdes87}
{Kwan} J and {Valdes} F (1987) {The Spatial and Mass Distributions of Molecular
  Clouds and Spiral Structures}. \apj 315:92

\bibitem[{{Lada} et~al.(2008){Lada}, {Muench}, {Rathborne}, {Alves}, and
  {Lombardi}}]{Lada+08}
{Lada} CJ, {Muench} AA, {Rathborne} J, et~al. (2008) {The Nature of the Dense
  Core Population in the Pipe Nebula: Thermal Cores Under Pressure}. \apj
  672(1):410--422

\bibitem[{{Lada} et~al.(2010){Lada}, {Lombardi}, and {Alves}}]{Lada+10}
{Lada} CJ, {Lombardi} M, and {Alves} JF (2010) {On the Star Formation Rates in
  Molecular Clouds}. ApJ 724:687--693

\bibitem[{{Lada} et~al.(2013){Lada}, {Lombardi}, {Roman-Zuniga}, {Forbrich},
  and {Alves}}]{lada13}
{Lada} CJ, {Lombardi} M, {Roman-Zuniga} C, et~al. (2013) {Schmidt's Conjecture
  and Star Formation in Molecular Clouds}. \apj 778(2):133

\bibitem[{{Ladjelate} et~al.(2020){Ladjelate}, {Andr{\'e}}, {K{\"o}nyves},
  {Ward-Thompson}, {Men'shchikov}, {Bracco}, {Palmeirim}, {Roy}, {Shimajiri},
  {Kirk}, {Arzoumanian}, {Benedettini}, {Di Francesco}, {Fiorellino},
  {Schneider}, and {Pezzuto}}]{Ladjelate+20}
{Ladjelate} B, {Andr{\'e}} P, {K{\"o}nyves} V, et~al. (2020) {The Herschel view
  of the dense core population in the Ophiuchus molecular cloud}. \aap, in
  press (arXiv:200111036)

\bibitem[{{Larson}(1969)}]{Larson69}
{Larson} RB (1969) {Numerical calculations of the dynamics of collapsing
  proto-star}. \mnras 145:271

\bibitem[{Larson(1981)}]{Larson81}
Larson RB (1981) {Turbulence and star formation in molecular clouds}. Mon Not R
  Astron Soc 194:809--826

\bibitem[{{Larson}(1985)}]{Larson85}
{Larson} RB (1985) {Cloud fragmentation and stellar masses.} \mnras
  214:379--398

\bibitem[{{Larson}(2005)}]{Larson05}
{Larson} RB (2005) {Thermal physics, cloud geometry and the stellar initial
  mass function}. \mnras 359:211--222

\bibitem[{{Lee} and {Hennebelle}(2018)}]{LeeHennebelle18}
{Lee} YN and {Hennebelle} P (2018) {Stellar mass spectrum within massive
  collapsing clumps. II. Thermodynamics and tidal forces of the first Larson
  core. A robust mechanism for the peak of the IMF}. \aap 611:A89

\bibitem[{{Lee} et~al.(2017){Lee}, {Hennebelle}, and {Chabrier}}]{Lee+17}
{Lee} YN, {Hennebelle} P, and {Chabrier} G (2017) {Analytical Core Mass
  Function (CMF) from Filaments: Under Which Circumstances Can Filament
  Fragmentation Reproduce the CMF?} \apj 847(2):114

\bibitem[{{Leisawitz} et~al.(1989){Leisawitz}, {Bash}, and
  {Thaddeus}}]{Leisawitz+89}
{Leisawitz} D, {Bash} FN, and {Thaddeus} P (1989) {A CO Survey of Regions
  around 34 Open Clusters}. \apjs 70:731

\bibitem[{Leroy et~al.(2013)Leroy, Walter, Sandstrom, Schruba, Munoz-Mateos,
  Bigiel, Bolatto, Brinks, de~Blok, Meidt, Rix, Rosolowsky, Schinnerer,
  Schuster, and Usero}]{Leroy+13}
Leroy AK, Walter F, Sandstrom K, et~al. (2013) {Molecular Gas and Star
  Formation in Nearby Disk Galaxies}. \aj 146(2):19

\bibitem[{{Leroy} et~al.(2015){Leroy}, {Bolatto}, {Ostriker}, {Rosolowsky},
  {Walter}, {Warren}, {Donovan Meyer}, {Hodge}, {Meier}, {Ott}, {Sandstrom},
  {Schruba}, {Veilleux}, and {Zwaan}}]{Leroy+15}
{Leroy} AK, {Bolatto} AD, {Ostriker} EC, et~al. (2015) {ALMA Reveals the
  Molecular Medium Fueling the Nearest Nuclear Starburst}. \apj 801(1):25

\bibitem[{{Leroy} et~al.(2017){Leroy}, {Schinnerer}, {Hughes}, {Kruijssen},
  {Meidt}, {Schruba}, {Sun}, {Bigiel}, {Aniano}, {Blanc}, {Bolatto},
  {Chevance}, {Colombo}, {Gallagher}, {Garcia-Burillo}, {Kramer}, {Querejeta},
  {Pety}, {Thompson}, and {Usero}}]{Leroy+17}
{Leroy} AK, {Schinnerer} E, {Hughes} A, et~al. (2017) {Cloud-scale ISM
  Structure and Star Formation in M51}. \apj 846:71

\bibitem[{{Levrier} et~al.(2012){Levrier}, {Le Petit}, {Hennebelle},
  {Lesaffre}, {Gerin}, and {Falgarone}}]{levrier2012}
{Levrier} F, {Le Petit} F, {Hennebelle} P, et~al. (2012) {UV-driven chemistry
  in simulations of the interstellar medium. I. Post-processed chemistry with
  the Meudon PDR code}. \aap 544:A22

\bibitem[{{Li} et~al.(2020){Li}, {Zhang}, {Liu}, {Beuther}, {Palau}, {Miquel.
  Girart}, {Smith}, {Hora}, {Lin}, {Qiu}, {Strom}, {Wang}, {Li}, and
  {Yue}}]{Li+20}
{Li} S, {Zhang} Q, {Liu} HB, et~al. (2020) {ALMA observations of NGC 6334S $-$
  I: Forming massive stars and cluster in subsonic and transonic filamentary
  clouds}. arXiv e-prints arXiv:2003.13534

\bibitem[{{Li} et~al.(2005){Li}, {Mac Low}, and {Klessen}}]{Li+05}
{Li} Y, {Mac Low} MM, and {Klessen} RS (2005) {Control of Star Formation in
  Galaxies by Gravitational Instability}. \apjl 620(1):L19--L22

\bibitem[{Lombardi et~al.(2010)Lombardi, Alves, and Lada}]{Lombardi+10}
Lombardi M, Alves JF, and Lada CJ (2010) {Larson's third law and the
  universality of molecular cloud structure}. Astron Astrophys 519:L7

\bibitem[{{Longmore} et~al.(2014){Longmore}, {Kruijssen}, {Bastian}, {Bally},
  {Rathborne}, {Testi}, {Stolte}, {Dale}, {Bressert}, and
  {Alves}}]{longmore2014}
{Longmore} SN, {Kruijssen} JMD, {Bastian} N, et~al. (2014) {The Formation and
  Early Evolution of Young Massive Clusters}. Protostars and Planets VI pp
  291--314

\bibitem[{{Lopez} et~al.(2011){Lopez}, {Krumholz}, {Bolatto}, {Prochaska}, and
  {Ramirez-Ruiz}}]{lopez11}
{Lopez} LA, {Krumholz} MR, {Bolatto} AD, et~al. (2011) {What Drives the
  Expansion of Giant H II Regions?: A Study of Stellar Feedback in 30 Doradus}.
  \apj 731:91

\bibitem[{{Lopez} et~al.(2014){Lopez}, {Krumholz}, {Bolatto}, {Prochaska},
  {Ramirez-Ruiz}, and {Castro}}]{lopez14}
{Lopez} LA, {Krumholz} MR, {Bolatto} AD, et~al. (2014) {The Role of Stellar
  Feedback in the Dynamics of H II Regions}. \apj 795:121

\bibitem[{{Loren}(1989)}]{Loren89b}
{Loren} RB (1989) {The Cobwebs of Ophiuchus. II. 13CO Filament Kinematics}.
  \apj 338:925

\bibitem[{{Marsh} et~al.(2016){Marsh}, {Kirk}, {Andr{\'e}}, {Griffin},
  {K{\"o}nyves}, {Palmeirim}, {Men'shchikov}, {Ward-Thompson}, {Benedettini},
  {Bresnahan}, {Francesco}, {Elia}, {Motte}, {Peretto}, {Pezzuto}, {Roy},
  {Sadavoy}, {Schneider}, {Spinoglio}, and {White}}]{Marsh+16}
{Marsh} KA, {Kirk} JM, {Andr{\'e}} P, et~al. (2016) {A census of dense cores in
  the Taurus L1495 cloud from the Herschel}. \mnras 459:342--356

\bibitem[{{Maschberger} et~al.(2014){Maschberger}, {Bonnell}, {Clarke}, and
  {Moraux}}]{Maschberger+14}
{Maschberger} T, {Bonnell} IA, {Clarke} CJ, et~al. (2014) {The relation between
  accretion rates and the initial mass function in hydrodynamical simulations
  of star formation}. \mnras 439(1):234--246

\bibitem[{{McKee} and {Ostriker}(2007)}]{McKeeOstriker07}
{McKee} CF and {Ostriker} EC (2007) {Theory of Star Formation}. \araa
  45(1):565--687

\bibitem[{{McKee} and {Ostriker}(1977)}]{mckee77}
{McKee} CF and {Ostriker} JP (1977) {A theory of the interstellar medium -
  Three components regulated by supernova explosions in an inhomogeneous
  substrate}. \apj 218:148--169

\bibitem[{{McKee} and {Zweibel}(1992)}]{McKeeZweibel92}
{McKee} CF and {Zweibel} EG (1992) {On the Virial Theorem for Turbulent
  Molecular Clouds}. \apj 399:551

\bibitem[{{McLeod} et~al.(2019{\natexlab{a}}){McLeod}, {Dale}, {Evans},
  {Ginsburg}, {Kruijssen}, {Pellegrini}, {Ramsay}, and {Testi}}]{McLeod+19b}
{McLeod} AF, {Dale} JE, {Evans} CJ, et~al. (2019{\natexlab{a}}) {Feedback from
  massive stars at low metallicities: MUSE observations of N44 and N180 in the
  Large Magellanic Cloud}. \mnras 486(4):5263--5288

\bibitem[{{McLeod} et~al.(2019{\natexlab{b}}){McLeod}, {Kruijssen}, {Weisz},
  {Zeidler}, {Schruba}, {Dalcanton}, {Longmore}, {Chevance}, {Faesi}, and
  {Byler}}]{McLeod+19}
{McLeod} AF, {Kruijssen} JMD, {Weisz} DR, et~al. (2019{\natexlab{b}}) {Stellar
  Feedback and Resolved Stellar IFU Spectroscopy in the nearby Spiral Galaxy
  NGC 300}. \apj in press arXiv:1910.11270

\bibitem[{{Meidt} et~al.(2013){Meidt}, {Schinnerer}, {Garc{\'\i}a-Burillo},
  {Hughes}, {Colombo}, {Pety}, {Dobbs}, {Schuster}, {Kramer}, {Leroy}, {Dumas},
  and {Thompson}}]{Meidt+13}
{Meidt} SE, {Schinnerer} E, {Garc{\'\i}a-Burillo} S, et~al. (2013) {Gas
  Kinematics on Giant Molecular Cloud Scales in M51 with PAWS: Cloud
  Stabilization through Dynamical Pressure}. \apj 779(1):45

\bibitem[{Meidt et~al.(2015)Meidt, Hughes, Dobbs, Pety, Thompson,
  Garcia-Burillo, Leroy, Schinnerer, Colombo, Querejeta, Kramer, Schuster, and
  Dumas}]{Meidt+15}
Meidt SE, Hughes A, Dobbs CL, et~al. (2015) {Short GMC lifetimes: an
  observational estimate with the PdBI Arcsecond Whirlpool Survey (PAWS)}. \apj
  806(1):72

\bibitem[{{Meidt} et~al.(2018){Meidt}, {Leroy}, {Rosolowsky}, {Kruijssen},
  {Schinnerer}, {Schruba}, {Pety}, {Blanc}, {Bigiel}, {Chevance}, {Hughes},
  {Querejeta}, and {Usero}}]{Meidt+18}
{Meidt} SE, {Leroy} AK, {Rosolowsky} E, et~al. (2018) {A Model for the Onset of
  Self-gravitation and Star Formation in Molecular Gas Governed by Galactic
  Forces. I. Cloud-scale Gas Motions}. \apj 854(2):100

\bibitem[{{Meidt} et~al.(2019){Meidt}, {Glover}, {Kruijssen}, {Leroy},
  {Rosolowsky}, {Hughes}, {Schinnerer}, {Schruba}, {Usero}, {Bigiel}, {Blanc},
  {Chevance}, {Pety}, and {Querejeta}}]{Meidt+19}
{Meidt} SE, {Glover} S, {Kruijssen} JMD, et~al. (2019) {A model for the onset
  of self-gravitation and star formation in molecular gas governed by galactic
  forces: II. a bottleneck set by cloud-environment decoupling}. \apj~submitted

\bibitem[{{Men'shchikov}(2013)}]{Menshchikov13}
{Men'shchikov} A (2013) {A multi-scale filament extraction method:
  getfilaments}. \aap 560:A63

\bibitem[{{Men'shchikov} et~al.(2012){Men'shchikov}, {Andr{\'e}}, {Didelon},
  {Motte}, {Hennemann}, and {Schneider}}]{Menshchikov+12}
{Men'shchikov} A, {Andr{\'e}} P, {Didelon} P, et~al. (2012) {A multi-scale,
  multi-wavelength source extraction method: getsources}. \aap 542:A81

\bibitem[{{Mestel} and {Spitzer}(1956)}]{MestelSpitzer56}
{Mestel} L and {Spitzer} J L (1956) {Star formation in magnetic dust clouds}.
  \mnras 116:503

\bibitem[{{Miesch} and {Scalo}(1995)}]{MieschScalo95}
{Miesch} MS and {Scalo} JM (1995) {Exponential Tails in the Centroid Velocity
  Distributions of Star-Forming Regions}. \apjl 450:L27

\bibitem[{{Miesch} et~al.(1999){Miesch}, {Scalo}, and {Bally}}]{Miesch+99}
{Miesch} MS, {Scalo} J, and {Bally} J (1999) {Velocity Field Statistics in
  Star-forming Regions. I. Centroid Velocity Observations}. \apj
  524(2):895--922

\bibitem[{{Miura} et~al.(2012){Miura}, {Kohno}, {Tosaki}, {Espada}, {Hwang},
  {Kuno}, {Okumura}, {Hirota}, {Muraoka}, {Onodera}, {Minamidani}, {Komugi},
  {Nakanishi}, {Sawada}, {Kaneko}, and {Kawabe}}]{Miura+12}
{Miura} RE, {Kohno} K, {Tosaki} T, et~al. (2012) {Giant Molecular Cloud
  Evolutions in the Nearby Spiral Galaxy M33}. \apj 761(1):37

\bibitem[{{Miville-Desch{\^e}nes} et~al.(2010){Miville-Desch{\^e}nes},
  {Martin}, {Abergel}, {Bernard}, {Boulanger}, {Lagache}, {Anderson},
  {Andr{\'e}}, {Arab}, {Baluteau}, {Blagrave}, {Bontemps}, {Cohen},
  {Compiegne}, {Cox}, {Dartois}, {Davis}, {Emery}, {Fulton}, {Gry}, {Habart},
  {Huang}, {Joblin}, {Jones}, {Kirk}, {Lim}, {Madden}, {Makiwa}, {Menshchikov},
  {Molinari}, {Moseley}, {Motte}, {Naylor}, {Okumura}, {Pinheiro Gon{\c
  c}alves}, {Polehampton}, {Rod{\'o}n}, {Russeil}, {Saraceno}, {Schneider},
  {Sidher}, {Spencer}, {Swinyard}, {Ward-Thompson}, {White}, and
  {Zavagno}}]{Miville-Deschenes+10}
{Miville-Desch{\^e}nes} MA, {Martin} PG, {Abergel} A, et~al. (2010)
  {Herschel-SPIRE observations of the Polaris flare: Structure of the diffuse
  interstellar medium at the sub-parsec scale}. \aap 518:L104

\bibitem[{{Miville-Desch{\^e}nes} et~al.(2016){Miville-Desch{\^e}nes}, {Duc},
  {Marleau}, {Cuillandre}, {Didelon}, {Gwyn}, and
  {Karabal}}]{Miville-Deschenes+16}
{Miville-Desch{\^e}nes} MA, {Duc} PA, {Marleau} F, et~al. (2016) {Probing
  interstellar turbulence in cirrus with deep optical imaging: no sign of
  energy dissipation at 0.01 pc scale}. \aap 593:A4

\bibitem[{Miville-Desch{\^{e}}nes et~al.(2017)Miville-Desch{\^{e}}nes, Murray,
  and Lee}]{Miville-Deschenes+17}
Miville-Desch{\^{e}}nes MA, Murray N, and Lee EJ (2017) {Physical properties of
  molecular clouds for the entire Milky Way disk}. Astrophys J 834(1):1--31

\bibitem[{{Miyama} et~al.(1987){Miyama}, {Narita}, and {Hayashi}}]{Miyama+87}
{Miyama} SM, {Narita} S, and {Hayashi} C (1987) {Fragmentation of Isothermal
  Sheet-Like Clouds. II ---Full Nonlinear Numerical Simulations---}. Progress
  of Theoretical Physics 78(6):1273--1287

\bibitem[{{Molinari} et~al.(2010){Molinari}, {Swinyard}, {Bally}, {Barlow},
  {Bernard}, {Martin}, {Moore}, {Noriega-Crespo}, {Plume}, {Testi}, {Zavagno},
  {Abergel}, {Ali}, {Anderson}, {Andr{\'e}}, {Baluteau}, {Battersby},
  {Beltr{\'a}n}, {Benedettini}, {Billot}, {Blommaert}, {Bontemps}, {Boulanger},
  {Brand}, {Brunt}, {Burton}, {Calzoletti}, {Carey}, {Caselli}, {Cesaroni},
  {Cernicharo}, {Chakrabarti}, {Chrysostomou}, {Cohen}, {Compiegne}, {de
  Bernardis}, {de Gasperis}, {di Giorgio}, {Elia}, {Faustini}, {Flagey},
  {Fukui}, {Fuller}, {Ganga}, {Garcia-Lario}, {Glenn}, {Goldsmith}, {Griffin},
  {Hoare}, {Huang}, {Ikhenaode}, {Joblin}, {Joncas}, {Juvela}, {Kirk},
  {Lagache}, {Li}, {Lim}, {Lord}, {Marengo}, {Marshall}, {Masi}, {Massi},
  {Matsuura}, {Minier}, {Miville-Desch{\^e}nes}, {Montier}, {Morgan}, {Motte},
  {Mottram}, {M{\"u}ller}, {Natoli}, {Neves}, {Olmi}, {Paladini}, {Paradis},
  {Parsons}, {Peretto}, {Pestalozzi}, {Pezzuto}, {Piacentini}, {Piazzo},
  {Polychroni}, {Pomar{\`e}s}, {Popescu}, {Reach}, {Ristorcelli}, {Robitaille},
  {Robitaille}, {Rod{\'o}n}, {Roy}, {Royer}, {Russeil}, {Saraceno}, {Sauvage},
  {Schilke}, {Schisano}, {Schneider}, {Schuller}, {Schulz}, {Sibthorpe},
  {Smith}, {Smith}, {Spinoglio}, {Stamatellos}, {Strafella}, {Stringfellow},
  {Sturm}, {Taylor}, {Thompson}, {Traficante}, {Tuffs}, {Umana}, {Valenziano},
  {Vavrek}, {Veneziani}, {Viti}, {Waelkens}, {Ward-Thompson}, {White},
  {Wilcock}, {Wyrowski}, {Yorke}, and {Zhang}}]{Molinari+10}
{Molinari} S, {Swinyard} B, {Bally} J, et~al. (2010) {Clouds, filaments, and
  protostars: The Herschel Hi-GAL Milky Way}. \aap 518:L100

\bibitem[{{Molinari} et~al.(2011){Molinari}, {Bally}, {Noriega-Crespo},
  {Compi{\`e}gne}, {Bernard}, {Paradis}, {Martin}, {Testi}, {Barlow}, {Moore},
  {Plume}, {Swinyard}, {Zavagno}, {Calzoletti}, {Di Giorgio}, {Elia},
  {Faustini}, {Natoli}, {Pestalozzi}, {Pezzuto}, {Piacentini}, {Polenta},
  {Polychroni}, {Schisano}, {Traficante}, {Veneziani}, {Battersby}, {Burton},
  {Carey}, {Fukui}, {Li}, {Lord}, {Morgan}, {Motte}, {Schuller},
  {Stringfellow}, {Tan}, {Thompson}, {Ward-Thompson}, {White}, and
  {Umana}}]{molinari2011a}
{Molinari} S, {Bally} J, {Noriega-Crespo} A, et~al. (2011) {A 100 pc Elliptical
  and Twisted Ring of Cold and Dense Molecular Clouds Revealed by Herschel
  Around the Galactic Center}. \apjl 735(2):L33

\bibitem[{{Mookerjea} et~al.(2004){Mookerjea}, {Kramer}, {Nielbock}, and
  {Nyman}}]{Mookerjea+04}
{Mookerjea} B, {Kramer} C, {Nielbock} M, et~al. (2004) {The Giant Molecular
  Cloud associated with RCW 106. A 1.2 mm continuum mapping study}. \aap
  426:119--129

\bibitem[{{Motte} et~al.(1998){Motte}, {Andre}, and {Neri}}]{Motte+98}
{Motte} F, {Andre} P, and {Neri} R (1998) {The initial conditions of star
  formation in the rho Ophiuchi main cloud: wide-field millimeter continuum
  mapping}. \aap 336:150--172

\bibitem[{{Motte} et~al.(2001){Motte}, {Andr{\'e}}, {Ward-Thompson}, and
  {Bontemps}}]{Motte+01}
{Motte} F, {Andr{\'e}} P, {Ward-Thompson} D, et~al. (2001) {A SCUBA survey of
  the NGC 2068/2071 protoclusters}. \aap 372:L41--L44

\bibitem[{{Motte} et~al.(2018){Motte}, {Nony}, {Louvet}, {Marsh}, {Bontemps},
  {Whitworth}, {Men'shchikov}, {Nguyen Luong}, {Csengeri}, {Maury}, {Gusdorf},
  {Chapillon}, {K{\"o}nyves}, {Schilke}, {Duarte-Cabral}, {Didelon}, and
  {Gaudel}}]{Motte+18}
{Motte} F, {Nony} T, {Louvet} F, et~al. (2018) {The unexpectedly large
  proportion of high-mass star-forming cores in a Galactic mini-starburst}.
  Nature Astronomy 2:478--482

\bibitem[{{Murray} and {Chang}(2015)}]{Murray_Chang15}
{Murray} N and {Chang} P (2015) {Star Formation in Self-gravitating Turbulent
  Fluids}. \apj 804(1):44

\bibitem[{{Myers}(2009)}]{Myers09}
{Myers} P (2009) {Filamentary Structure of Star-forming Complexes}. \apj
  700:1609--1625

\bibitem[{{Myers}(1983)}]{Myers83}
{Myers} PC (1983) {Dense cores in dark clouds. III. Subsonic turbulence.} \apj
  270:105--118

\bibitem[{{Nagasawa}(1987)}]{Nagasawa87}
{Nagasawa} M (1987) {Gravitational Instability of the Isothermal Gas Cylinder
  with an Axial magnetic Field}. Progress of Theoretical Physics 77(3):635--652

\bibitem[{{Nakamura} and {Li}(2005)}]{NakamuraLi05}
{Nakamura} F and {Li} ZY (2005) {Quiescent Cores and the Efficiency of
  Turbulence-accelerated, Magnetically Regulated Star Formation}. \apj
  631(1):411--428

\bibitem[{{Naranjo-Romero} et~al.(2015){Naranjo-Romero},
  {V{\'a}zquez-Semadeni}, and {Loughnane}}]{NaranjoRomero+15}
{Naranjo-Romero} R, {V{\'a}zquez-Semadeni} E, and {Loughnane} RM (2015)
  {Hierarchical Gravitational Fragmentation. I. Collapsing Cores within
  Collapsing Clouds}. \apj 814(1):48

\bibitem[{{Nguyen-Luong} et~al.(2020){Nguyen-Luong}, {Nakamura}, {Sugitani},
  {Shimoikura}, {Dobashi}, {Kinoshita}, {Kim}, {Kang}, {Sanhueza}, {Evans}, and
  {White}}]{Nguyen-Luong+20}
{Nguyen-Luong} Q, {Nakamura} F, {Sugitani} K, et~al. (2020) {Large-scale
  Molecular Gas Distribution in the M17 Cloud Complex: Dense Gas Conditions of
  Massive Star Formation?} \apj 891(1):66

\bibitem[{{Ntormousi} et~al.(2016){Ntormousi}, {Hennebelle}, {Andr{\'e}}, and
  {Masson}}]{Ntormousi+16}
{Ntormousi} E, {Hennebelle} P, {Andr{\'e}} P, et~al. (2016) {The effect of
  ambipolar diffusion on low-density molecular ISM filaments}. \aap 589:A24

\bibitem[{{Offner} and {Liu}(2018)}]{offner2018}
{Offner} SSR and {Liu} Y (2018) {Turbulent action at a distance due to stellar
  feedback in magnetized clouds}. Nature Astronomy 2:896--900

\bibitem[{{Ohashi} et~al.(2016){Ohashi}, {Sanhueza}, {Chen}, {Zhang},
  {Busquet}, {Nakamura}, {Palau}, and {Tatematsu}}]{Ohashi+16}
{Ohashi} S, {Sanhueza} P, {Chen} HRV, et~al. (2016) {Dense Core Properties in
  the Infrared Dark Cloud G14.225-0.506 Revealed by ALMA}. \apj 833(2):209

\bibitem[{{Onishi} et~al.(1998){Onishi}, {Mizuno}, {Kawamura}, {Ogawa}, and
  {Fukui}}]{Onishi+98}
{Onishi} T, {Mizuno} A, {Kawamura} A, et~al. (1998) {A C$^{18}$O Survey of
  Dense Cloud Cores in Taurus: Star Formation}. \apj 502:296--314

\bibitem[{{Oort}(1954)}]{Oort54}
{Oort} JH (1954) {Outline of a theory on the origin and acceleration of
  interstellar clouds and O associations}. \bain 12:177

\bibitem[{{Ossenkopf} and {Mac Low}(2002)}]{OssenkopfMacLow02}
{Ossenkopf} V and {Mac Low} MM (2002) {Turbulent velocity structure in
  molecular clouds}. \aap 390:307--326

\bibitem[{{Ostriker}(1964)}]{Ostriker64}
{Ostriker} J (1964) {The Equilibrium of Polytropic and Isothermal Cylinders.}
  \apj 140:1056

\bibitem[{{Padoan} and {Nordlund}(2011)}]{padoan2011}
{Padoan} P and {Nordlund} {\AA} (2011) {The Star Formation Rate of Supersonic
  Magnetohydrodynamic Turbulence}. \apj 730:40

\bibitem[{{Padoan} et~al.(2016){Padoan}, {Pan}, {Haugb{\o}lle}, and
  {Nordlund}}]{Padoan+16}
{Padoan} P, {Pan} L, {Haugb{\o}lle} T, et~al. (2016) {Supernova Driving. I. The
  Origin of Molecular Cloud Turbulence}. \apj 822(1):11

\bibitem[{{Padoan} et~al.(2019){Padoan}, {Pan}, {Juvela}, {Haugb{\o}lle}, and
  {Nordlund}}]{Padoan+19}
{Padoan} P, {Pan} L, {Juvela} M, et~al. (2019) {The Origin of Massive Stars:
  The Inertial-Inflow Model}. arXiv e-prints arXiv:1911.04465

\bibitem[{{Palau} et~al.(2014){Palau}, {Estalella}, {Girart}, {Fuente},
  {Fontani}, {Commer{\c{c}}on}, {Busquet}, {Bontemps}, {S{\'a}nchez-Monge},
  {Zapata}, {Zhang}, {Hennebelle}, and {di Francesco}}]{Palau+14}
{Palau} A, {Estalella} R, {Girart} JM, et~al. (2014) {Fragmentation of Massive
  Dense Cores Down to \&lt;\raisebox{-0.5ex}\textasciitilde 1000 AU: Relation
  between Fragmentation and Density Structure}. \apj 785(1):42

\bibitem[{{Palmeirim} et~al.(2013){Palmeirim}, {Andr{\'e}}, {Kirk},
  {Ward-Thompson}, {Arzoumanian}, {K{\"o}nyves}, {Didelon}, {Schneider},
  {Benedettini}, {Bontemps}, {Di Francesco}, {Elia}, {Griffin}, {Hennemann},
  {Hill}, {Martin}, {Men'shchikov}, {Molinari}, {Motte}, {Nguyen Luong},
  {Nutter}, {Peretto}, {Pezzuto}, {Roy}, {Rygl}, {Spinoglio}, and
  {White}}]{Palmeirim+13}
{Palmeirim} P, {Andr{\'e}} P, {Kirk} J, et~al. (2013) {Herschel view of the
  Taurus B211/3 filament and striations: evidence of filamentary growth?} \aap
  550:A38

\bibitem[{{Pan} et~al.(2016){Pan}, {Fujimoto}, {Tasker}, {Rosolowsky},
  {Colombo}, {Benincasa}, and {Wadsley}}]{Pan+16}
{Pan} HA, {Fujimoto} Y, {Tasker} EJ, et~al. (2016) {Effects of galactic disc
  inclination and resolution on observed GMC properties and Larson's scaling
  relations}. \mnras 458(3):2443--2453

\bibitem[{{Pan} et~al.(2019){Pan}, {Padoan}, and {Nordlund}}]{Pan+19}
{Pan} L, {Padoan} P, and {Nordlund} {\AA} (2019) {The Probability Distribution
  of Density Fluctuations in Supersonic Turbulence}. \apj 881:155

\bibitem[{{Panopoulou} et~al.(2014){Panopoulou}, {Tassis}, {Goldsmith}, and
  {Heyer}}]{Panopoulou+14}
{Panopoulou} GV, {Tassis} K, {Goldsmith} PF, et~al. (2014) {$^{13}$CO filaments
  in the Taurus molecular cloud}. \mnras 444:2507--2524

\bibitem[{{Panopoulou} et~al.(2017){Panopoulou}, {Psaradaki}, {Skalidis},
  {Tassis}, and {Andrews}}]{Panopoulou+17}
{Panopoulou} GV, {Psaradaki} I, {Skalidis} R, et~al. (2017) {A closer look at
  the `characteristic' width of molecular cloud filaments}. \mnras
  466:2529--2541

\bibitem[{{Parker}(1979)}]{Parker79}
{Parker} EN (1979) {Cosmical magnetic fields. Their origin and their activity}.
  Oxford University Press

\bibitem[{{Passot} and {V{\'a}zquez-Semadeni}(1998)}]{PassotVazquezSemadeni98}
{Passot} T and {V{\'a}zquez-Semadeni} E (1998) {Density probability
  distribution in one-dimensional polytropic gas dynamics}. \pre
  58(4):4501--4510

\bibitem[{{Passot} et~al.(1988){Passot}, {Pouquet}, and {Woodward}}]{Passot+88}
{Passot} T, {Pouquet} A, and {Woodward} P (1988) {The plausibility of
  Kolmogorov-type spectra in molecular clouds}. \aap 197(1-2):228--234

\bibitem[{{Passot} et~al.(1995){Passot}, {Vazquez-Semadeni}, and
  {Pouquet}}]{Passot+95}
{Passot} T, {Vazquez-Semadeni} E, and {Pouquet} A (1995) {A Turbulent Model for
  the Interstellar Medium. II. Magnetic Fields and Rotation}. \apj 455:536

\bibitem[{{Peretto} et~al.(2012){Peretto}, {Andr{\'e}}, {K{\"o}nyves},
  {Schneider}, {Arzoumanian}, {Palmeirim}, {Didelon}, {Attard}, {Bernard}, {Di
  Francesco}, {Elia}, {Hennemann}, {Hill}, {Kirk}, {Men'shchikov}, {Motte},
  {Nguyen Luong}, {Roussel}, {Sousbie}, {Testi}, {Ward-Thompson}, {White}, and
  {Zavagno}}]{Peretto+12}
{Peretto} N, {Andr{\'e}} P, {K{\"o}nyves} V, et~al. (2012) {The Pipe Nebula as
  seen with Herschel: formation of filamentary structures by large-scale
  compression?} \aap 541:A63

\bibitem[{{Peretto} et~al.(2013){Peretto}, {Fuller}, {Duarte-Cabral}, {Avison},
  {Hennebelle}, {Pineda}, {Andr{\'e}}, {Bontemps}, {Motte}, {Schneider}, and
  {Molinari}}]{Peretto+13}
{Peretto} N, {Fuller} GA, {Duarte-Cabral} A, et~al. (2013) {Global collapse of
  molecular clouds as a formation mechanism for the most massive stars}. \aap
  555:A112

\bibitem[{{Peretto} et~al.(2014){Peretto}, {Fuller}, {Andr{\'e}},
  {Arzoumanian}, {Rivilla}, {Bardeau}, {Duarte Puertas}, {Guzman Fernandez},
  {Lenfestey}, {Li}, {Olguin}, {R{\"o}ck}, {de Villiers}, and
  {Williams}}]{Peretto+14}
{Peretto} N, {Fuller} GA, {Andr{\'e}} P, et~al. (2014) {SDC13 infrared dark
  clouds: Longitudinally collapsing filaments?} \aap 561:A83

\bibitem[{{Phillips} et~al.(1979){Phillips}, {Huggins}, {Wannier}, and
  {Scoville}}]{Phillips+79}
{Phillips} TG, {Huggins} PJ, {Wannier} PG, et~al. (1979) {Observations of CO(J
  = 2-1) emission from molecular clouds.} \apj 231:720--731

\bibitem[{{Pineda} et~al.(2010){Pineda}, {Goodman}, {Arce}, {Caselli},
  {Foster}, {Myers}, and {Rosolowsky}}]{Pineda+10}
{Pineda} JE, {Goodman} AA, {Arce} HG, et~al. (2010) {Direct Observation of a
  Sharp Transition to Coherence in Dense Cores}. \apjl 712(1):L116--L121

\bibitem[{{Pineda} et~al.(2011){Pineda}, {Goodman}, {Arce}, {Caselli},
  {Longmore}, and {Corder}}]{Pineda+11}
{Pineda} JE, {Goodman} AA, {Arce} HG, et~al. (2011) {Expanded Very Large Array
  Observations of the Barnard 5 Star-forming Core: Embedded Filaments
  Revealed}. \apjl 739:L2

\bibitem[{{Pineda} et~al.(2013){Pineda}, {Langer}, {Velusamy}, and
  {Goldsmith}}]{Pineda+13}
{Pineda} JL, {Langer} WD, {Velusamy} T, et~al. (2013) {A Herschel [C ii]
  Galactic plane survey. I. The global distribution of ISM gas components}.
  \aap 554:A103

\bibitem[{{Planck int. res. XXXV}(2016)}]{PlanckXXXV+16}
{Planck int res XXXV} (2016) {\textit{Planck} intermediate results. XXXV.
  Probing the role of the magnetic field in the formation of structure in
  molecular clouds}. \aap 586:A138

\bibitem[{{Plume} et~al.(1997){Plume}, {Jaffe}, {Evans}, {Mart{\'\i}n-Pintado},
  and {G{\'o}mez-Gonz{\'a}lez}}]{Plume+97}
{Plume} R, {Jaffe} DT, {Evans} I Neal~J, et~al. (1997) {Dense Gas and Star
  Formation: Characteristics of Cloud Cores Associated with Water Masers}. \apj
  476(2):730--749

\bibitem[{{Pokhrel} et~al.(2016){Pokhrel}, {Gutermuth}, {Ali}, {Megeath},
  {Pipher}, {Myers}, {Fischer}, {Henning}, {Wolk}, {Allen}, and
  {Tobin}}]{Pokhrel+16}
{Pokhrel} R, {Gutermuth} R, {Ali} B, et~al. (2016) {A Herschel-SPIRE survey of
  the Mon R2 giant molecular cloud: analysis of the gas column density
  probability density function}. \mnras 461:22--35

\bibitem[{{Pon} et~al.(2011){Pon}, {Johnstone}, and {Heitsch}}]{Pon+11}
{Pon} A, {Johnstone} D, and {Heitsch} F (2011) {Modes of Star Formation in
  Finite Molecular Clouds}. \apj 740(2):88

\bibitem[{{Pon} et~al.(2012){Pon}, {Toal{\'a}}, {Johnstone},
  {V{\'a}zquez-Semadeni}, {Heitsch}, and {G{\'o}mez}}]{Pon+12}
{Pon} A, {Toal{\'a}} JA, {Johnstone} D, et~al. (2012) {Aspect Ratio Dependence
  of the Free-fall Time for Non-spherical Symmetries}. \apj 756(2):145

\bibitem[{{Rathborne} et~al.(2014){Rathborne}, {Longmore}, {Jackson},
  {Kruijssen}, {Alves}, {Bally}, {Bastian}, {Contreras}, {Foster}, {Garay},
  {Testi}, and {Walsh}}]{rathborne14}
{Rathborne} JM, {Longmore} SN, {Jackson} JM, et~al. (2014) {Turbulence Sets the
  Initial Conditions for Star Formation in High-pressure Environments}. \apjl
  795:L25

\bibitem[{{Reid} and {Wilson}(2005)}]{ReidWilson05}
{Reid} MA and {Wilson} CD (2005) {High-Mass Star Formation. I. The Mass
  Distribution of Submillimeter Clumps in NGC 7538}. \apj 625(2):891--905

\bibitem[{{Reid} and {Wilson}(2006{\natexlab{a}})}]{ReidWilson06a}
{Reid} MA and {Wilson} CD (2006{\natexlab{a}}) {High-Mass Star Formation. II.
  The Mass Function of Submillimeter Clumps in M17}. \apj 644(2):990--1005

\bibitem[{{Reid} and {Wilson}(2006{\natexlab{b}})}]{ReidWilson06b}
{Reid} MA and {Wilson} CD (2006{\natexlab{b}}) {High-Mass Star Formation. III.
  The Functional Form of the Submillimeter Clump Mass Function}. \apj
  650(2):970--984

\bibitem[{{Reina-Campos} and {Kruijssen}(2017)}]{reinacampos17}
{Reina-Campos} M and {Kruijssen} JMD (2017) {A unified model for the maximum
  mass scales of molecular clouds, stellar clusters and high-redshift clumps}.
  \mnras 469:1282--1298

\bibitem[{{Rivera-Ingraham} et~al.(2016){Rivera-Ingraham}, {Ristorcelli},
  {Juvela}, {Montillaud}, {Men'shchikov}, {Malinen}, {Pelkonen}, {Marston},
  {Martin}, {Pagani}, {Paladini}, {Paradis}, {Ysard}, {Ward-Thompson},
  {Bernard}, {Marshall}, {Montier}, and {T{\'o}th}}]{Rivera-Ingraham+16}
{Rivera-Ingraham} A, {Ristorcelli} I, {Juvela} M, et~al. (2016) {Galactic cold
  cores. VII. Filament formation and evolution: Methods and observational
  constraints}. \aap 591:A90

\bibitem[{Roman-Duval et~al.(2010)Roman-Duval, Jackson, Heyer, Rathborne, and
  Simon}]{Roman-Duval+10}
Roman-Duval J, Jackson JM, Heyer M, et~al. (2010) {Physical Properties and
  Galactic Distribution of Molecular Clouds Identified in the Galactic Ring
  Survey}. Astrophys J 723(1):492--507

\bibitem[{{Roman-Duval} et~al.(2016){Roman-Duval}, {Heyer}, {Brunt}, {Clark},
  {Klessen}, and {Shetty}}]{roman-duval2016a}
{Roman-Duval} J, {Heyer} M, {Brunt} CM, et~al. (2016) {Distribution and Mass of
  Diffuse and Dense CO Gas in the Milky Way}. \apj 818(2):144

\bibitem[{{Rom{\'a}n-Z{\'u}{\~n}iga} et~al.(2010){Rom{\'a}n-Z{\'u}{\~n}iga},
  {Alves}, {Lada}, and {Lombardi}}]{RomanZuniga+10}
{Rom{\'a}n-Z{\'u}{\~n}iga} CG, {Alves} JF, {Lada} CJ, et~al. (2010) {Deep
  Near-infrared Survey of the Pipe Nebula. II. Data, Methods, and Dust
  Extinction Maps}. \apj 725(2):2232--2250

\bibitem[{{Rosolowsky} et~al.(2010){Rosolowsky}, {Dunham}, {Ginsburg},
  {Bradley}, {Aguirre}, {Bally}, {Battersby}, {Cyganowski}, {Dowell},
  {Drosback}, {Evans}, {Glenn}, {Harvey}, {Stringfellow}, {Walawender}, and
  {Williams}}]{Rosolowsky+10}
{Rosolowsky} E, {Dunham} MK, {Ginsburg} A, et~al. (2010) {The Bolocam Galactic
  Plane Survey. II. Catalog of the Image Data}. \apjs 188(1):123--138

\bibitem[{{Rosolowsky} et~al.(2008){Rosolowsky}, {Pineda}, {Kauffmann}, and
  {Goodman}}]{Rosolowsky+08}
{Rosolowsky} EW, {Pineda} JE, {Kauffmann} J, et~al. (2008) {Structural Analysis
  of Molecular Clouds: Dendrograms}. \apj 679(2):1338--1351

\bibitem[{{Rowles} and {Froebrich}(2009)}]{RowlesFroebrich09}
{Rowles} J and {Froebrich} D (2009) {The structure of molecular clouds - I.
  All-sky near-infrared extinction maps}. \mnras 395(3):1640--1648

\bibitem[{{Roy} et~al.(2019){Roy}, {Andr{\'e}}, {Arzoumanian},
  {Miville-Desch{\^e}nes}, {K{\"o}nyves}, {Schneider}, {Pezzuto}, {Palmeirim},
  and {Kirk}}]{Roy+19}
{Roy} A, {Andr{\'e}} P, {Arzoumanian} D, et~al. (2019) {How the power spectrum
  of dust continuum images may hide the presence of a characteristic filament
  width}. \aap 626:A76

\bibitem[{{Russeil} et~al.(2019){Russeil}, {Figueira}, {Zavagno}, {Motte},
  {Schneider}, {Men'shchikov}, {Bontemps}, {Andr{\'e}}, {Anderson},
  {Benedettini}, {Didelon}, {Di Francesco}, {Elia}, {K{\"o}nyves}, {Nguyen
  Luong}, {Nony}, {Pezzuto}, {Rygl}, {Schisano}, {Spinoglio}, {Tig{\'e}}, and
  {White}}]{Russeil+19}
{Russeil} D, {Figueira} M, {Zavagno} A, et~al. (2019) {Herschel-HOBYS study of
  the earliest phases of high-mass star formation in NGC 6357}. \aap 625:A134

\bibitem[{{Salji} et~al.(2015){Salji}, {Richer}, {Buckle}, {Francesco},
  {Hatchell}, {Hogerheijde}, {Johnstone}, {Kirk}, {Ward-Thompson}, and {JCMT
  GBS Consortium}}]{Salji+15}
{Salji} CJ, {Richer} JS, {Buckle} JV, et~al. (2015) {The JCMT Gould Belt
  Survey: properties of star-forming filaments in Orion A North}. \mnras
  449:1782--1796

\bibitem[{{Salpeter}(1955)}]{Salpeter55}
{Salpeter} EE (1955) {The Luminosity Function and Stellar Evolution.} \apj
  121:161

\bibitem[{{Sanders} et~al.(1985){Sanders}, {Scoville}, and
  {Solomon}}]{Sanders+85}
{Sanders} DB, {Scoville} NZ, and {Solomon} PM (1985) {Giant molecular clouds in
  the galaxy. II. Characteristics of discretefeatures.} \apj 289:373--387

\bibitem[{{Scalo}(1990)}]{Scalo90}
{Scalo} J (1990) {Perception of interstellar structure - Facing complexity}.
  In: {Capuzzo-Dolcetta} R, {Chiosi} C, and {di Fazio} A (eds) Physical
  Processes in Fragmentation and Star Formation, Astrophysics and Space Science
  Library, vol 162, pp 151--176

\bibitem[{{Schisano} et~al.(2014){Schisano}, {Rygl}, {Molinari}, {Busquet},
  {Elia}, {Pestalozzi}, {Polychroni}, {Billot}, {Carey}, {Paladini},
  {Noriega-Crespo}, {Moore}, {Plume}, {Glover}, and
  {V{\'a}zquez-Semadeni}}]{Schisano+14}
{Schisano} E, {Rygl} KLJ, {Molinari} S, et~al. (2014) {The Identification of
  Filaments on Far-infrared and Submillimiter Images: Morphology, Physical
  Conditions and Relation with Star Formation of Filamentary Structure}. \apj
  791:27

\bibitem[{{Schneider} et~al.(2010){Schneider}, {Csengeri}, {Bontemps}, {Motte},
  {Simon}, {Hennebelle}, {Federrath}, and {Klessen}}]{Schneider+10}
{Schneider} N, {Csengeri} T, {Bontemps} S, et~al. (2010) {Dynamic star
  formation in the massive DR21 filament}. \aap 520:A49

\bibitem[{{Schneider} and {Elmegreen}(1979)}]{Schneider+79}
{Schneider} S and {Elmegreen} BG (1979) {A catalog of dark globular filaments}.
  ApJs 41:87--95

\bibitem[{Schruba et~al.(2010)Schruba, Leroy, Walter, Sandstrom, and
  Rosolowsky}]{Schruba+10}
Schruba A, Leroy AK, Walter F, et~al. (2010) {the Scale Dependence of the
  Molecular Gas Depletion Time in M33}. \apj 722(2):1699--1706

\bibitem[{{Schruba} et~al.(2019){Schruba}, {Kruijssen}, and
  {Leroy}}]{Schruba+19}
{Schruba} A, {Kruijssen} JMD, and {Leroy} AK (2019) {How Galactic Environment
  Affects the Dynamical State of Molecular Clouds and Their Star Formation
  Efficiency}. \apj 883(1):2

\bibitem[{Scoville and Hersh(1979)}]{ScovilleHersh79}
Scoville NZ and Hersh K (1979) {Collisional growth of Giant Molecular Clouds}.
  \apj 229:578--582

\bibitem[{{Scoville} and {Wilson}(2004)}]{ScovilleWilson04}
{Scoville} NZ and {Wilson} CD (2004) {Molecular Gas forming Massive Star
  Clusters and Starbursts}. In: {Lamers} HJGLM, {Smith} LJ, and {Nota} A (eds)
  The Formation and Evolution of Massive Young Star Clusters, Astronomical
  Society of the Pacific Conference Series, vol 322, p 245

\bibitem[{{Seifried} et~al.(2011){Seifried}, {Schmidt}, and
  {Niemeyer}}]{seifried2011}
{Seifried} D, {Schmidt} W, and {Niemeyer} JC (2011) {Forced turbulence in
  thermally bistable gas: a parameter study}. \aap 526:A14

\bibitem[{{Seifried} et~al.(2018){Seifried}, {Walch}, {Haid}, {Girichidis}, and
  {Naab}}]{Seifried+18}
{Seifried} D, {Walch} S, {Haid} S, et~al. (2018) {Is Molecular Cloud Turbulence
  Driven by External Supernova Explosions?} \apj 855(2):81

\bibitem[{{Shetty} and {Ostriker}(2008)}]{ShettyOstriker08}
{Shetty} R and {Ostriker} EC (2008) {Cloud and Star Formation in Disk Galaxy
  Models with Feedback}. \apj 684(2):978--995

\bibitem[{{Shetty} et~al.(2011){Shetty}, {Glover}, {Dullemond}, and
  {Klessen}}]{shetty2011}
{Shetty} R, {Glover} SC, {Dullemond} CP, et~al. (2011) {Modelling CO emission -
  I. CO as a column density tracer and the X factor in molecular clouds}.
  \mnras 412(3):1686--1700

\bibitem[{{Shetty} et~al.(2012){Shetty}, {Beaumont}, {Burton}, {Kelly}, and
  {Klessen}}]{Shetty+12}
{Shetty} R, {Beaumont} CN, {Burton} MG, et~al. (2012) {The linewidth-size
  relationship in the dense interstellar medium of the Central Molecular Zone}.
  \mnras 425(1):720--729

\bibitem[{{Shimajiri} et~al.(2019{\natexlab{a}}){Shimajiri}, {Andr{\'e}},
  {Ntormousi}, {Men'shchikov}, {Arzoumanian}, and {Palmeirim}}]{Shimajiri+19a}
{Shimajiri} Y, {Andr{\'e}} P, {Ntormousi} E, et~al. (2019{\natexlab{a}})
  {Probing fragmentation and velocity sub-structure in the massive NGC 6334
  filament with ALMA}. \aap 632:A83

\bibitem[{{Shimajiri} et~al.(2019{\natexlab{b}}){Shimajiri}, {Andr{\'e}},
  {Palmeirim}, {Arzoumanian}, {Bracco}, {K{\"o}nyves}, {Ntormousi}, and
  {Ladjelate}}]{Shimajiri+19b}
{Shimajiri} Y, {Andr{\'e}} P, {Palmeirim} P, et~al. (2019{\natexlab{b}})
  {Probing accretion of ambient cloud material into the Taurus B211/B213
  filament}. \aap 623:A16

\bibitem[{{Shirley} et~al.(2003){Shirley}, {Evans}, {Young}, {Knez}, and
  {Jaffe}}]{Shirley+03}
{Shirley} YL, {Evans} I Neal~J, {Young} KE, et~al. (2003) {A CS J=5--\&gt;4
  Mapping Survey Toward High-Mass Star-forming Cores Associated with Water
  Masers}. \apjs 149(2):375--403

\bibitem[{{Shu}(1977)}]{Shu77}
{Shu} FH (1977) {Self-similar collapse of isothermal spheres and star
  formation.} \apj 214:488--497

\bibitem[{{Shu} et~al.(1987){Shu}, {Adams}, and {Lizano}}]{Shu+87}
{Shu} FH, {Adams} FC, and {Lizano} S (1987) {Star formation in molecular
  clouds: observation and theory.} \araa 25:23--81

\bibitem[{Silk(1997)}]{silk97}
Silk J (1997) {Feedback, Disk Self-Regulation, and Galaxy Formation}. \apj
  481(2):703--709

\bibitem[{{Smith} et~al.(2014){Smith}, {Glover}, and {Klessen}}]{Smith+14}
{Smith} RJ, {Glover} SCO, and {Klessen} RS (2014) {On the nature of
  star-forming filaments - I. Filament morphologies}. \mnras 445(3):2900--2917

\bibitem[{{Smith} et~al.(2016){Smith}, {Glover}, {Klessen}, and
  {Fuller}}]{Smith+16}
{Smith} RJ, {Glover} SCO, {Klessen} RS, et~al. (2016) {On the nature of
  star-forming filaments - II. Subfilaments and velocities}. \mnras
  455(4):3640--3655

\bibitem[{{Soler}(2019)}]{Soler19}
{Soler} JD (2019) {Using Herschel and Planck observations to delineate the role
  of magnetic fields in molecular cloud structure}. \aap 629:A96

\bibitem[{{Soler} and {Hennebelle}(2017)}]{soler2017}
{Soler} JD and {Hennebelle} P (2017) {What are we learning from the relative
  orientation between density structures and the magnetic field in molecular
  clouds?} \aap 607:A2

\bibitem[{{Soler} et~al.(2013){Soler}, {Hennebelle}, {Martin},
  {Miville-Desch{\^e}nes}, {Netterfield}, and {Fissel}}]{Soler+13}
{Soler} JD, {Hennebelle} P, {Martin} PG, et~al. (2013) {An Imprint of Molecular
  Cloud Magnetization in the Morphology of the Dust Polarized Emission}. \apj
  774(2):128

\bibitem[{{Solomon} et~al.(1987){Solomon}, {Rivolo}, {Barrett}, and
  {Yahil}}]{Solomon+87}
{Solomon} PM, {Rivolo} AR, {Barrett} J, et~al. (1987) {Mass, Luminosity, and
  Line Width Relations of Galactic Molecular Clouds}. \apj 319:730

\bibitem[{{Stahler} and {Palla}(2005)}]{StahlerPalla05}
{Stahler} SW and {Palla} F (2005) {The Formation of Stars}. Wiley-VCH

\bibitem[{{Stanke} et~al.(2006){Stanke}, {Smith}, {Gredel}, and
  {Khanzadyan}}]{Stanke+06}
{Stanke} T, {Smith} MD, {Gredel} R, et~al. (2006) {An unbiased search for the
  signatures of protostars in the {\ensuremath{\rho}} Ophiuchi molecular cloud
  . II. Millimetre continuum observations}. \aap 447(2):609--622

\bibitem[{{Stod{\'o}lkiewicz}(1963)}]{Stodolkiewicz63}
{Stod{\'o}lkiewicz} JS (1963) {On the Gravitational Instability of Some
  Magneto-Hydrodynamical Systems of Astrophysical Interest. Part III.} Acta
  Astron 13:30--54

\bibitem[{{Storm} et~al.(2016){Storm}, {Mundy}, {Lee},
  {Fern{\'a}ndez-L{\'o}pez}, {Looney}, {Teuben}, {Arce}, {Rosolowsky},
  {Meisner}, {Isella}, {Kauffmann}, {Shirley}, {Kwon}, {Plunkett}, {Pound},
  {Segura-Cox}, {Tassis}, {Tobin}, {Volgenau}, {Crutcher}, and
  {Testi}}]{Storm+16}
{Storm} S, {Mundy} LG, {Lee} KI, et~al. (2016) {CARMA Large Area Star Formation
  Survey: Dense Gas in the Young L1451 Region of Perseus}. \apj 830(2):127

\bibitem[{{Stutzki} and {Guesten}(1990)}]{StutzkiGuesten90}
{Stutzki} J and {Guesten} R (1990) {High Spatial Resolution Isotopic CO and CS
  Observations of M17 SW: The Clumpy Structure of the Molecular Cloud Core}.
  \apj 356:513

\bibitem[{{Stutzki}(1993)}]{Stutzki93}
{Stutzki} R (1993) {The Small Scale Structure of Molecular Clouds.} Reviews in
  Modern Astronomy 6:209--232

\bibitem[{{Sun} et~al.(2018){Sun}, {Leroy}, {Schruba}, {Rosolowsky}, {Hughes},
  {Kruijssen}, {Meidt}, {Schinnerer}, {Blanc}, {Bigiel}, {Bolatto}, {Chevance},
  {Groves}, {Herrera}, {Hygate}, {Pety}, {Querejeta}, {Usero}, and
  {Utomo}}]{Sun+18}
{Sun} J, {Leroy} AK, {Schruba} A, et~al. (2018) {Cloud-scale Molecular Gas
  Properties in 15 Nearby Galaxies}. \apj 860(2):172

\bibitem[{{Sun} et~al.(2020){Sun}, {Leroy}, {Ostriker}, {Hughes}, {Rosolowsky},
  {Schruba}, {Schinnerer}, {Blanc}, {Faesi}, {Kruijssen}, {Meidt}, {Utomo},
  {Bigiel}, {Bolatto}, {Chevance}, {Chiang}, {Dale}, {Emsellem}, {Glover},
  {Grasha}, {Henshaw}, {Herrera}, {Jimenez-Donaire}, {Lee}, {Pety},
  {Querejeta}, {Saito}, {Sandstrom}, and {Usero}}]{Sun+20}
{Sun} J, {Leroy} AK, {Ostriker} EC, et~al. (2020) {Dynamical Equilibrium in the
  Molecular ISM in 28 Nearby Star-Forming Galaxies }. \apj in press

\bibitem[{{Tafalla} and {Hacar}(2015)}]{TafallaHacar15}
{Tafalla} M and {Hacar} A (2015) {Chains of dense cores in the Taurus
  L1495/B213 complex}. \aap 574:A104

\bibitem[{{Tafalla} et~al.(2004){Tafalla}, {Myers}, {Caselli}, and
  {Walmsley}}]{Tafalla+04}
{Tafalla} M, {Myers} PC, {Caselli} P, et~al. (2004) {On the internal structure
  of starless cores. I. Physical conditions and the distribution of CO, CS,
  N$_{2}$H$^{+}$, and NH$_{3}$ in L1498 and L1517B}. \aap 416:191--212

\bibitem[{{Taff} and {Savedoff}(1973)}]{]TaffSavedoff73}
{Taff} LG and {Savedoff} MP (1973) {The mass distribution of objects
  under-going collisions with applications to interstellar HI clouds}. \mnras
  164:357

\bibitem[{{Tasker} and {Tan}(2009)}]{TaskerTan09}
{Tasker} EJ and {Tan} JC (2009) {Star Formation in Disk Galaxies. I. Formation
  and Evolution of Giant Molecular Clouds via Gravitational Instability and
  Cloud Collisions}. \apj 700(1):358--375

\bibitem[{{Tassis} et~al.(2010){Tassis}, {Christie}, {Urban}, {Pineda},
  {Mouschovias}, {Yorke}, and {Martel}}]{Tassis+10}
{Tassis} K, {Christie} DA, {Urban} A, et~al. (2010) {Do lognormal
  column-density distributions in molecular clouds imply supersonic
  turbulence?} \mnras 408(2):1089--1094

\bibitem[{{Tenorio-Tagle} and {Bodenheimer}(1988)}]{TenorioTagleBodenheimer88}
{Tenorio-Tagle} G and {Bodenheimer} P (1988) {Large-scale expanding
  superstructures in galaxies.} \araa 26:145--197

\bibitem[{{Testi} and {Sargent}(1998)}]{TestiSargent98}
{Testi} L and {Sargent} AI (1998) {Star Formation in Clusters: A Survey of
  Compact Millimeter-Wave Sources in the Serpens Core}. \apjl 508(1):L91--L94

\bibitem[{{Tielens}(2010)}]{tielens2010}
{Tielens} AGGM (2010) {The Physics and Chemistry of the Interstellar Medium}.
  Cambridge University Press

\bibitem[{{Toal{\'a}} et~al.(2012){Toal{\'a}}, {V{\'a}zquez-Semadeni}, and
  {G{\'o}mez}}]{Toala+12}
{Toal{\'a}} JA, {V{\'a}zquez-Semadeni} E, and {G{\'o}mez} GC (2012) {The
  Free-fall Time of Finite Sheets and Filaments}. \apj 744(2):190

\bibitem[{{Tohline}(1982)}]{Tohline82}
{Tohline} JE (1982) {Hydrodynamic Collapse}. Fundamentals of Cosmic Physics
  8:1--82

\bibitem[{{Tomisaka}(1984)}]{Tomisaka84}
{Tomisaka} K (1984) {Coagulation of interstellar clouds in spiral gravitational
  potential and formation of giant molecular clouds}. \pasj 36(3):457--475

\bibitem[{{Tomisaka}(2014)}]{Tomisaka14}
{Tomisaka} K (2014) {Magnetohydrostatic Equilibrium Structure and Mass of
  Filamentary Isothermal Cloud Threaded by Lateral Magnetic Field}. \apj 785:24

\bibitem[{{Traficante} et~al.(2018{\natexlab{a}}){Traficante}, {Duarte-Cabral},
  {Elia}, {Fuller}, {Merello}, {Molinari}, {Peretto}, {Schisano}, and {Di
  Giorgio}}]{Traficante+18b}
{Traficante} A, {Duarte-Cabral} A, {Elia} D, et~al. (2018{\natexlab{a}})
  {Testing the Larson relations in massive clumps}. \mnras 477(2):2220--2242

\bibitem[{{Traficante} et~al.(2018{\natexlab{b}}){Traficante}, {Fuller},
  {Smith}, {Billot}, {Duarte-Cabral}, {Peretto}, {Molinari}, and
  {Pineda}}]{Traficante+18c}
{Traficante} A, {Fuller} GA, {Smith} RJ, et~al. (2018{\natexlab{b}}) {Massive
  70 {\ensuremath{\mu}}m quiet clumps - II. Non-thermal motions driven by
  gravity in massive star formation?} \mnras 473(4):4975--4985

\bibitem[{{Traficante} et~al.(2018{\natexlab{c}}){Traficante}, {Lee},
  {Hennebelle}, {Molinari}, {Kauffmann}, and {Pillai}}]{Traficante+18a}
{Traficante} A, {Lee} YN, {Hennebelle} P, et~al. (2018{\natexlab{c}}) {A
  possible observational bias in the estimation of the virial parameter in
  virialized clumps}. \aap 619:L7

\bibitem[{{Tritsis} and {Tassis}(2016)}]{TritsisTassis16}
{Tritsis} A and {Tassis} K (2016) {Striations in molecular clouds: streamers or
  MHD waves?} \mnras 462:3602--3615

\bibitem[{{Troland} and {Crutcher}(2008)}]{TrolandCrutcher08}
{Troland} TH and {Crutcher} RM (2008) {Magnetic Fields in Dark Cloud Cores:
  Arecibo OH Zeeman Observations}. \apj 680(1):457--465

\bibitem[{{Tsitali} et~al.(2015){Tsitali}, {Belloche}, {Garrod}, {Parise}, and
  {Menten}}]{Tsitali+15}
{Tsitali} AE, {Belloche} A, {Garrod} RT, et~al. (2015) {Star formation in
  Chamaeleon I and III: a molecular line study of the starless core
  population}. \aap 575:A27

\bibitem[{{Utomo} et~al.(2018){Utomo}, {Sun}, {Leroy}, {Kruijssen},
  {Schinnerer}, {Schruba}, {Bigiel}, {Blanc}, {Chevance}, {Emsellem},
  {Herrera}, {Hygate}, {Kreckel}, {Ostriker}, {Pety}, {Querejeta},
  {Rosolowsky}, {Sandstrom}, and {Usero}}]{Utomo+18}
{Utomo} D, {Sun} J, {Leroy} AK, et~al. (2018) {Star Formation Efficiency per
  Free-fall Time in nearby Galaxies}. \apjl 861(2):L18

\bibitem[{{Valdivia} et~al.(2016){Valdivia}, {Hennebelle}, {G{\'e}rin}, and
  {Lesaffre}}]{valdivia2016}
{Valdivia} V, {Hennebelle} P, {G{\'e}rin} M, et~al. (2016) {H$_{2}$
  distribution during the formation of multiphase molecular clouds}. \aap
  587:A76

\bibitem[{{Valdivia} et~al.(2017){Valdivia}, {Godard}, {Hennebelle}, {Gerin},
  {Lesaffre}, and {Le Bourlot}}]{valdivia2017}
{Valdivia} V, {Godard} B, {Hennebelle} P, et~al. (2017) {Origin of CH$^{+}$ in
  diffuse molecular clouds. Warm H$_{2}$ and ion-neutral drift}. \aap 600:A114

\bibitem[{{Vazquez-Semadeni}(1994)}]{VazquezSemadeni94}
{Vazquez-Semadeni} E (1994) {Hierarchical Structure in Nearly Pressureless
  Flows as a Consequence of Self-similar Statistics}. \apj 423:681

\bibitem[{{V{\'a}zquez-Semadeni}(1999)}]{VazquezSemadeni99}
{V{\'a}zquez-Semadeni} E (1999) {Turbulence in Molecular Clouds}. In: {Wall}
  WF, {Carrami{\~n}ana} A, and {Carrasco} L (eds) Millimeter-Wave Astronomy:
  Molecular Chemistry \&amp;amp: Physics in Space, Proceedings of the 1996
  INAOE Summer School of Millimeter-Wave Astronomy held at INAOE, Tonantzintla,
  Puebla, Mexico, 15-31 July 1996. Edited by W. F. Wall, A. Carrami{\~n}ana,
  and L. Carrasco. Kluwer Academic Publishers, 1999., p.161, Astrophysics and
  Space Science Library, vol 241, p 161

\bibitem[{{V{\'a}zquez-Semadeni} and
  {Garc{\'\i}a}(2001)}]{VazquezSemadeniGarcia01}
{V{\'a}zquez-Semadeni} E and {Garc{\'\i}a} N (2001) {The Probability
  Distribution Function of Column Density in Molecular Clouds}. \apj
  557(2):727--735

\bibitem[{{Vazquez-Semadeni} et~al.(1995){Vazquez-Semadeni}, {Passot}, and
  {Pouquet}}]{VazquezSemadeni+95}
{Vazquez-Semadeni} E, {Passot} T, and {Pouquet} A (1995) {A Turbulent Model for
  the Interstellar Medium. I. Threshold Star Formation and Self-Gravity}. \apj
  441:702

\bibitem[{{V{\'a}zquez-Semadeni} et~al.(2006){V{\'a}zquez-Semadeni}, {Ryu},
  {Passot}, {Gonz{\'a}lez}, and {Gazol}}]{vazquez2006}
{V{\'a}zquez-Semadeni} E, {Ryu} D, {Passot} T, et~al. (2006) {Molecular Cloud
  Evolution. I. Molecular Cloud and Thin Cold Neutral Medium Sheet Formation}.
  \apj 643:245--259

\bibitem[{{V{\'a}zquez-Semadeni} et~al.(2007){V{\'a}zquez-Semadeni},
  {G{\'o}mez}, {Jappsen}, {Ballesteros-Paredes}, {Gonz{\'a}lez}, and
  {Klessen}}]{VazquezSemadeni+07}
{V{\'a}zquez-Semadeni} E, {G{\'o}mez} GC, {Jappsen} AK, et~al. (2007)
  {Molecular Cloud Evolution. II. From Cloud Formation to the Early Stages of
  Star Formation in Decaying Conditions}. \apj 657(2):870--883

\bibitem[{{V{\'a}zquez-Semadeni} et~al.(2017){V{\'a}zquez-Semadeni},
  {Gonz{\'a}lez-Samaniego}, and {Col{\'{\i}}n}}]{vazquez2017}
{V{\'a}zquez-Semadeni} E, {Gonz{\'a}lez-Samaniego} A, and {Col{\'{\i}}n} P
  (2017) {Hierarchical star cluster assembly in globally collapsing molecular
  clouds}. \mnras 467:1313--1328

\bibitem[{{V{\'a}zquez-Semadeni} et~al.(2018){V{\'a}zquez-Semadeni},
  {Zamora-Avil{\'e}s}, {Galv{\'a}n-Madrid}, and {Forbrich}}]{vazquez2018}
{V{\'a}zquez-Semadeni} E, {Zamora-Avil{\'e}s} M, {Galv{\'a}n-Madrid} R, et~al.
  (2018) {Molecular cloud evolution - VI. Measuring cloud ages}. \mnras
  479(3):3254--3263

\bibitem[{{V{\'a}zquez-Semadeni}
  et~al.(2019{\natexlab{a}}){V{\'a}zquez-Semadeni}, {Palau},
  {Ballesteros-Paredes}, {G{\'o}mez}, and
  {Zamora-Avil{\'e}s}}]{VazquezSemadeni+19}
{V{\'a}zquez-Semadeni} E, {Palau} A, {Ballesteros-Paredes} J, et~al.
  (2019{\natexlab{a}}) {Global Hierarchical Collapse In Molecular Clouds.
  Towards a Comprehensive Scenario}. \mnras p 2348

\bibitem[{{V{\'a}zquez-Semadeni}
  et~al.(2019{\natexlab{b}}){V{\'a}zquez-Semadeni}, {Palau},
  {Ballesteros-Paredes}, {G{\'o}mez}, and {Zamora-Avil{\'e}s}}]{Vazquez+19}
{V{\'a}zquez-Semadeni} E, {Palau} A, {Ballesteros-Paredes} J, et~al.
  (2019{\natexlab{b}}) {Global hierarchical collapse in molecular clouds.
  Towards a comprehensive scenario}. \mnras 490(3):3061--3097

\bibitem[{{Veltchev} et~al.(2018){Veltchev}, {Ossenkopf-Okada}, {Stanchev},
  {Schneider}, {Donkov}, and {Klessen}}]{Veltchev+18}
{Veltchev} TV, {Ossenkopf-Okada} V, {Stanchev} O, et~al. (2018) {Spatially
  associated clump populations in Rosette from CO and dust maps}. \mnras
  475:2215--2235

\bibitem[{{Walch} and {Naab}(2015)}]{walch2015}
{Walch} S and {Naab} T (2015) {The energy and momentum input of supernova
  explosions in structured and ionized molecular clouds}. \mnras 451:2757--2771

\bibitem[{{Wang} et~al.(2015){Wang}, {Testi}, {Ginsburg}, {Walmsley},
  {Molinari}, and {Schisano}}]{Wang+15}
{Wang} K, {Testi} L, {Ginsburg} A, et~al. (2015) {Large-scale filaments
  associated with Milky Way spiral arms}. \mnras 450:4043--4049

\bibitem[{Ward et~al.(2019)Ward, Chevance, Kruijssen, Hygate, Schruba, and
  Longmore}]{Ward+19}
Ward JL, Chevance M, Kruijssen JMD, et~al. (2019) Towards a multi-tracer
  timeline of star formation in the lmc i: Deriving the lifetimes of
  h\,\textsc{i} clouds. \mnras~submitted

\bibitem[{{Whitworth} et~al.(1996){Whitworth}, {Bhattal}, {Francis}, and
  {Watkins}}]{Whitworth+96}
{Whitworth} AP, {Bhattal} AS, {Francis} N, et~al. (1996) {Star formation and
  the singular isothermal sphere}. \mnras 283(3):1061--1070

\bibitem[{{Williams} et~al.(1994){Williams}, {de Geus}, and
  {Blitz}}]{Williams+94}
{Williams} JP, {de Geus} EJ, and {Blitz} L (1994) {Determining structure in
  molecular clouds}. \apj 428:693--712

\bibitem[{{Wilson} et~al.(1970){Wilson}, {Jefferts}, and {Penzias}}]{Wilson+70}
{Wilson} RW, {Jefferts} KB, and {Penzias} AA (1970) {Carbon Monoxide in the
  Orion Nebula}. \apjl 161:L43

\bibitem[{{Wolfire} et~al.(2003){Wolfire}, {McKee}, {Hollenbach}, and
  {Tielens}}]{wolfire2003}
{Wolfire} MG, {McKee} CF, {Hollenbach} D, et~al. (2003) {Neutral Atomic Phases
  of the Interstellar Medium in the Galaxy}. \apj 587:278--311

\bibitem[{{Wu} et~al.(2010){Wu}, {Evans}, {Shirley}, and {Knez}}]{Wu+10}
{Wu} J, {Evans} I Neal~J, {Shirley} YL, et~al. (2010) {The Properties of
  Massive, Dense Clumps: Mapping Surveys of HCN and CS}. \apjs 188(2):313--357

\bibitem[{{Xu} and {Lazarian}(2020)}]{Xu_Laz20}
{Xu} S and {Lazarian} A (2020) {Turbulence in a Self-gravitating Molecular
  Cloud Core}. \apj 890(2):157

\bibitem[{{Ysard} et~al.(2013){Ysard}, {Abergel}, {Ristorcelli}, {Juvela},
  {Pagani}, {K{\"o}nyves}, {Spencer}, {White}, and {Zavagno}}]{Ysard+13}
{Ysard} N, {Abergel} A, {Ristorcelli} I, et~al. (2013) {Variation in dust
  properties in a dense filament of the Taurus molecular complex (L1506)}. \aap
  559:A133

\bibitem[{{Zamora-Avil{\'e}s} et~al.(2012){Zamora-Avil{\'e}s},
  {V{\'a}zquez-Semadeni}, and {Col{\'\i}n}}]{Zamora2012}
{Zamora-Avil{\'e}s} M, {V{\'a}zquez-Semadeni} E, and {Col{\'\i}n} P (2012) {An
  Evolutionary Model for Collapsing Molecular Clouds and Their Star Formation
  Activity}. \apj 751(1):77

\bibitem[{{Zamora-Avil{\'e}s} et~al.(2017){Zamora-Avil{\'e}s},
  {Ballesteros-Paredes}, and {Hartmann}}]{ZamoraAviles+17}
{Zamora-Avil{\'e}s} M, {Ballesteros-Paredes} J, and {Hartmann} LW (2017) {Are
  fibres in molecular cloud filaments real objects?} \mnras 472(1):647--656

\bibitem[{{Zamora-Avil{\'e}s} et~al.(2018){Zamora-Avil{\'e}s},
  {V{\'a}zquez-Semadeni}, {K{\"o}rtgen}, {Banerjee}, and
  {Hartmann}}]{Zamora-Aviles+18}
{Zamora-Avil{\'e}s} M, {V{\'a}zquez-Semadeni} E, {K{\"o}rtgen} B, et~al. (2018)
  {Magnetic suppression of turbulence and the star formation activity of
  molecular clouds}. \mnras 474(4):4824--4836

\bibitem[{{Zhang} and {Li}(2017)}]{ZhangLi17}
{Zhang} CP and {Li} GX (2017) {Mass-size scaling $M\propto r^1.67$ of massive
  star-forming clumps - evidences of turbulence-regulated gravitational
  collapse}. \mnras 469(2):2286--2291

\bibitem[{{Zhang} et~al.(2018){Zhang}, {Xu}, {Vasyunin}, {Semenov}, {Wang},
  {Dib}, {Liu}, {Liu}, {Zhang}, {Liu}, {Wang}, {Li}, {Wu}, {Yuan}, {Li}, and
  {Gao}}]{Zhang+18}
{Zhang} GY, {Xu} JL, {Vasyunin} AI, et~al. (2018) {Physical properties and
  chemical composition of the cores in the California molecular cloud}. \aap
  620:A163

\bibitem[{{Zhang} et~al.(2020){Zhang}, {Zavagno}, {Yuan}, {Liu}, {Figueira},
  {Russeil}, {Schuller}, {Marsh}, and {Wu}}]{Zhang+20b}
{Zhang} S, {Zavagno} A, {Yuan} J, et~al. (2020) {HII regions and high-mass
  starless clump candidates I: Catalogs and properties}. arXiv e-prints
  arXiv:2003.11433

\end{thebibliography}

\end{document}